\documentclass[a4paper,fleqn,usenatbib,useAMS]{mnras}

\usepackage{graphicx}	
\usepackage{amsmath}	
\usepackage{amssymb}	
\usepackage{bm}		
\usepackage{pdflscape}	
\usepackage[english]{babel}
\usepackage{selinput}
\usepackage{multicol, lipsum, graphicx, caption}
\usepackage[toc]{appendix}
\usepackage{xcolor}
\usepackage[framemethod=tikz]{mdframed}
\usepackage{adjustbox}
\SelectInputMappings{%
  aacute={á},
  ntilde={ñ},
  Euro={€}
}
\usepackage{babel}
\usepackage{stfloats}
\usepackage{caption}

\usepackage[T1]{fontenc}
\usepackage{ae,aecompl}

\usepackage{newtxtext,newtxmath}
\usepackage{zref-xr}
\zexternaldocument{bk_compression}

\title[Power spectrum and bispectrum compression]
{Maximal compression of the redshift space galaxy power spectrum and bispectrum}

\author[D. Gualdi et al.]{Davide Gualdi$^{1}$\thanks{Contact e-mail: \href{davide.gualdi.14@ucl.ac.uk}{davide.gualdi.14@ucl.ac.uk}}, Marc Manera$^{3,2,1}$, 
Benjamin Joachimi$^{1}$,
Ofer Lahav$^{1}$
\\
$^{1}$Department of Physics and Astronomy, University College London, Gower Street, London WC1E 6BT, UK \\
$^{2}$Centre for Mathematical Sciences, DAMTP, Cambridge University, Wilberforce Rd, Cambridge CB3 0WA\\
$^{3}$Institut de Física d’Altes Energies (IFAE), The Barcelona Institute of
Science and Technology, Campus UAB, 08193 Bellaterra (Barcelona) Spain}

\pubyear{2017}

\begin{document}

\label{firstpage}
\pagerange{\pageref{firstpage}--\pageref{lastpage}}
\maketitle

\begin{abstract}
We explore two methods of compressing the redshift space galaxy power spectrum and bispectrum with respect to a chosen set of cosmological parameters. 
Both methods involve reducing the dimension of the original data-vector ( e.g. $\,1000\,$ elements ) to the number of cosmological parameters considered ( e.g. $\,$seven$\,$ ) using the Karhunen-Loève algorithm.
In the first case, we run MCMC sampling on the compressed data-vector in order to recover the one-dimensional (1D) and two-dimensional (2D) posterior distributions.
The second option, approximately  $2000$ times faster, works by orthogonalising the parameter space through diagonalisation of the Fisher information  matrix before the compression, obtaining the posterior distributions without the need of MCMC sampling.
Using these methods for future spectroscopic redshift surveys like DESI,  EUCLID and  PFS would drastically reduce the number of simulations needed to compute accurate covariance matrices with minimal loss of constraining power.
We consider a redshift bin of a DESI-like experiment. Using the power spectrum combined with the bispectrum as a data-vector, both compression methods on average recover the $68\%$ credible regions to within $0.7\%$ and $2\%$ of those resulting from standard MCMC sampling respectively.
These confidence intervals are also smaller than the ones obtained using only the power spectrum by $\left(81\%, 80\%, 82\%\right)$ respectively for the bias parameter $b_1$, the growth rate $f$ and the scalar amplitude parameter $A_{\mathrm{s}}$.

\end{abstract}




\begingroup
\let\clearpage\relax
\endgroup

\begin{multicols*}{2}

\section{Introduction}

In recent years the  number of available large data sets  to be used for cosmological analysis has drastically increased ({\it PLANCK}, \footnote{\url{http://sci.esa.int/planck/}} \citealp{Ade:2013zuv} ; Sloan Digital Sky Survey \footnote{\url{http://www.sdss3.org/surveys/boss.php}}, \citealp{2011AJ....142...72E}; DES, \citealp{2005astro.ph.10346T} \footnote{\url{https://www.darkenergysurvey.org}})
and will do even more so in the near future (DESI\footnote{\url{http://desi.lbl.gov}}, \citealp{Levi:2013gra}; EUCLID \footnote{\url{http://sci.esa.int/euclid/}}, \citealp{2011arXiv1110.3193L}; PFS \footnote{\url{http://pfs.ipmu.jp}}, \citealp{2014PASJ...66R...1T}). Improving current analysis techniques  to extract as much information as possible from these catalogues has  become highly relevant.
Up to now most work has been done using two points statistics (2pt) like the two points correlation function or its Fourier transform, the power spectrum.
However, gravity increases the level of non-linearity in the matter distribution field, creating non-Gaussian features which are not constrained by the sole use of 2pt statistics. 
Higher-order statistics like the three points (3pt) correlation function or its Fourier transform, the bispectrum, have already been studied in the past in order to capture the non-Gaussian part of the statistical information contained in the large scale structure galaxy field. The first measurements of the 3pt correlation function and the bispectrum on a galaxy catalogue were performed by \citet{1975ApJ...196....1P}, \citet{1977ApJ...217..385G} and \citet{1982ApJ...259..474F}. 
\citet{1984ApJ...279..499F} studied the relation between the cosmological and bias parameters, modelling the relation between luminous and dark matter, which affects the amplitude and the shape of the bispectrum.
Integrated 3pt statistics like the skewness were introduced in order to reduce the complexity and number of modelling parameters \citep{1980lssu.book.....P,1994ApJ...429...36F,1994ApJ...433....1B,2013MNRAS.429.1206J}. The modelling of redshift-space distortions into the 3pt statistics was later introduced and studied by \citet{Matarrese:1997sk}, \citet{Verde:1998zr}, \citet{Heavens:1998es}, \citet{Scoccimarro:1997st}, \citet{2000ApJ...544..597S}. Different 
3pt statistics have also been proposed as useful tools to quantify deviations from GR \citep{PhysRevD.79.103506,Bernardeau:2011sf} and to measure primordial non-Gaussianities \citep{1994ApJ...429...36F,1994ApJ...430..447G,doi:10.1046/j.1365-8711.2000.03191.x, 2010AdAst2010E..73L, 2016JCAP...06..014T}.
In the last twenty years, these statistics have received increasing attention especially for the benefit of lifting degeneracies present in 2pt statistics between cosmological and nuisance parameters in datasets produced by spectroscopic redshift surveys like BOSS. The most recent measurement on this dataset for the bispectrum was made by \citet{Gil-Marin:2016wya} and for the 3pt correlation function by \citet{Slepian:2015hca}.

When studying the power spectrum and the bispectrum, it is necessary to subdivide the range of possible $\bm{k}$-vectors in Fourier space into bins. By doing this, one defines the number of possible elements for both the power spectrum and bispectrum data-vectors. While in Fourier space the number of possible elements for the power spectrum data-vector is the same as the number of bins, for the bispectrum it corresponds to the number of triangles that can be formed by triplets of the available $\bm{k}$-vectors (less than $\sim n_{\mathrm{bins}}^3$). 
This difference becomes even larger when redshift-space distortions \citep{1987MNRAS.227....1K}, described in section \ref{sub_sec:rsd_formalism},  are included in the analysis. In addition, as explained in Appendix \ref{sec:est_def}, not only the modulus of the three $\bm{k}$-vectors but also the orientation of the triangle with respect to the line of sight becomes relevant. Consequently, a fine binning in $\bm{k}$-space corresponds to a very large number of possible triangles. While for the power spectrum in redshift space there are $\sim n_{\mathrm{bins}}^2$ possible data-vector elements, for the bispectrum there are slightly less than $\sim n_{\mathrm{bins}}^5$ possible triplets since the vectors need to satisfy the triangle condition.

The problem is that when doing parameter estimation using a data-vector for a given statistic, the corresponding covariance matrix is needed.
This can either be computed analytically, which requires the evaluation of several multidimensional integral expressions, or it can be estimated from simulations. However, in order to obtain a precise and accurate estimate of the covariance using numerical simulations, the number of realisations must be larger than the number of elements of the data-vector \citep{2007A&A...464..399H,2014MNRAS.442.2728T}.
Therefore, numerically estimating the covariance matrix of a 3pt statistic from simulations could become very expensive in terms of simulations required to predict it accurately.
Indeed, one usually sacrifices the stronger constraining power achievable by considering a larger number of triangle configurations for a more accurate estimate of the covariance matrix obtained using only a subset of all the possible triangles.
Compressing the original data-vector is an efficient method to avoid losing access to part of the constraining power contained in the 3pt statistics.
Alternatives to the full bispectrum have been proposed in the last few years. For example \citet{2015PhRvD..91d3530S} substituted to the tree level matter bispectrum, functions of the cross-power spectrum between $\delta^2(x)$ and $\delta$.
The performances of some of them, including the modal decomposition proposed by \citet{2013PhRvD..88f3512S}, have been recently studied by \citet{Byun:2017fkz}. We will compare their results with ours in the conclusions.


Therefore one can reduce the original number of data-vector elements either by performing a selection of triangles based on some criteria or by compressing the data-vector. This paper is about this second option, where we present an application of a linear compression mechanism by Karhunen and Lo\`eve (K-L) used in \citet{Tegmark:1996bz} for the first time in a cosmological context to compress with respect to multiple parameters. The underlying principle is to assign a vector, such that the weights are proportional to the sensitivity of each element to the variation of a model parameter.
When the covariance matrix can be assumed to be parameter independent, the dimension of the compressed data-vector corresponds to the number of considered model parameters.

 \begin{eqnarray}
\notag
\end{eqnarray}

\noindent
\begin{minipage}[c]{\linewidth}
\centering
\label{fig:2_meths}
\includegraphics[width=1.0\linewidth]{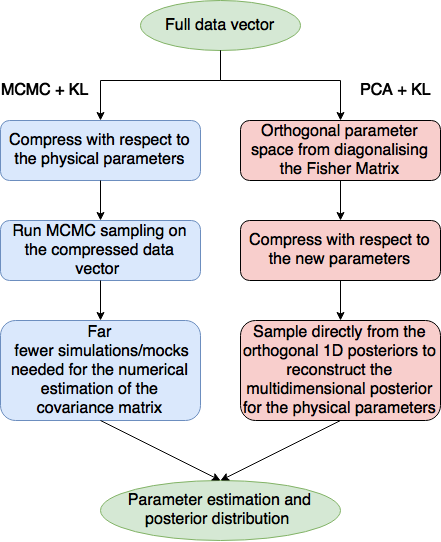}
\captionof{figure}{Diagram highlighting the two compression methods presented in this work, including the respective advantages with respect to the uncompressed data-vector.}
\end{minipage}

\begin{eqnarray}
\notag
\end{eqnarray}

We use this prescription in two ways to do parameter inference, which are summarised in Figure \ref{fig:2_meths}. In our first method, we run an MCMC sampling for the compressed data-vector (hereafter MCMC + KL). This has the immediate and appealing benefit for a cosmological survey of requiring fewer simulations to estimate precisely the covariance  matrix in the case of a long data-vector like the bispectrum (with $ \gtrsim 10^3$ triangles) to just $1$ data-vector element for each cosmological parameter. Indeed, in order to have a reliable estimate of the covariance  matrix, one needs a significantly larger number of mocks than of data-vector elements. For example, in the DR12 BOSS paper studying the bispectrum \citep{Gil-Marin:2016wya}, the number of triangles used was limited to 825 out of the 6391 possible ones, since only 2048 mocks were available \citep{Kitaura:2015uqa}. 

In our second method, before compressing, we orthogonalise the parameter space by diagonalising the Fisher information  matrix (principal component analysis, hereafter PCA + KL). This enables sampling from 1D posterior distributions of the new set of parameters, to recover the full multi-dimensional posterior distribution for the original physical set of parameters without the need of an MCMC sampling. This proves to be $\sim10^3$ times faster than an MCMC run, in particular because the linear matter power spectrum needs to be recomputed far fewer times. 

In addition to being competitive with the standard likelihood approach, this method could be also very useful for future cosmological surveys in order to test and forecast the constraining power of the expected data set for several different combinations of cosmological and nuisance (e.g. bias) parameters.
We apply our analysis to the redshift space galaxy bispectrum and to the joint data-vector formed by the galaxy power spectrum and bispectrum.

The paper is structured as follows: section \ref{pert_theory} describes the basis of perturbation theory upon which the data-vector estimators and covariance  matrix terms are computed. In section \ref{cov_estimators} we present the covariance terms. Section \ref{analysis_setup} contains the specifics of the analysis performed. In section \ref{compress_formalism} the compression formalism is presented. In section \ref{mcmc_comp_dv} we present the results of applying the MCMC + KL method. Section \ref{pca_kl} describes the performance of the PCA + KL compression in obtaining constraints on the cosmological parameters. In section \ref{sec:joint_dv} we comment on the added value given by jointly using the power spectrum to the bispectrum. We conclude in section \ref{conclusions}. All detailed derivations are deferred to the appendices \ref{sec:png} (redshift space kernels), \ref{sec:est_def} (estimator definition),  \ref{sec:cov_terms} (covariance matrix terms), \ref{sec:comp_cov} (compressed covariance formalism), \ref{sec:weights_ort} (weights orthogonalisation).

\section{Perturbation Theory with redshift-space distortions} \label{pert_theory}

\subsection{Bias model}
Since luminous objects like galaxies are not exact tracers of the underlying dark matter distribution, it is necessary to model their relationship.
This is commonly referred to as `galaxy biasing'; the relationship could be linear/non-linear, deterministic/stochastic, local/non-local,  and a function of scale and cosmic time. 
Understanding biasing is important in its own right  as a probe of galaxy formation and evolution.
Galaxy biasing was recognised when it was noticed that different populations of galaxies (e.g. spirals, ellipticals) have different clustering strengths.

A physical mechanism for galaxy biasing was suggested by \citet{Kaiser1984} and developed by \citet{Bardeen1986}, namely that galaxies would tend to form in peaks in the matter density distribution thus being more clustered than the underlying matter distribution.  In this model, more massive (and thus rarer) tracers are naturally more highly biased. Biasing can be non-linear \citep{McDonald2009} and stochastic \citep{Denkel1999}.  
Galaxy biasing also evolves with redshift (\citealp{Clerkin2015} and references therein), being larger at higher redshift.
It is also scale dependent at small physical scales where the non-linear effects of galaxy formation
are important, although there is almost no scale dependence above $20-40$ Mpc $h^{-1}$
\citep{2011MNRAS.415..383M,Crocce2013}.  Another popular approach, known as the Halo Model,  is to parametrise the relationship
between galaxies and the dark matter distribution by assuming that all galaxies reside in dark matter haloes or
sub-haloes, discussed e.g. in \citealp{Tinker2010} and  references therein. 
We note that biasing and the relationship between galaxies and haloes can also be constrained through combinations of various observables, e.g. galaxy positions and weak gravitational lensing.

Here we adopt a biasing model which is a Taylor expansion of the galaxy density fluctuations $\delta_\mathrm{g}$  in terms of the matter fluctuations $\delta_\mathrm{m}$ \citep{Fry:1992vr}:

\begin{eqnarray}
\delta_{\mathrm{\mathrm{g}}} = \sum_{i=0}^\infty \dfrac{b_i}{i!}\,\delta_{i,\mathrm{m}}^{\mathrm{n.l.}}.
\end{eqnarray}

\noindent where $\delta_{i,\mathrm{m}}^{\mathrm{n.l.}}$ is the non-linear matter density fluctuation.

In this work, only the first two terms of the above expansion are considered, $b_1$ and $b_2$.
The above bias model considers only the local relationship between $\delta_{\mathrm{g}}$ and $\delta_{\mathrm{m}}$; non-local bias terms may be included in a more accurate modelling
\citep{Chan:2012jj,Baldauf:2012hs,Bel:2015jla}.

\subsection{Redshift space formalism}
\label{sub_sec:rsd_formalism}
The conversion of the galaxy redshifts from surveys like BOSS or DESI to proper distances is a cosmological model-dependent operation.
In addition, the local gravitational field influences the peculiar velocities of galaxies producing redshift-space distortions \citep{1987MNRAS.227....1K} which affect the observed redshift.
\noindent \citet{Heavens:1998es} have shown how to express the Fourier transform of the redshift space galaxy overdensity fluctuation, measured in cosmological surveys, $\delta^{\mathrm{\mathrm{s}}}_{\mathrm{g}}\left(\bm{k}\right)$  in relation to the real space linear matter fluctuation $\delta_{\mathrm{m}}\left(\bm{k}\right)$

\begin{align}
\label{delta_s}
\delta^{\mathrm{\mathrm{s}}}_{\mathrm{g}}\left[\bm{k}\right]\,&=\,\mathrm{F}^{(1)}_{\mathrm{\mathrm{s}}}\left[\bm{k}\right]\delta_{\mathrm{m}}\left(\bm{k}\right) \notag \\
&+\,\dfrac{1}{(2\pi)^3}\,\int\,d^3k_1d^3k_2\,\delta_{\mathrm{D}}\left(\bm{k}-\bm{k}_2-\bm{k}_1\right)
\notag \\
&\times\,\mathrm{F}^{(2)}_{\mathrm{\mathrm{s}}}\left[\bm{k}_1,\bm{k}_2\right]\,\delta_{\mathrm{m}}\left(\bm{k}_1\right)\delta_{\mathrm{m}}\left(\bm{k}_2\right)\notag\\
&+\,\dfrac{1}{(2\pi)^6}\,\int\,d^3k_1d^3k_2d^3k_3\,\delta_{\mathrm{D}}\left(\bm{k}-\bm{k}_3-\bm{k}_2-\bm{k}_1\right) \notag \\
&\times \mathrm{F}^{(3)}_{\mathrm{\mathrm{s}}}\left[\bm{k}_1,\bm{k}_2,\bm{k}_3\right]\,\delta_{\mathrm{m}}\left(\bm{k}_1\right)\delta_{\mathrm{m}}\left(\bm{k}_2\right)\delta_{\mathrm{m}}\left(\bm{k}_2\right) , 
\end{align}

\noindent where the redshift-space distortion kernels $\mathrm{\mathrm{F}^{(1,2,3)}_{\mathrm{\mathrm{s}}}}$ are given in the Appendix \ref{sec:png}. In this work all $\bm{k}$-vectors are described in terms of their components parallel $k^\parallel$ and perperdicular  $k^\perp$ to the line of sight. We define $\mu = k_i^\parallel/k_i$.  The galaxy redshift space power spectrum is defined as

\begin{eqnarray}
\label{rsd_pk}
\langle\delta^{\mathrm{\mathrm{s}}}_{\mathrm{g}}\left(\bm{k_1}\right)\delta^{\mathrm{\mathrm{s}}}_{\mathrm{g}}\left(\bm{k_2}\right)\rangle\,=\,\left(2\pi\right)^3\,\mathrm{P}^{\mathrm{\mathrm{s}}}_{\mathrm{\mathrm{g}}}\left(\bm{k}_1\right)\delta_{\mathrm{D}}\left(\bm{k}_1\,+\,\bm{k}_2\right) . 
\end{eqnarray}

\noindent Substituting Eq. \eqref{delta_s} into Eq. \eqref{rsd_pk} and applying Wick's theorem assuming that the initial perturbations are Gaussian the power spectrum is given by:

\begin{align}
&\mathrm{P}^{\mathrm{\mathrm{s}}}_{\mathrm{\mathrm{g}}}\left(\bm{k}\right)\,\equiv\, \mathrm{P}^{\mathrm{s},(1)}_{\mathrm{g}}\,+\,\mathrm{P}^{\mathrm{s},(2)}_{\mathrm{g}}\,+\,\mathrm{P}^{\mathrm{s},(3)}_{\mathrm{g}}
\notag \\
&=\left(b_1 + f\mu^2\right)^2\mathrm{P}^{\mathrm{m}}\left( k \right)
\notag \\
&+2\int \dfrac{d^3q}{\left(2\pi\right)^3}\mathrm{P}^{\mathrm{m}}\left(q\right)\mathrm{P}^{\mathrm{m}}\left(|\bm{k}-\bm{q}|\right)\left(\mathrm{F}^{(2)}_{\mathrm{\mathrm{s}}}\left[\bm{q},\bm{k}-\bm{q}\right]\right)^2\notag\\
&+\,6\left(b_1\,+\,f\mu^2\right)\mathrm{P}^{\mathrm{m}}(k)\,\int\, \dfrac{d^3q}{\left(2\pi\right)^3}\mathrm{P}^{\mathrm{m}}\left(q\right)\mathrm{F}^{(3)}_{\mathrm{\mathrm{s}}}\left[\bm{q},-\bm{q},\bm{k}\right],
\end{align}

\noindent
where the growth rate is defined and parametrised as $f(z)\equiv d \ln D/ d\ln a\backsimeq \Omega_{\mathrm{m}}^{\gamma}(z)$  where $\Omega_{\mathrm{m}}(z)$ is the mass density parameter function at a given redshift z, the growth index $\gamma \simeq 0.55$ for a standard cosmology \citep{1980lssu.book.....P,Lahav:1991wc,Linder:2005in}, $D(a)$ is the growing mode of the amplitude fluctuation, and  scale factor $a = (1+z)^{-1}$ . $\mathrm{P}^{\mathrm{m}}(k)$ is the linear matter power spectrum defined analogously to Eq. \eqref{rsd_pk}. In this work when considering the power spectrum data-vector we stop at tree level using only $\mathrm{P}^{\mathrm{s}}_{\mathrm{g}}=\mathrm{P}^{\mathrm{s},(1)}_{\mathrm{g}}$. This choice is consistent with previous analyses \citep{Scoccimarro:1999ed,Sefusatti:2006pa,Song:2015gca, Gagrani:2016rfy}. The error arising from excluding the 1loop terms \citep{1994ApJ...431..495J,2002PhR...367....1B,2008PhRvD..78l3534T,2016PhRvD..93h3517L} is less than $10\%$ up to $k_{\mathrm{max}} = 0.2 h \mathrm{Mpc}^{-1}$ increasing to $30\%$ for $k_{\mathrm{max}} = 0.3 h \mathrm{Mpc}^{-1}$. An accurate modelling of the redshift space galaxy power spectrum was introduced by \citet{2010PhRvD..82f3522T}. This translates approximately into an error on the diagonal of the power spectrum covariance matrix computed in section \ref{cov_pk} of $\sim 1\%$ up to $k_{\mathrm{max}} = 0.2 h \mathrm{Mpc}^{-1}$ and $\sim 9\%$ for $k_{\mathrm{max}} = 0.3 h \mathrm{Mpc}^{-1}$ when including cosmic variance terms.  For brevity, sometimes we write $\mathrm{P}^{\mathrm{s}}_{\mathrm{g}}\left(k_i^\perp,k_i^\parallel\right) \equiv\mathrm{P}^{\mathrm{s}}_{\mathrm{g}}\left(\bm{k}_i\right)$, keeping in mind that the galaxy power spectrum depends on the parallel and perpendicular components of the wave-vector. Therefore $\mathrm{P}^{\mathrm{s}}_{\mathrm{g}}$ will not depend on the second order bias parameter $b_2$.

Analogously, the bispectrum is defined as \citep{1984ApJ...279..499F}:

\begin{align}
\langle\delta^{\mathrm{\mathrm{s}}}_{\mathrm{g}}\left(\bm{k_1}\right)\delta^{\mathrm{\mathrm{s}}}_{\mathrm{g}}\left(\bm{k_2}\right)\delta^{\mathrm{\mathrm{s}}}_{\mathrm{g}}\left(\bm{k_3}\right)\rangle&=\left(2\pi\right)^3\mathrm{B}^{\mathrm{\mathrm{s}}}_{\mathrm{\mathrm{g}}}\left(\bm{k}_1,\bm{k}_2,\bm{k}_3\right)
\notag\\
&\times\,\delta_{\mathrm{D}}\left(\bm{k}_1+\bm{k}_2+\bm{k}_3\right).
\end{align}

\noindent The relation to the linear matter density power spectrum is at first order:

\begin{align}
\mathrm{B}^{\mathrm{\mathrm{s}}}_{\mathrm{\mathrm{g}}}\left(\bm{k}_1,\bm{k}_2,\bm{k}_3\right)\,&=\,2\mathrm{P}^{\mathrm{m}}\left(k_1\right)\mathrm{P}^{\mathrm{m}}\left(k_2\right)\mathrm{F}^{(1)}_{\mathrm{\mathrm{s}}}\left[\bm{k}_1\right]\mathrm{F}^{(1)}_{\mathrm{\mathrm{s}}}\left[\bm{k}_2\right]\mathrm{F}^{(2)}_{\mathrm{\mathrm{s}}}\left[\bm{k}_1,\bm{k}_2\right]\notag \\
&+\,\mathrm{two\,cyclic\, terms}.
\end{align}

\noindent For the redshift space galaxy bispectrum, the most recent and accurate models (up to the mild non-linear regime) were introduced by \citet{2017arXiv170502574H} and \citet{2017arXiv170509306B}.

\section{Covariance}\label{cov_estimators}

In this section we summarise the covariance terms at tree level for the power spectrum and the bispectrum, including also the cross covariance between the two. Here are written only the final analytical expressions, while we show the full derivations in Appendix \ref{sec:cov_terms}.

\subsection{$\mathrm{\textbf{C}^{\mathrm{PP}}}$ : power spectrum covariance matrix}\label{cov_pk}
The power spectrum covariance matrix is given by two terms,

\begin{eqnarray}
\label{pp_cov}
\mathrm{C}^{\mathrm{P}^{\mathrm{\mathrm{s}}}_{\mathrm{\mathrm{g}}}} = 
\mathrm{C^{\mathrm{PP}}_{P}} + \mathrm{C^{\mathrm{PP}}_{T}},
\end{eqnarray}

\noindent where $\mathrm{C^{\mathrm{PP}}_{P}}$ is proportional to the square of the power spectrum and $\mathrm{C^{\mathrm{PP}}_{T}}$ is proportional to the trispectrum.
The first term is given by:

\begin{align}
\mathrm{C^{\mathrm{PP}}_{P}} \left(\bar{\bm{k}}_1; \bar{\bm{k}}_3\right)&=  \dfrac{1}{V_{\mathrm{\mathrm{s}}}^2\,V^{\mathrm{c}}_1\,V^{\mathrm{c}}_3}\,\prod_{i=0}^4\int dV^{\mathrm{c}}_i\, \delta_{\mathrm{D}}\left(\bm{k}_1+\bm{k}_2\right)\delta_{\mathrm{D}}\left(\bm{k}_3+\bm{k}_4\right)
\notag\\
&\times\langle\delta^{\mathrm{\mathrm{s}}}_{\mathrm{\mathrm{g}}}\left(\bm{k}_1\right)\delta^{\mathrm{\mathrm{s}}}_{\mathrm{\mathrm{g}}}\left(\bm{k}_3\right)\rangle\langle\delta^{\mathrm{\mathrm{s}}}_{\mathrm{\mathrm{g}}}\left(\bm{k}_2\right)\delta^{\mathrm{\mathrm{s}}}_{\mathrm{\mathrm{g}}}\left(\bm{k}_4\right)\rangle\,\,+\,1\mathrm{p.}
\notag \\
& \approx \dfrac{2\pi}{V_{\mathrm{\mathrm{s}}} \bar{k}_1^\perp \Delta k^2}\mathrm{D}^\mathrm{P}_{1234} \, \mathrm{P}^{\mathrm{\mathrm{s}}}_{\mathrm{\mathrm{g}}}\left(\bar{\bm{k}}_1\right)^2
\,=\,\dfrac{4\pi}{V_{\mathrm{\mathrm{s}}} \bar{k}_1^\perp \Delta k^2}\delta^{\mathrm{K}}_{13} \, \mathrm{P}^{\mathrm{\mathrm{s}}}_{\mathrm{\mathrm{g}}}\left(\bar{\bm{k}}_1\right)^2,
\end{align}

\noindent where the integrals are over a cylindrical shell $V^{\mathrm{c}}_i$ centered at each $\bar{\bm{k}}_i$ with integral limits on: $\bar{k}_i^\parallel - \Delta k^\parallel/2\leq k_i^\parallel\leq\bar{k}_i^\parallel + \Delta k^\parallel/2$, $\bar{k}_i^\perp - \Delta k^\perp/2\leq k_i^\perp\leq\bar{k}_i^\perp + \Delta k^\perp/2$ and $0 \leq \phi\leq 2\pi$. From the definition of the power spectrum estimator given in Appendix \ref{sec:est_def}, the cylindrical bins for $\bm{k}_2$ and $\bm{k}_4$ are centered respectively on $\bm{\bar{k}}_1$ and $\bm{\bar{k}}_3$.
When a Dirac delta $\delta_\mathrm{D}\left(\bm{k}_i+\bm{k}_j\right)$ is used to simplify one of the integrals over the cylindrical shells, it produces a Kronecker delta $\delta^{\mathrm{K}}_{ij}$ which is equal to one and not to zero only when $k^\parallel_i=k^\parallel_j$ and $k^\perp_i=k^\perp_j$.

We have defined a combination of Kronecker deltas $\mathrm{D}^\mathrm{P}_{1234} = \delta^{\mathrm{K}}_{13}\delta^{\mathrm{K}}_{24} + \delta^{\mathrm{K}}_{14}\delta^{\mathrm{K}}_{23} $ in order to take into account the additional permutation,  which, given the initial  Dirac deltas conditions, reduces to  $\mathrm{D}^\mathrm{P}_{1234} = 2 \times \delta^{\mathrm{K}}_{13} $. $V_{\mathrm{\mathrm{s}}}$ is the survey volume.
The second term in Eq.\ref{pp_cov} is proportional to the trispectrum and is given by

\begin{align}
\mathrm{C^{\mathrm{PP}}_{T}}\left(\bar{\bm{k}}_1; \bar{\bm{k}}_3\right) \approx \dfrac{1}{2\pi \, V_{\mathrm{\mathrm{s}}}}\,\int_0^{2\pi} d\phi_{13}\,
\mathrm{T}^{\mathrm{\mathrm{s}}}_{\mathrm{\mathrm{g}}}\left(\bar{k}^\perp_1, \bar{k}^\parallel_1,\bar{k}^\perp_3,\bar{k}^\parallel_3, \phi_{13}\right).
\end{align}

\noindent where $\phi_{13}$ is the difference between the azimuthal angles of $\bar{\bm{k}}_1$ and $\bar{\bm{k}}_3$. The full expanded expression is written in Appendix \ref{sec:cov_terms}, while the trispectrum definition is in Appendix \ref{sec:est_def}.

\subsection{$\mathrm{\textbf{C}^{\mathrm{BB}}}$: bispectrum covariance matrix}

For the bispectrum covariance  matrix, at leading order, only the diagonal term proportional to the product of three power spectra is required:

\begin{align}
\mathrm{C^{\mathrm{BB}}_{P}}\left(\bar{\bm{k}}_1,\bar{\bm{k}}_2, \bar{\bm{k}}_3;\bar{\bm{k}}_4,\bar{\bm{k}}_5, \bar{\bm{k}}_6\right)
&=\dfrac{(2\pi)^{5}\Delta k^\parallel_3}{V_{\mathrm{\mathrm{s}}}\,\bar{k}^\perp_1\bar{k}^\perp_2\bar{k}^\perp_3\Delta k^6}\Lambda^{-1}_{123}\,
\notag \\
&\times\mathrm{D}_{123456}\prod_{i=1}^3\mathrm{P}^{\mathrm{\mathrm{s}}}_{\mathrm{\mathrm{g}}}\left(\bar{\bm{k}}_i\right).
\end{align}

\noindent where $\Lambda$ is a function defined in Appendix \ref{sec:est_def} which is related to the fraction of wave-vector triplets allowed by the triangle condition such that the bispectrum estimator \eqref{bk_rsd_est} is unbiased.
$\mathrm{D}_{123456}$ is a shorthand notation for the sum of all the possible permutations of pairings of $\bm{k}$-vectors between the first and second triplets, encoded in Kronecker deltas, e.g. $\delta^{\mathrm{K}}_{14}\,\delta^{\mathrm{K}}_{25}\,\delta^{\mathrm{K}}_{36}$. This is the symmetry factor which is equal respectively to $1$, $2$ and $6$ in the case of  scalene, isosceles and equilateral triangles. Full computations can be found in Appendix \ref{sec:cov_terms}.

\subsection{$\mathrm{\textbf{C}^{\mathrm{BP}}}$: cross-variance matrix}
The cross-variance term is also given by the sum of two parts. The first part is proportional to the product between the power spectrum and bispectrum. The second part is proportional to the tetraspectrum

\begin{eqnarray}
    \mathrm{C}^{\mathrm{BP}} = \mathrm{C}^{\mathrm{BP}}_{\mathrm{m1}} + \mathrm{C}^{\mathrm{BP}}_{\mathrm{Te}}.
\end{eqnarray}

\noindent The expression of the first term as derived in Appendix \ref{sec:cov_terms} is:

\begin{align}
\mathrm{C^{\mathrm{BP}}_{m1}}\left(\bar{\bm{k}}_1;\bar{\bm{k}}_3, \bar{\bm{k}}_4,\bar{\bm{k}}_5\right)& =2\times  \dfrac{(2\pi)^2}{V_{\mathrm{s}}\,V_1^{\rm{c}}}\Lambda^{-1}_{345}\,\left(\delta^{\mathrm{K}}_{13} + \delta^{\mathrm{K}}_{14} + \delta^{\mathrm{K}}_{15}\right)\,
\notag\\ 
&\times \mathrm{P}^{\mathrm{\mathrm{s}}}_{\mathrm{\mathrm{g}}}\left(\bar{\bm{k}}_1\right)\mathrm{B}^{\mathrm{\mathrm{s}}}_{\mathrm{\mathrm{g}}}\left(\bar{\bm{k}}_3,\bar{\bm{k}}_4,\bar{\bm{k}}_5\right),
\end{align}

\noindent Where the semicolon separates the wave vector relative to the power spectrum from the ones of the bispectrum.

The last non-vanishing term is the one proportional to the 
tetraspectrum $\mathrm{Te}^{\mathrm{\mathrm{g}}}_{\mathrm{s}}$ given by the connected part of the five-points correlator

\begin{align}
\mathrm{C^{\mathrm{BP}}_{Te}}\left(\bar{\bm{k}}_1;\bar{\bm{k}}_3, \bar{\bm{k}}_4,\bar{\bm{k}}_5\right) &=
\dfrac{(2\pi)^2\Delta k_5^\parallel}{V_{\mathrm{\mathrm{s}}}}\Lambda^{-1}_{345}\,\prod_{i=1,3,4,5}\int \dfrac{dV^{\mathrm{c}}_i}{V^{\mathrm{c}}_i}
\notag \\
&\times
\,\delta_{\mathrm{D}}\left(\bm{k}_3 + \bm{k}_4 + \bm{k}_5\right)\mathrm{T}^{\mathrm{\mathrm{s}}}_{e,g}\left(\bm{k}_1,\bm{k}_3,\bm{k}_4,\bm{k}_5\right).
\end{align}

\noindent The definition and analytical expression for the tetraspectrum are in Appendix \ref{sec:est_def}. In the Appendix it is shown that at leading order the tetraspectrum is proportional to the fourth power of the linear matter power spectrum. Being this one order higher than all the terms considered in the paper, $\mathrm{C^{\mathrm{BP}}_{Te}}$ has not been included in the numerical computations.

All the terms rederived above agree with the ones derived in the literature when considering redshift-space distortions \citep{Sefusatti:2006pa, Scoccimarro:1997st}.

\subsection{Shot noise contribution}

All terms of the covariance  matrix have been corrected in order to account for the shot noise, which is given by the average number density of galaxies. In our analysis we consider a DESI-like ELGs sample with  $\bar{n}_\mathrm{\mathrm{g}} \approx 1.175 \times 10^{-3} h^3\mathrm{Mpc}^{-3}$ and a survey volume $V_{\mathrm{\mathrm{s}}} \approx 12.3 h^{-3}\mathrm{Gpc}^3$ \citep{Levi:2013gra,Song:2015gca, 2016arXiv161100036D}. The only correction corresponds to substituting the galaxy power spectrum with:

\begin{eqnarray}
\mathrm{P}^{\mathrm{\mathrm{s}}}_{\mathrm{\mathrm{g}}}\left(\bm{k}\right) \longrightarrow \mathrm{P}^{\mathrm{\mathrm{s}}}_{\mathrm{\mathrm{g}}}\left(\bm{k}\right) + \dfrac{1}{\bar{n}_\mathrm{\mathrm{g}} } . 
\end{eqnarray}

\noindent We made the assumption that the shot noise is well approximated by a Gaussian distribution (which is reasonable if the galaxy number density is fairly high). In that case only second moments exist, and thus contributions appear only in the 2pt correlators, and those yield power spectrum contributions to the cosmic variance. If the noise were Poisson-distributed, the corrected expressions would become much more complicated \citep{1997MNRAS.290..651M}.

\section{analysis setup} \label{analysis_setup}

In this section we describe the pipeline and codes for the compression and the MCMC analyses.
We compressed the power spectrum and the bispectrum with respect to the set of parameters: $\{b_1,\,b_2,\,f,\,\Omega_{\mathrm{m}},\,\Omega_{\mathrm{b}},\,A_{\mathrm{\mathrm{s}}},\,n_{\mathrm{\mathrm{s}}}\}$ where the first two are galaxy bias parameters, $f$ is the linear growth rate, $\Omega_{\mathrm{m}}$ and $\Omega_{\mathrm{b}}$ are the density parameters of total matter and baryonic matter, $A_\mathrm{s}$ is the scalar amplitude of the primordial perturbations and $n_{\mathrm{s}}$ is the scalar spectral index.
For both cases the fiducial cosmology was fixed to $b_1 =1.90$, $b_2=0.20$, $f =\Omega_{\mathrm{m}}(z=0.81)^{0.55}$, $\Omega_{\mathrm{m}}(z=0) = 0.307$, $\Omega_{\mathrm{b}}(z=0) = 0.0482$ and $A_{\mathrm{\mathrm{s}}}=2.9\times 10^{-9}$ similar to the one used in \citet{Kitaura:2015uqa} and close to the one reported in the {\it PLANCK} 15 results \citep{Ade:2015xua}. The redshift used is that of the effective redshift-bin ($0.6<z<1.0$) of a ELG sample of a DESI-like survey with $z_{\mathrm{eff}} = 0.81$.

The $k$-range chosen is $0.01 < k < 0.2\, \mathrm{Mpc}^{-1}\,h$, $12$ bins for both parallel (linear binning) and perpendincular (logarithmic binning) to the line of sight components. 
We have adopted a logarithmic binning for the perpendicular components of the wave-vectors in order to better capture the different features at different scales. A linear binnings has been used for parallel to the line of sight. 
For the perpendicular component, $\Delta \log_{10}k^\perp = 0.11827$ while for the parallel one $\Delta k^\parallel = 0.0182 \, \mathrm{Mpc}^{-1}\,h$. 
With these settings, the resulting number of configurations satisfying the triangle condition is $\mathrm{N}_{\mathrm{tr}} = 1333$ for the bispectrum and $\mathrm{N}_{\mathrm{pairs}} = 132$ combinations of parallel and perpendicular components for the power spectrum .

The triangle configurations for the bispectrum are generated by a five dimensional loop choosing first the three perpendicular components of the sides of the triangle and secondly two of the parallel ones. The third parallel component is chosen such that the final triangle satisfies the triangle condition. All sides and projections must be in the range given above.

For the MCMC sampling we have used 64 samplers together with the same number of Xeon E5-2650 processors connected through MPI \citep{gabriel04:_open_mpi}, each with 2000 burn-in steps followed by 10000 steps for the actual posterior sampling.
The Python package Emcee was used as MCMC sampler \citep{2013PASP..125..306F}.
Uninformative flat priors have been used both in the compression code and MCMC sampler. 
We employed the CAMB code \citep{Lewis:1999bs} in order to generate the linear matter power spectrum for different cosmological parameters.

\section{Compression Formalism}\label{compress_formalism}


\subsection{Fisher information  matrix}

The log-likelihood $\mathcal{L}$ for a Gaussian probability distribution relative to an $n$-dimensional data-vector $\bm{x}$, can be written as

\begin{align}
-2\mathcal{L} \,&=\,n\ln2\pi\,+\,\ln\det\mathrm{\textbf{Cov}} \,+\,\left(\bm{x}\,-\,\langle\bm{x}\rangle\right)^\intercal\mathrm{\textbf{Cov}}^{-1}\left(\bm{x}\,-\,\langle\bm{x}\rangle\right),
\end{align}

\noindent where $\mathrm{\textbf{Cov}}\,=\,\langle\left(\bm{x}\,-\,\langle\bm{x}\rangle\right)\left(\bm{x}\,-\,\langle\bm{x}\rangle\right)^\intercal\rangle$ is the covariance  matrix and  $\langle\bm{x}\rangle$ is the mean of the data-vector.
\noindent From this quantity the Fisher information matrix can be defined as

\begin{eqnarray}
\mathrm{F}_{ij}\,=\,-\bigg\langle \dfrac{\partial^2\mathcal{L}}{\partial\theta_i\partial\theta_j}\bigg\rangle\bigg\rvert_{\bm{\theta}_{ML} }\,\equiv\,-\langle\mathcal{L}_{,ij}\rangle,
\end{eqnarray}

\noindent which is a measure of the curvature around the maximum likelihood point $\bm{\theta}_{ML} = (\theta^1_{ML},\theta^2_{ML},....\theta^m_{ML})$ where the $\theta$'s are the $m$ model parameters and the comma notation indicates the derivatives with respect to them.
In the case of a Gaussian likelihood, the Fisher matrix can be expressed as

\begin{eqnarray}
\mathrm{F}_{ij} =  \dfrac{1}{2}\mathrm{Tr}\left[\bm{\mathrm{A}}_i\bm{\mathrm{A}}_j\,+\,\mathrm{\textbf{Cov}}^{-1}\bm{\mathrm{M}}_{ij}\right],
\end{eqnarray}

\noindent where the matrices $\bm{\mathrm{A}}_i$ and $\bm{\mathrm{M}}_{ij}$ are defined as $\bm{\mathrm{A}}_i\equiv \mathrm{\textbf{Cov}}^{-1}\mathrm{\textbf{Cov}}_{,i}$ and $\bm{\mathrm{M}}_{ij}\equiv \langle\bm{x}\rangle_{,i}\langle\bm{x}\rangle_{,j}^\intercal + \langle\bm{x}\rangle_{,i}^\intercal\langle\bm{x}\rangle_{,j}$.

The diagonal entries of the Fisher  matrix are related to the minimum error attainable in estimating a parameter $\theta_i$. In particular, in the case of single parameter estimation the minimum attainable error is $\Delta\theta_i^{\mathrm{min}} = 1/(\mathrm{F}_{ii})^{\frac{1}{2}}$. When more than one parameter is considered, the full Fisher  matrix is needed to compute the minimum marginalised error for each parameter, which is given by $\Delta\theta_i^{\mathrm{min}} = (\mathrm{F}_{ii}^{-1})^{\frac{1}{2}}$.
The target of the compression is to obtain a new data vector such that for each parameter $\theta_i$, $\Delta\theta_i^{\mathrm{min}}$ is minimised. In other words, the compression algorithm will be obtained by maximising $\mathrm{F}_{ii}$ for each parameter $\theta_i$.

\begin{figure*}
\centering
\includegraphics[width=\textwidth]{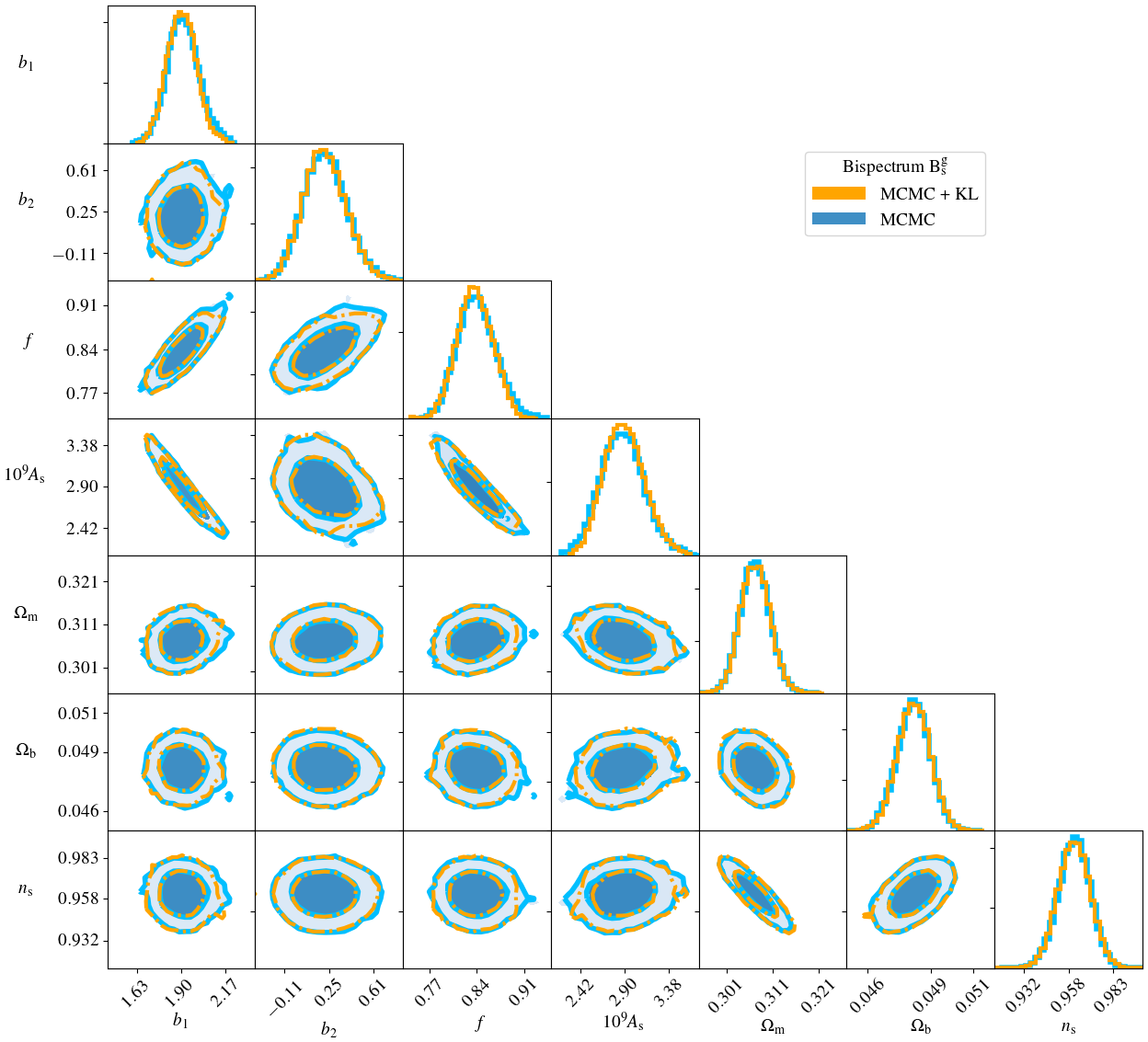}
\caption{ Bispectrum case: 1D and 2D posterior distributions for the MCMC sampling done using the full data-vector (blue) and the compressed data-vector obtained through the K-L compression (orange). The contours correspond to the $68\%$ and $95\%$ confidence intervals. We see that there is no substantial loss of information despite the presence of strong degeneracies between the parameters; this is quantified in Figure \ref{fig:err_comp} and Table \ref{tab:recap_table} . The 2D posterior distributions have been smoothed using the Gaussian kernel density estimation package provided by Scipy.}
\label{fig:com_vs_or_bk_tr}
\end{figure*}

\begin{figure*}
\centering
\includegraphics[width=\textwidth]{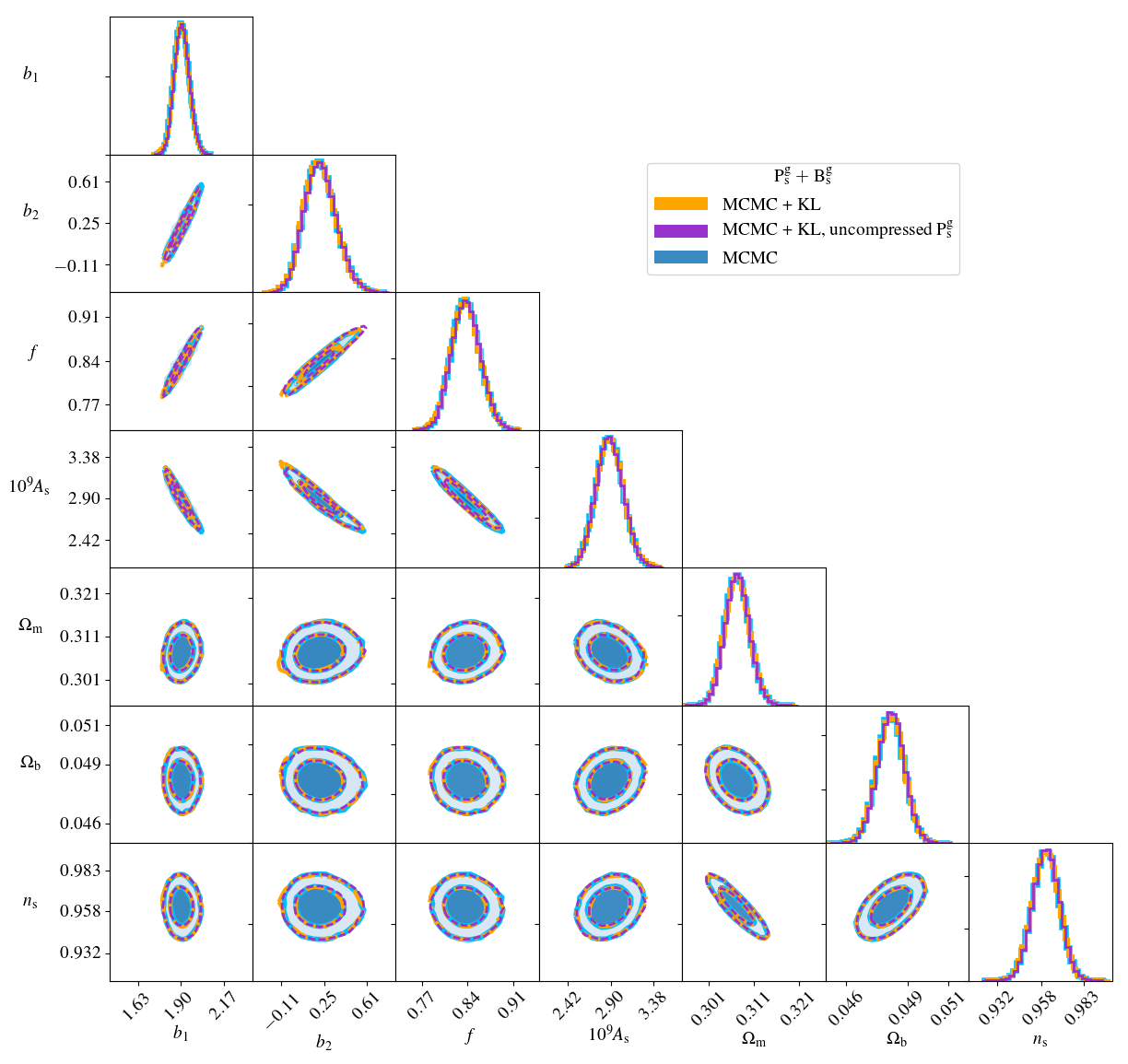}
\caption{  Joint data-vector $\left[\mathrm{P}^{\mathrm{\mathrm{s}}}_{\mathrm{\mathrm{g}}},\mathrm{B}^{\mathrm{\mathrm{s}}}_{\mathrm{\mathrm{g}}}\right]$ posteriors: the colours for the 1D and 2D posterior distributions are: the full data-vector (blue), the compressed data-vector (orange) and the uncompressed power spectrum plus the compressed bispectrum (purple). The contours correspond to the $68\%$ and $95\%$ confidence intervals. There is no substantial loss of information due to compression, which again is quantified in Figure \ref{fig:err_comp} and Table \ref{tab:recap_table} }
\label{fig:com_vs_or_pkbk_tr}
\end{figure*}

\subsection{Karhunen-Loève compression method}

A general linear transformation of the data-vector $\bm{x}$ with a transformation  matrix $\bm{\mathrm{B}}$ is given by 

\begin{eqnarray}
\bm{y}\,=\,\bm{\mathrm{B}}\,\bm{x}.
\end{eqnarray}

\noindent The mean and the covariance matrix for $\bm{y}$ become respectively $\langle\bm{y}\rangle = \bm{\mathrm{B}}\langle\bm{x}\rangle$ and $\mathrm{\textbf{Cov}}_{\bm{y}} = \bm{\mathrm{B}}^\intercal\mathrm{\textbf{Cov}}_{\bm{x}}\bm{\mathrm{B}}$.


In the case in which only one of the linear combinations of the data is considered, $\bm{\mathrm{B}}$ has only one row, $\bm{\mathrm{B}}=\bm{b}^\intercal$. Therefore the diagonal entries of the Fisher  matrix are given by:

\begin{eqnarray}
\label{fisher_diag}
\mathrm{F}_{ii}\,=\,\dfrac{1}{2}\left(\dfrac{\bm{b}^\intercal\,\mathrm{\textbf{Cov}}_{,i}\,\bm{b}}{\bm{b}^\intercal\,\mathrm{\textbf{Cov}}\,\bm{b}}\right)^2 \,+\,\dfrac{\left(\bm{b}^\intercal\,\langle\bm{x}\rangle_{,i}\right)^2}{\left(\bm{b}^\intercal\,\mathrm{\textbf{Cov}}\,\bm{b}\right)},
\end{eqnarray}

\noindent where again the comma notation followed by the index $i$ stands for the derivative with respect to the model parameter $\theta_i$.
As explained by \citet{2000MNRAS.317..965H}, who also derive weights based on minimizing the Fisher matrix diagonal elements, it is a very complex problem to find an analytical solution for $\bm{b}$ from the full Eq.\ref{fisher_diag}. 
Only by ignoring then the first term of Eq. \ref{fisher_diag} it is possible to find an analytic solution for $\bm{b}$. 
This implies assuming that the derivatives of the covariance matrix with respect to model parameters are neglible when compared to the data vector ones.
Recently \cite{2017MNRAS.472.4244H} presented a method to relax this approximation.

For both the power spectrum and bispectrum we numerically checked that it is reasonable to assume it to be valid, by taking the ratio between the diagonal elements of $\mathrm{\textbf{Cov}}_{,i} /\sqrt{\mathrm{\textbf{Cov}}}$ with $\langle\bm{x}\rangle_{,i}$ (corresponding to the ratio of the first and second term in Eq. \ref{fisher_diag}, respectively). This ratio for the bispectrum case results to be on average $\leq 10\%$ for the considered parameters. There are peaks for few triangle configurations ($\leq10$ for more than a thousand triangles) for which the ratio is around $70-80\%$ while the minima are around $2\%$.

By maximising $\mathrm{F}_{ii}$ as described in \citet{Tegmark:1996bz} using a Lagrange multiplier, it follows that the compressed data-vector is given by a scalar

\begin{eqnarray}
y_i\, = \,\langle\bm{x}\rangle_{,i}^\intercal\,\mathrm{\textbf{Cov}}^{-1}\,\bm{x} \,\equiv\, \bm{b}^\intercal\,\bm{x},
\end{eqnarray}

\noindent where we have defined the weighting vector $\bm{b} = \mathrm{\textbf{Cov}}^{-1} \langle\bm{x}\rangle_{,i}$. 

For the compression it is acceptable to use an analytic approximation of the covariance matrix (in our case as it is described in section \ref{cov_estimators}), as any systematic error in the covariance would mostly degrade the compression; it would not bias the parameter inference. The compression enables then to use an accurate simulated covariance for the actual inference.
This has the advantage of being able to compute weights for a data vector with an arbitrary large dimension (e.g. number of triangles).
The Fisher matrix diagonal element can be rewritten as

\begin{eqnarray}
\label{fmatrix_comp}
\mathrm{F}_{ii}\,=\,\langle\bm{x}\rangle_{,i}^\intercal\mathrm{\textbf{Cov}}^{-1}\langle\bm{x}\rangle_{,i} . 
\end{eqnarray}

\noindent In order to apply the KL compression method, one needs to choose a fiducial set of parameters at which to compute (analytically or numerically) the derivatives of the mean. In our case the fiducial values are reported in section \ref{analysis_setup}. 
We compute the numerical derivatives using the five point method for the first derivative in one dimension \citep{Abramowitz:1974:HMF:1098650}.

\begin{figure*}
\centering
\includegraphics[width=0.95\textwidth]{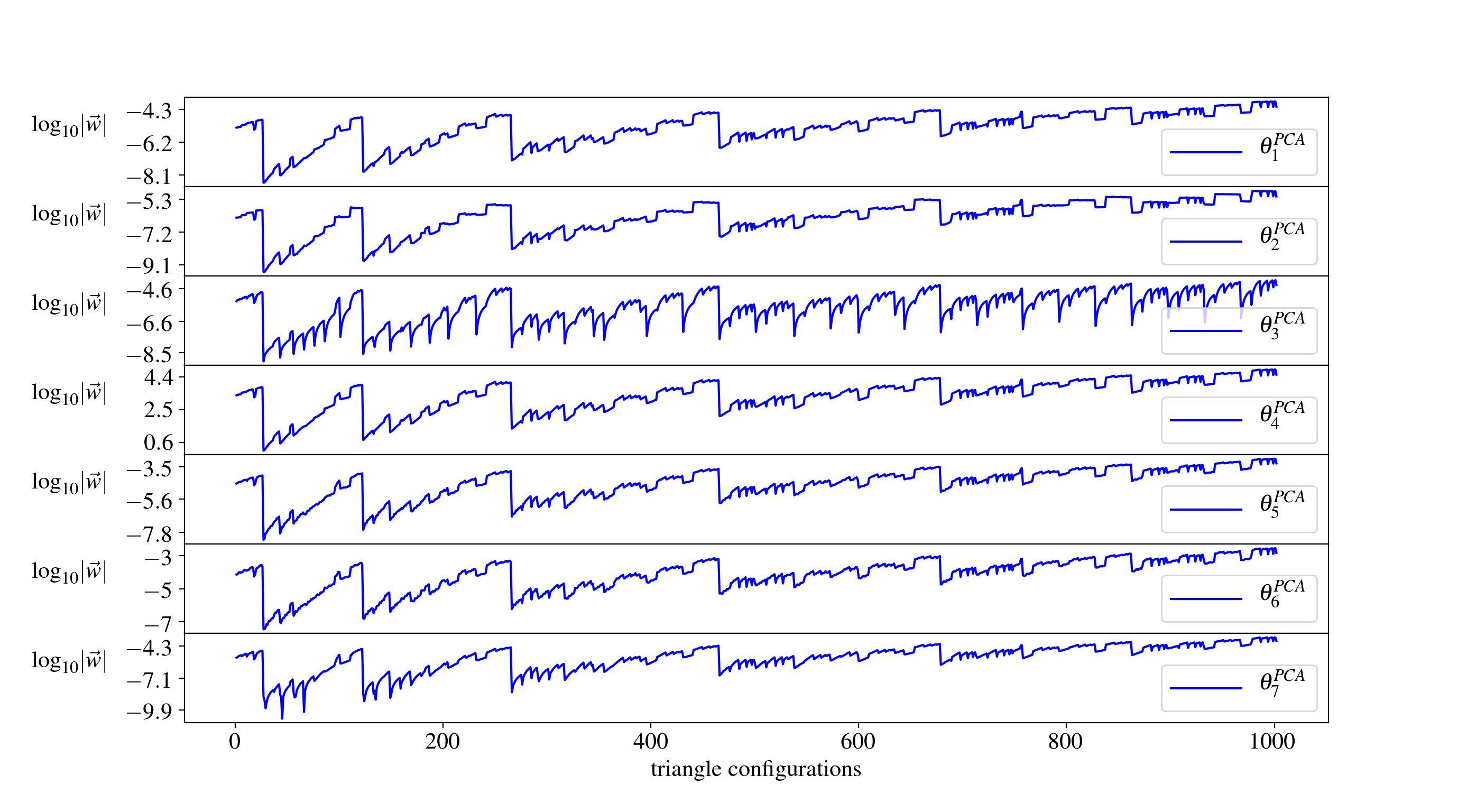}
\caption{ Logarithm of the absolute value of the weights for all the triangle configurations used in the bispectrum  data-vector. Each row corresponds to the weights for the bispectrum $\mathrm{B}^{\mathrm{\mathrm{s}}}_{\mathrm{\mathrm{g}}}$ with respect to a specific linear combination $\theta_{\mathrm{PCA}}$ of the original cosmological parameters obtained by diagonalising the Fisher information  matrix. The discontinuities observed reflect the five loops used to produce the set of triangles. From left to right the average size of the sides of the triangles increases, from the smallest triangle up to the largest. 
The amplitude of the weights slightly depends on the size of the triangle (increasing from left to right). The fluctuations are far more influenced by the shape and orientation of the triangles (which can be seen within each loop). 
}
\label{fig:weights_pa}
\end{figure*}

\begin{figure*}
\centering
\includegraphics[width=\textwidth]{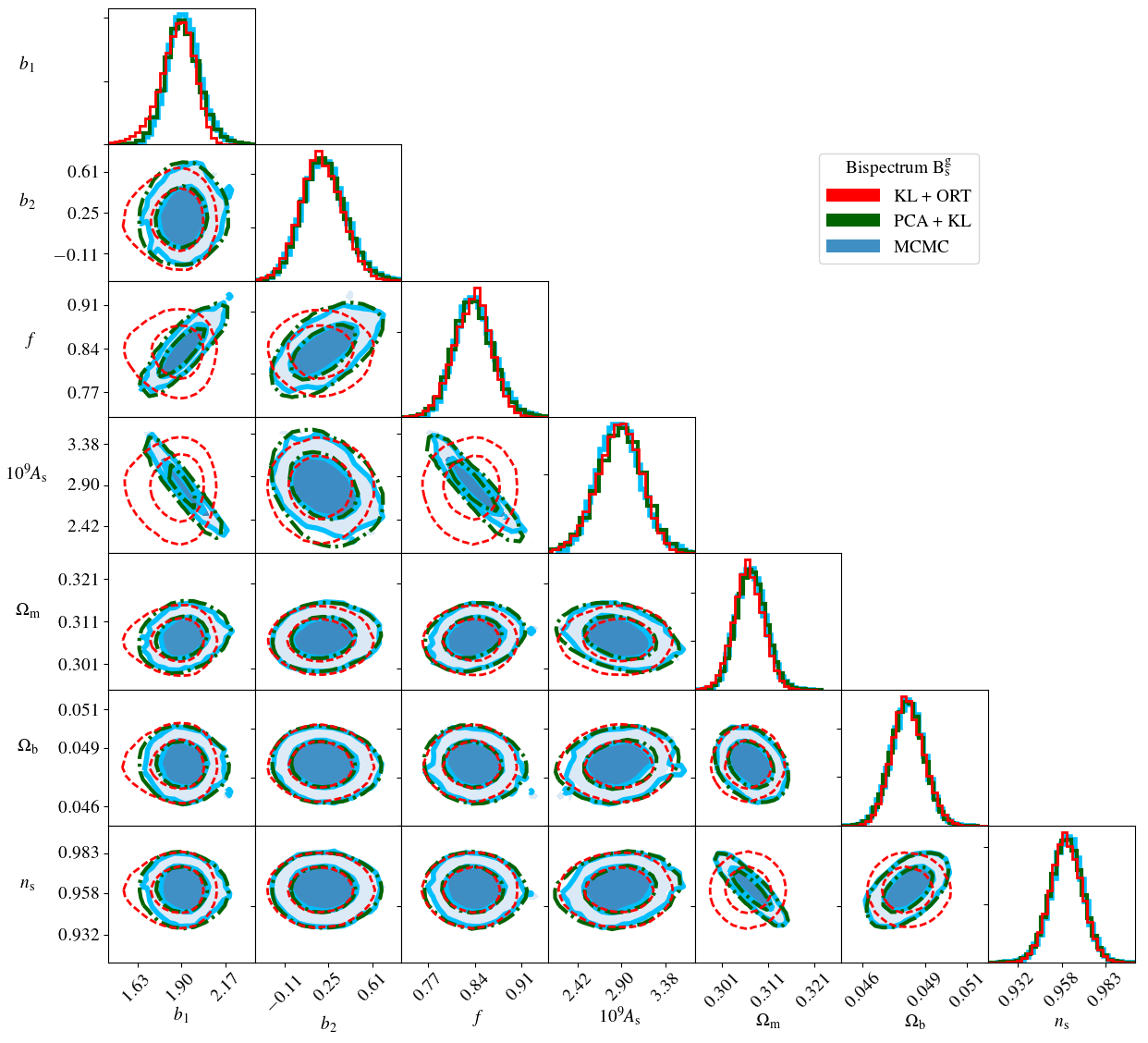}
\caption{Comparison between the MCMC-derived posteriors and the ones obtained using only the compression. The 1D and 2D posterior distributions are relative to the data-vector $\mathrm{B}^{\mathrm{\mathrm{s}}}_{\mathrm{\mathrm{g}}}$, for the three different cases: MCMC (blue), KL + ORT compression (red), PCA + KL compression (green). The contours correspond to the $68\%$ and $95\%$ confidence intervals (Figure \ref{fig:err_comp} and Table \ref{tab:recap_table} for numerical values).}
\label{fig:bk_kl-pc-mc}
\end{figure*}

\begin{figure*}
\centering
\includegraphics[width=\textwidth]{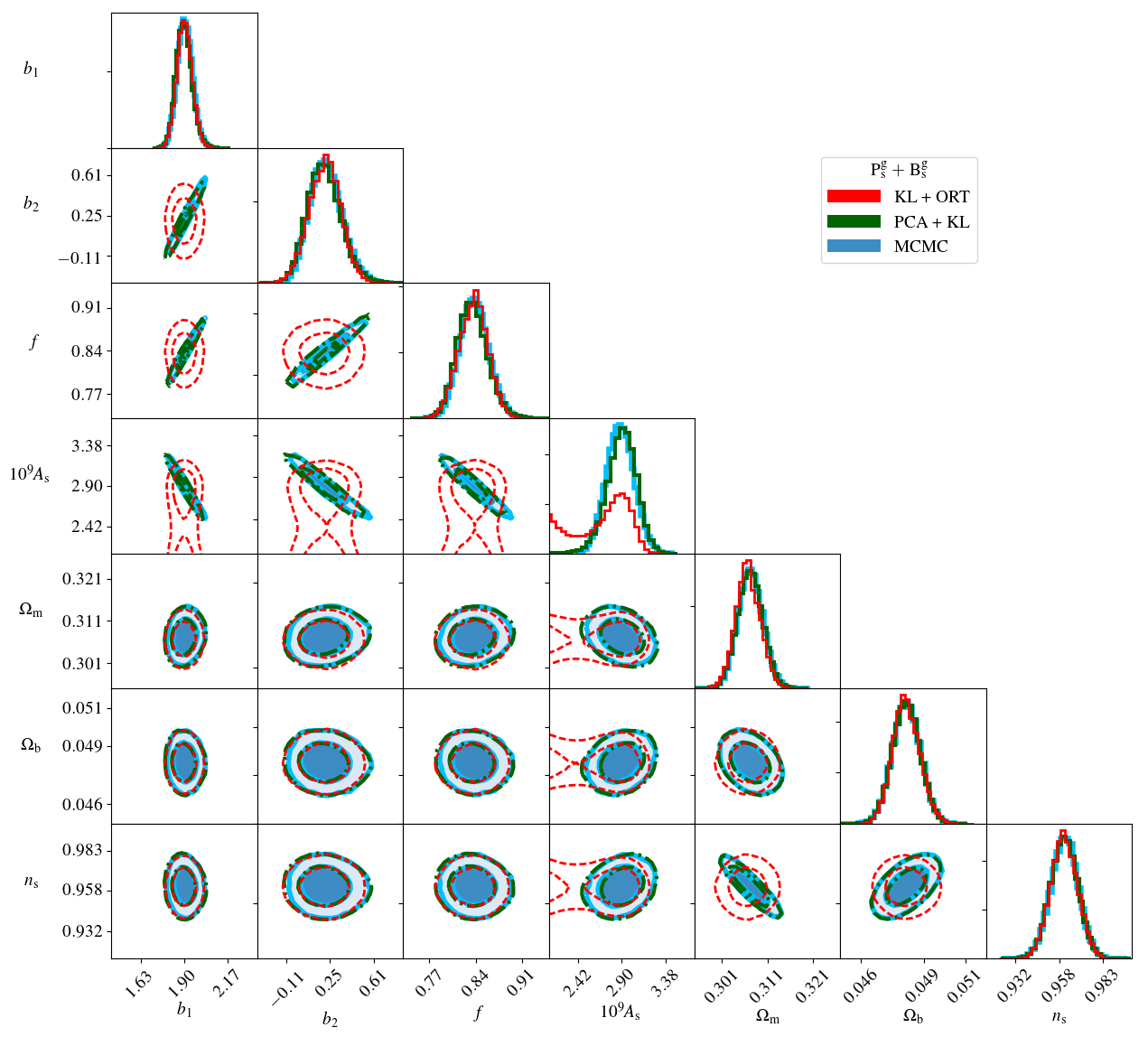}
\caption{  1D and 2D posterior distributions using as data-vector $\left[\mathrm{P}^{\mathrm{\mathrm{s}}}_{\mathrm{\mathrm{g}}},\mathrm{B}^{\mathrm{\mathrm{s}}}_{\mathrm{\mathrm{g}}}\right]$, for the three different cases: MCMC (blue), KL + ORT compression (red), PCA + KL compression (green). The contours correspond to the $68\%$ and $95\%$ confidence intervals(Figure \ref{fig:err_comp} and Table \ref{tab:recap_table} for numerical values). The KL method fails to reproduce the correct contours, both 1D and 2D, since for the used value of $k_{\mathrm{max}}$ it fails to work, producing an unphysical secondary peak for the parameter $A_\mathrm{s}$. On the contrary the PCA + KL method recovers with very good agreement the MCMC contours. }
\label{fig:pkbk_kl-pc-mc}
\end{figure*}

\section{MCMC of Compressed data-vectors}\label{mcmc_comp_dv}

In this section we compare the results obtained by running the MCMC algorithm for both uncompressed and compressed data-vectors. The goal is to check whether it would be  possible to substitute the original data-vector for its compressed version. Even if this would not bring any relevant advantage in terms of speed when computing the data-vector from the theoretical model, in the case of a real cosmological survey, it will be much easier and less expensive in terms of required simulations / mock catalogues to estimate the covariance matrix of the compressed data-vector rather than the full data-vector. Moreover, in the case of the bispectrum, a much larger number of triangles could be used for the original data-vector, allowing more information to be captured, since the dimension of the compressed covariance matrix would be reduced to the number of parameters considered in the analysis.
In order to use the compression, it is necessary to convert the covariance  matrix for the full data-vector to the one for the compressed data-vector. This is shown in Appendix \ref{sec:comp_cov}. 

In this paper, we consider two cases for the data-vector: the galaxy bispectrum $\mathrm{B}^{\mathrm{\mathrm{s}}}_{\mathrm{\mathrm{g}}}$ and the joint data-vector $\left[\mathrm{P}^{\mathrm{\mathrm{s}}}_{\mathrm{\mathrm{g}}},\mathrm{B}^{\mathrm{\mathrm{s}}}_{\mathrm{\mathrm{g}}}\right]$. For the latter, we include two further cases depending on whether or not we compress the power spectrum $\mathrm{P}^{\mathrm{\mathrm{s}}}_{\mathrm{\mathrm{g}}}$.

Figure \ref{fig:com_vs_or_bk_tr} shows the 1D and 2D marginalised posterior distributions when only the bispectrum data-vector is considered. For the considered set of parameters there is no substantial loss of information when the compressed data-vector is used, even if some degeneracies are present. In this case, the compressed vector has seven elements instead of the $\sim$ 1000 triangles for the uncompressed bispectrum. The 1D and 2D posteriors have been smoothed using a Gaussian kernel density estimation procedure, for clearer visualisation. 
Figure  \ref{fig:com_vs_or_pkbk_tr} shows the 1D and 2D marginalised posterior distributions when we consider both the power spectrum and the bispectrum. As can be seen, there is no qualitative or quantitative difference on the posterior distributions between compressing or not compressing the power spectrum together with the bispectrum. The precise numbers can be found in the summarising Table \ref{tab:recap_table}.

\begin{figure*}
\centering
\includegraphics[width=\textwidth]{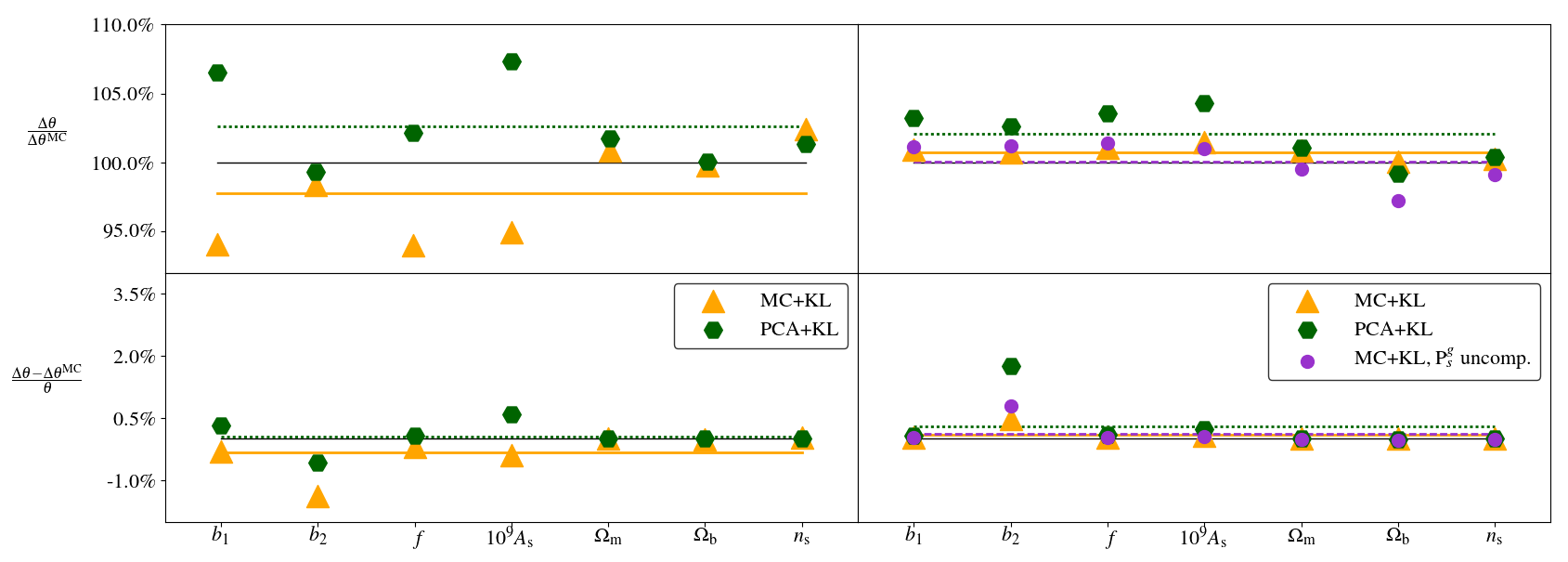}
\caption{Ratio (first panel) and relative difference (second panel) between the $68\%$ confidence intervals of the 1D marginalised posteriors for all the cosmological parameters obtained using either the compression plus the MCMC sampling (MCMC + KL, orangle triangles) or just the compression (PCA + KL, green hexagons) with the ones obtained running the MCMC on the full data-vector.
The black lines in all plots represent the reference values given by running a MCMC sampling using the full data-vector.
The ratios $\Delta\theta / \Delta\theta^{\mathrm{MC}}$ give an idea of how much the $68\%$ confidence intervals obtained through compression differ from the one given when using the full data-vector. The relative difference $(\Delta\theta - \Delta\theta^{\mathrm{MC}})/\theta$  is helpful since it scales the difference between the compressed and full data-vector results with respect to the value of the chosen parameters.
The first column shows the results when just the bispectrum $\mathrm{B}^{\mathrm{\mathrm{s}}}_{\mathrm{\mathrm{g}}}$ is considered, where the largest discrepancy happens for the parameters most degenerate between each others ($b_1$, $f$, $A_{\mathrm{\mathrm{s}}}$). Using the MCMC + KL method on the bispectrum proves to be with negligible loss of information. In the second column the ratios and relative differences in the case of the joint data-vector $\left[\mathrm{P}^{\mathrm{\mathrm{s}}}_{\mathrm{\mathrm{g}}}+\mathrm{B}^{\mathrm{\mathrm{s}}}_{\mathrm{\mathrm{g}}}\right]$ are shown. In all the subplots, the horizontal lines show the averages of the ratios and relative diferrences of the same colour. For $\left[\mathrm{P}^{\mathrm{\mathrm{s}}}_{\mathrm{\mathrm{g}}}+\mathrm{B}^{\mathrm{\mathrm{s}}}_{\mathrm{\mathrm{g}}}\right]$ the compression is optimal, both using MCMC + KL and PCA + KL. Compressing or not the power spectrum together with the bispectrum seems to produce no relevant statistical difference. The bottom right subplot $b_2$ appears to be the parameters whose divergence from the MCMC result is greater with respect to its fiducial value.
%
}
\label{fig:err_comp}
\end{figure*}

\section{Posterior Distributions Directly From Compressed data-vectors}\label{pca_kl}

It is possible to compute the 1D likelihood for each linear combination $y_{\theta_i}$ of the original data-vector obtained compressing with respect to the parameter $\theta_i$ as done in \citet{Zablocki:2015zcm}:

\begin{eqnarray}
\ln\mathcal{L}\,=\,-\dfrac{(y_{\theta_i}-\bar{y}_{\theta_i})^2}{2\sigma^2_{\langle y_{\theta_i}\rangle}} \quad
\mathrm{with}\quad\sigma^2_{\langle y_{\theta_i}\rangle}\,=\,\sum_{j=1}^{n_x}b_{j}^2\,\mathrm{\textbf{Cov}}_{jj},
\end{eqnarray}

\noindent where $\bar{y}_{\theta_i} \equiv \langle \bar{y}_{\theta_i}\rangle$ is the mean of the compressed linear combination.  Nevertheless, these 1D likelihoods would not be realistic since they are obtained using the compressed data-scalar $y_{\theta_i}$ by varying only one parameter at a time. In reality $y_{\theta_i}$ is sensitive to all the other parameters. 

In order to account for this fact, we transform the set of $y_{\theta_i}$ scalars in such a way that, at linear order, they are sensitive only to the parameter with respect to which the original data-vector was compressed. 
One method to do so is to orthogonalise the parameter space by diagonalising the Fisher Information  matrix. This has the advantage of obtaining realistic multidimensional posteriors directly from the 1D posteriors of the new set of model parameters (linear combinations of the original, physical parameters).

For comparison we include another method presented in \citet{Zablocki:2015zcm} which consists of orthogonalising the weights in such a way that the Fisher  matrix for the compressed data-vector would become diagonal; we call this weights-orthogonalisation and it is described in Appendix \ref{sec:weights_ort}.
The idea behind these two slightly different orthogonalisation approaches is the same, namely to diagonalise the Fisher Information  matrix. The difference is that while the diagonalisation in the first method is the starting point independently from compressing (or not compressing) the data-vector later, in the second method the diagonalisation is a consequence of the procedure used. 

Both methods are approximations at linear level. Therefore as anticipated earlier they both fail (even if at different levels as it is shown later) when non-linear degeneracies are present. For example, cases where the 2D-posterior distribution of a pair of parameters can no longer be approximated by an ellipse but has instead a `banana'-shape. This break-down of the above procedures is in agreement with the fact that the compression method relies on the assumption that the multidimensional posteriors are Gaussian.

\subsection{Parameter space orthogonalisation - PCA}

As anticipated above, the compression returns only 1D posterior distributions for each one of the parameters. Therefore an additional step is required in order to be able to assume that these distributions correspond to 1D marginalisations from the original multidimensional distribution. This is because the KL compression with respect one model parameter returns a linear combination of the original data vector which is still sensitive to the variation of the other model parameters. 
\citet{Zablocki:2015zcm} obtained marginalised 1D posterior distributions by orthogonalising the weighting vectors for all the model parameters through a Gram-Schmidt like procedure described in Appendix \ref{sec:weights_ort}. We label this method as ORT + KL.

In order to test whether the compression results match the ones obtained running a MCMC  sampling algorithm, we also compare the 2D posterior distributions  for different pairs of parameters. 
In the case of ORT + KL the only possibility to reconstruct the multidimensional posterior is to take the outer product of the 1D marginalised posteriors.
However doing so would be misleading: it would return ellipses with axes oriented at different angles to the ones given by the MCMC sampling as it can be seen in Figures \ref{fig:bk_kl-pc-mc} and \ref{fig:pkbk_kl-pc-mc} looking at the difference between the 2D red (ORT + KL) and blue (MCMC) contours. These axes orientation reflects the different degree of degeneracy between the parameters.

Therefore in order to avoid this difference, instead of orthogonalising the weights we perform a principal component analysis (PCA) transformation of our parameter space before applying the KL compression. This is done by diagonalising the Fisher information matrix using the eigenvalue decompositions

\begin{eqnarray}
\mathrm{F}_{\bm{\theta}_{\mathrm{phys.}}} =  \,\bm{\mathrm{P}}\, \;\mathrm{F}_{\bm{\theta}_{\mathrm{PCA}}}\;\bm{\mathrm{P}}^\intercal
\quad
\mathrm{where}
\quad
\bm{\theta}_{\mathrm{PCA}}\,\,=\,\bm{\mathrm{P}}^\intercal\,\bm{\theta}_{\mathrm{phys.}},
\end{eqnarray}

\noindent where $\bm{\mathrm{P}}$ is the linear transformation matrix. After having diagonalised the Fisher  matrix we compress the data-vector with respect to this new set of parameters $\bm{\theta}_{\mathrm{PCA}}$. 
The weights obtained doing so are displayed in Figure \ref{fig:weights_pa} for the case of the bispectrum as data-vector.
The effect of a PCA decomposition is to rotate the parameter space to the axes corresponding to the degeneracies between the original set of parameters. Therefore taking the outer product of the 1D posteriors of the parameters
$\bm{\theta}_{\mathrm{PCA}}$ in order to get the multidimensional posterior distribution should return with good approximation the one sampled by the MCMC code. Once the multidimensional posterior has been reconstructed by taking the outer product of the 1D posterior distributions for the $\bm{\theta}_{\mathrm{PCA}}$ set of parameters, this can be randomly sampled in terms of the physical parameters $\bm{\theta}_{\mathrm{phys.}}$ using the rotation matrix $\bm{\mathrm{P}}$ in order to get the 1D and 2D marginalised posterior distributions for the $\bm{\theta}_{\mathrm{phys.}}$ parameters. 
The results for the PCA + KL method just described can be seen looking at the green (PCA + KL) and blue (MCMC) 2D contours still in Figures \ref{fig:bk_kl-pc-mc} and \ref{fig:pkbk_kl-pc-mc}. The PCA + KL method recovers tilted ellipses in good agreement with the MCMC ones.

\subsection{Comparison with MCMC sampling}

Figure ~\ref{fig:err_comp}
shows the ratio and relative difference of the 68$\%$ confidence intervals of the 1D marginalised posteriors between the compression methods MCMC + KL and PCA + KL and the standard MCMC sampling. In the bispectrum case, while MCMC + KL tends to underestimate the $68\%$ confidence intervals obtained by the MCMC, PCA + KL tends to overestimate them by approximately the same amount ($\sim 2.5\%$). For the joint data-vector, MCMC + KL returns equivalent confidence intervals to the MCMC's ones while PCA + KL overestimates them in average by $\sim2.5\%$ (first panel) but still less than $\sim1\%$ in terms of relative difference (second panel).
We consider the averages of these ratios since at the same time the compression methods overestimate the 68$\%$ confidence intervals for some parameters while underestimating them for others.
In terms of the individual parameters 68$\%$ confidence intervals, PCA + KL diverges at most from the MCMC's ones respectively by $\sim7\%$ and $\sim6\%$ in the case of the bispectrum and the joint data-vector. MCMC + KL diverges at most respectively by $\sim6\%$ and $\sim4\%$ in the case of the bispectrum and the joint data-vector.

Both in the case of the bispectrum and in the case of the joint data-vector $\left[\mathrm{P}^{\mathrm{\mathrm{s}}}_{\mathrm{\mathrm{g}}},\mathrm{B}^{\mathrm{\mathrm{s}}}_{\mathrm{\mathrm{g}}}\right]$ the compression for both MCMC + KL (Figures \ref{fig:com_vs_or_bk_tr} and  \ref{fig:com_vs_or_pkbk_tr} ) and PCA + KL (Figures \ref{fig:bk_kl-pc-mc} and  \ref{fig:pkbk_kl-pc-mc} )  methods well matches the 1-2D contours derived from the MCMC sampling.
This shows that in the bispectrum case there is no relevant difference between MCMC + KL and PCA + KL methods while there is a very small one in the case of the joint data-vector.

In the bispectrum case (Figure \ref{fig:bk_kl-pc-mc}) when the 2D posterior distributions are considered, only the contours derived by orthogonalising the parameter space before the compression (PCA + KL) have elliptical shapes with the right inclination.
In the case of the joint data-vector instead (Figure \ref{fig:pkbk_kl-pc-mc}), the KL method alone fails to recover even the 1D posterior for the parameter $A_{\mathrm{s}}$, producing an unphysical secondary peak, while using PCA + KL returns 1D and 2D contours equivalent to the MCMC's ones.
This shows that the orthogonalisation of the parameter space (PCA step) performs better than the weights orthogonalisation for combinations of parameters presenting strong non-linear degeneracies.  As described in the following subsection, the compression methods PCA + KL and KL + ORT break down when the $k_{\mathrm{max}}$ used drops below a certain threshold value. For the KL + ORT method this is $\sim 0.2\, \mathrm{Mpc}^{-1}h$ as shown in Figure \ref{fig:pkbk_kl-pc-mc}, while for PCA + KL the threshold value is lower. 
Applying the compression up to those mildly non-linear scales on real data like the BOSS DR12 requires to modify the model in order to take into account non-local bias terms. However this would not require the introduction of additional bias parameters. This is because assuming that even if the galaxy bias is non-local in Eulerian space, it is in Lagrangian space, and hence the additional non-local bias terms can be related at first order to the linear bias $b_1$ as done in \citet{2015MNRAS.451..539G}. 

The parameter set used in our analyis has been chosen to have strong degeneracies in order to test the applicabilty of the MCMC + KL and PCA + KL methods. When working with real data the parameter set is usually designed to be less degenerate, for example considering only $\sigma_8$ instead of $\Omega_m$ and $A_{\mathrm{s}}$ or also combining $\sigma_8$ with $b_1$, $b_2$ and $f$.

In absence of non-linear (banana shape) degeneracies, the compression PCA + KL can be a valid and much faster substitute to the standard MCMC sampling since it returns realistic mmultidimensional posterior distributions. In particular, compression could be used to accurately forecast the constraints for different sets of parameters. It would also be possible to qualitatively study the degeneracies present in each set.

Running both compression algorithms (KL or PCA) takes approximately $20$ minutes for seven parameters, with the time depending on the number of intervals for each parameter range and therefore how many times the linear matter power spectrum is computed. As a rough estimate, if we consider $100$ intervals for each of the seven parameter ranges, $\mathrm{P}^{\mathrm{m}}$ will be recomputed $300$ times (no need to recompute the matter power spectrum when varying only one of the parameters $b_1$, $b_2$, $f$ or $A_{\mathrm{s}}$). For a standard MCMC with $64$ samplers and $10^4$ steps instead of 300 times $\mathrm{P}^{\mathrm{m}}$ needs to be recomputed $64\times10^4$ times. Therefore the compression is more than $\sim 2100$ times faster than the MCMC.

The compression PCA + KL is also much less demanding in terms of computing resources. It can be run in few minutes on a laptop with a single Intel i7 processor while for the MCMC we have used 64 threads working in parallel on 64 Xeon E5-2650 processors connected using MPI for $\sim72$ hours.

\begin{figure*}
\centering
\includegraphics[width=\textwidth]{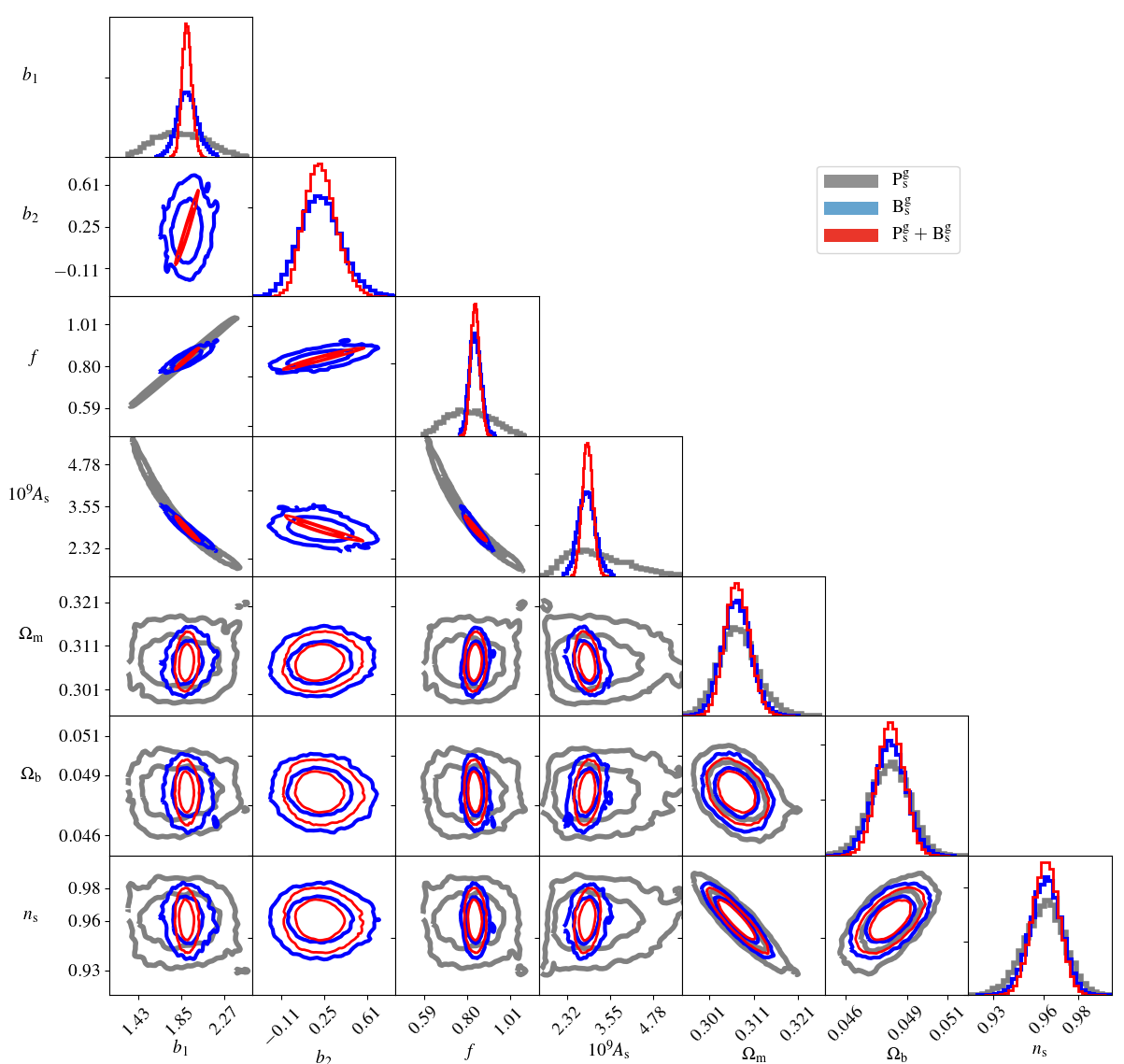}
\caption{Forecasted posteriors for the redsshift bin of q DESI-like survey for what concerns the combined two and three points statistics. 1D and 2D contours are derived from MCMC sampling using for the three data-vectors: $\mathrm{P}^{\mathrm{\mathrm{s}}}_{\mathrm{\mathrm{g}}}$ (grey), $\mathrm{B}^{\mathrm{\mathrm{s}}}_{\mathrm{\mathrm{g}}}$ (blue) and $\left[\mathrm{P}^{\mathrm{\mathrm{s}}}_{\mathrm{\mathrm{g}}}+\mathrm{B}^{\mathrm{\mathrm{s}}}_{\mathrm{\mathrm{g}}}\right]$ (red). 
Combining two and three points statistics allows to drastically reduce and break the degeneracies in the parameter space, in particular the ones between $b_1$, $A_{\mathrm{\mathrm{s}}}$ and $b_2$. This improvement obtained using the joint data-vector is therefore particularly evident for all these amplitude-like parameters (Table \ref{tab:recap_table}). In particular it's qualitatively evident from this plot the benefit of adding the bispectrum to the power spectrum. It is important to notice that for the adopted model the galaxy power spectrum does not depend on the parameter $b_2$. Even with one parameter less to constrain than the bispectrum, as expected, the MCMC for the galaxy power spectrum alone does not well converge for most of the degenerate parameters.  }
\label{fig:3stats_mcmc}
\end{figure*}

\subsection{Limitations of the compression}

Unfortunately the orthogonalisation prescriptions do not work when "strong" non-linear degeneracies are present in the parameter space and the Fisher matrix can no longer be diagonalised in practice. This happens because the multidimensional posterior distribution can no longer be approximated by a multi-variate Gaussian distribution. In our work, this happens when $k_{\mathrm{max}}$ is lowered, reducing the information accessible through the power spectrum and the bispectrum. 

This failure of the method manifests itself with the appearence of unphysical secondary peaks in the posterior distributions. 
For example, when the MCMC (or MCMC + KL) returns 2D posterior contours with degeneracies that are banana-shaped, like in the case of the joint data-vector $\left[\mathrm{P}^{\mathrm{\mathrm{s}}}_{\mathrm{\mathrm{g}}},\mathrm{B}^{\mathrm{\mathrm{s}}}_{\mathrm{\mathrm{g}}}\right]$ for the parameters $(b_1,b_2,f,\sigma_8)$, the compression alone (KL + ORT or PCA + KL) fails to recover the 1D and 2D posteriors. In these cases both the KL + ORT and PCA + KL produce unphysical secondary peaks in the posterior distribution. However, as shown in Figure \ref{fig:pkbk_kl-pc-mc} PCA + KL still works considering only larger scales ($k_{\mathrm{max}}$ lower), where standard perturbation theory gives more accurate predictions, than when using only the KL + ORT method.

In the case of the bispectrum the breakdown of the PCA + KL method happens for $k_{\mathrm{max}} < 0.18\, \mathrm{Mpc}^{-1}h$ while for the joint data-vector $\left[\mathrm{P}^{\mathrm{\mathrm{s}}}_{\mathrm{\mathrm{g}}}, \mathrm{B}^{\mathrm{\mathrm{s}}}_{\mathrm{\mathrm{g}}}\right]$  for $k_{\mathrm{max}} < 0.17\, \mathrm{Mpc}^{-1}h$.   A potential solution to this limitation could be the application of the Gaussianisation method proposed by \citet{Schuhmann:2015dma} which we plan to include in a future work.
This failure also happens when $V_{\mathrm{s}}$ or $n_{\mathrm{g}}$ are one order of magnitude smaller. However, these scenarios are below the specifications of current and future cosmological surveys. 
Another solutions is to consider less parameters by either fixing to a fiducial value one or more of the most degenerate ones or by rewriting them in terms of combinations which absorb the degeneracies (e.g. $f\sigma_8$).
It is important to note that in this paper we have on purpose considered an extreme case of degenerate parameter space in order to test the method applicability. In a realistic case a much less degenerate parameter space is usually considered when analysing data.
In the considered parameter space the main cause of parameters degeneracy is due to having both $A_{\rm{s}}$, $\Omega_{\rm{m}}$ and $f$ instead of using only $\sigma_8$ and $f$. It is then reasonable to assume that the method would still work when a more complex biasing model is considered by adding a further bias parameter $b_3$ or the tidal bias $b_{\rm{s}_2}$, provided that $A_{\rm{s}}$ and $\Omega_{\rm{m}}$ are substituted with $\sigma_8$. A more accurate biasing model would be needed in an application to real data in order to describe the bias non local nature observed in simulations \citep{2013PhRvD..87h3002S, 2017MNRAS.472.3959M}.

\begin{table*}
    \centering
    \begin{tabular}{clccccccc}
            \hline
            & & & & & $\Delta\theta\, = \, \frac{\theta_{\mathrm{max}}^{68\%} - \theta_{\mathrm{min}}^{68\%}}{2}$ & & &  \\
            \hline
            & &  $\Delta b_1$ &  $\Delta b_2$ &  $\Delta f$   &  $\Delta 10^9A_{\mathrm{\mathrm{s}}}$ & $\Delta \Omega_{\mathrm{m}}$ & $\Delta \Omega_{\mathrm{b}}$ &  $\Delta n_{\mathrm{\mathrm{s}}}$   \\
            \hline
            $\mathrm{P}^{\mathrm{\mathrm{s}}}_{\mathrm{\mathrm{g}}}$  
            & MCMC                           & 0.251 & -     & 0.111 & 0.856 & 0.0041 & 0.0009 & 0.0116 \\
            \hline
            & MCMC                           & 0.095 & 0.174 & 0.028 & 0.232 & 0.0031 & 0.0008 & 0.0091 \\
            $\mathrm{B}^{\mathrm{\mathrm{s}}}_{\mathrm{\mathrm{g}}}$  
            & MCMC + KL                      & 0.090 & 0.172 & 0.027 & 0.220 & 0.0032 & 0.0008 & 0.0093 \\
            & PCA + KL                        & 0.101 & 0.173 & 0.029 & 0.249 & 0.0032 & 0.0008 & 0.0092 \\
            \hline
            & MCMC                           & 0.047 & 0.132 & 0.022 & 0.151 & 0.0027 & 0.0007 & 0.0079 \\
            $\left[\mathrm{P}^{\mathrm{\mathrm{s}}}_{\mathrm{\mathrm{g}}},\mathrm{B}^{\mathrm{\mathrm{s}}}_{\mathrm{\mathrm{g}}}\right]$ 
            & MCMC + KL                      & 0.047 & 0.134 & 0.022 & 0.153 & 0.0027 & 0.0007 & 0.0079 \\
            & " " ($\mathrm{P}^{\mathrm{\mathrm{s}}}_{\mathrm{\mathrm{g}}}$ uncomp.)
                                            & 0.048 & 0.134 & 0.022 & 0.152 & 0.0027 & 0.0006 & 0.0078 \\
            & PCA + KL                      & 0.049 & 0.136 & 0.022 & 0.157 & 0.0027 & 0.0007 & 0.0079 \\
            \hline
            & MCMC                           & 51$\%$& 24$\%$ & 23$\%$& 35$\%$ & 13$\%$& 15$\%$  & 13$\%$  \\
             $\% \left(\Delta\theta^{\mathrm{B}}-\Delta\theta^{\mathrm{P+B}}\right)/{\Delta\theta^{\mathrm{B}}}$
            & MCMC + KL                      & 47$\%$& 22$\%$& 17$\%$& 31$\%$ & 13$\%$& 14$\%$  & 15$\%$\\
            & " " ($\mathrm{P}^{\mathrm{\mathrm{s}}}_{\mathrm{\mathrm{g}}}$ uncomp.) 
                                            & 47$\%$& 22$\%$  & 17$\%$&31$\%$& 14$\%$ & 17$\%$  & 16$\%$  \\
            & PCA + KL                       &52$\%$& 22$\%$& 22$\%$& 37$\%$ & 14$\%$& 15$\%$  & 14$\%$\\
            \hline
            & MCMC                      & 81$\%$& -$\%$& 80$\%$& 82$\%$ & 34$\%$& 27$\%$  & 32$\%$\\
            $\% \left(\Delta\theta^{\mathrm{P}}-\Delta\theta^{\mathrm{P+B}}\right)/{\Delta\theta^{\mathrm{P}}}$
            & MCMC + KL                 & 81$\%$& -$\%$& 80$\%$& 82$\%$ & 33$\%$& 27$\%$  & 31$\%$\\
            & " " ($\mathrm{P}^{\mathrm{\mathrm{s}}}_{\mathrm{\mathrm{g}}}$ uncomp.)
                                        & 81$\%$& -$\%$& 80$\%$& 82$\%$ & 34$\%$& 29$\%$  & 33$\%$\\
            & PCA + KL                  & 81$\%$& -$\%$& 80$\%$& 82$\%$ & 33$\%$& 27$\%$  & 32$\%$\\
     \end{tabular}  
         \caption{ The $68\%$ confidence intervals of the 1D posteriors for the bispectrum and the joint data-vector $\left[\mathrm{P}^{\mathrm{\mathrm{s}}}_{\mathrm{\mathrm{g}}},\mathrm{B}^{\mathrm{\mathrm{s}}}_{\mathrm{\mathrm{g}}}\right]$ obtained from the full data-vectors and from the two possible compression applications, KL + MCMC sampling or PCA + KL compression. Since the marginalised 1D distributions are not perfectly Gaussian (symmetric with respect to the maximum likelihood point) in the Table is reported for each parameter the halved difference between the maximum and minimum values included in the $68\%$ confidence interval. The four rows in the third panel show the improvement in percentage on the constraints gained by using the joint data-vector instead of the bispectrum alone for all the compression methods considered. The percentage shows indicates how much smaller are the constraints obtained using $\left[\mathrm{P}^{\mathrm{\mathrm{s}}}_{\mathrm{\mathrm{g}}},\mathrm{B}^{\mathrm{\mathrm{s}}}_{\mathrm{\mathrm{g}}}\right]$ as data-vector than the ones computed using just $\mathrm{B}^{\mathrm{\mathrm{s}}}_{\mathrm{\mathrm{g}}}$. 
         In the last panel the percentages are relative to the improvement obtained by adding the bispectrum to the power spectrum. The added value of the bispectrum with respect to power spectrum alone consists in $68\%$ confidence intervals $\sim4$ times smaller for the amplitude-like parameters $b_1$, $f$ and $A_{\mathrm{s}}$.}
    \label{tab:recap_table}
\end{table*}

\section{joint data-vector added value}
\label{sec:joint_dv}

In both cases presented in this paper, either running an MCMC sampling on the compressed data-vector or doing the parameter estimation directly from the compression (MCMC + KL or PCA + KL), the added value of combining the power spectrum with the bispectrum is qualitatively (Figure \ref{fig:3stats_mcmc}) and quantitatively (Figure \ref{fig:err_comp} and Table  \ref{tab:recap_table}) evident. This improvement obtained using the joint data-vector is particularly relevant for all the degenerate amplitude-like parameters: $b_1$, $b_2$, $f$ and $A_{\mathrm{\mathrm{s}}}$. For the redshift bin of a DESI-like survey considered in our analysis, the improvement  for the above mentioned parameters obtained using the joint data-vector with respect using only the bispectrum is respectively of 52$\%$, 22$\%$, 22$\%$ and 37$\%$ (using PCA + KL). The improvement with respect to the power spectrum alone is even greater, in particular: 81$\%$, 80$\%$ and 82$\%$ for $b_1$, $f$ and $A_{\mathrm{\mathrm{s}}}$.

Comparing our results to \citet{Sefusatti:2006pa} and in particular their Table 8 for $k_{\rm{max}} = 0.2 \mathrm{Mpc}^{-1} h$,  it is possible to see a similar effect due to including the bispectrum in the analysis together with the power spectrum.
The similarity consists in the fact that the parameters that more benefit from this addition are the bias ones ($b_1$, $b_2$) and the amplitude ones ($A_{\rm{s}}$, $\sigma_8$). The greater improvement that we find for certain parameters by adding the bispectrum to the power spectrum with respect to Sefusatti et al. can be explained by the fact that we consider redshift space distortions, which are encoded in the growth rate parameter $f$. This increases the degeneracies already present between $b_1$, $b_2$, $\sigma_8$, $A_{\rm{s}}$ and $\Omega_m$. Therefore the added value of the bispectrum in lifting the degeneracies increases.Together with this, in Sefusatti et al.’s paper the covariance matrix of the bispectrum includes also off-diagonal terms, which decrease the constraining power of the bispectrum since these terms describe the correlation between different triangle configurations.

For what concerns the most recent bispectrum measurements from BOSS DR12 data by \citet{Gil-Marin:2016wya}, it is important to point out that the data vector considered is given by the power spectrum monopole and quadrupoles together with the bispectrum monopole. Using the bispectrum monopole implies washing out some of its information when integrating over the azimuthal and polar angles. This could explain the lower impact of the bispectrum monopole with respect to the power spectrum monopole and quadrupole in constraining the parameters.
Moreover in this case the covariance matrix used to do parameter inference is estimated from galaxy mocks, therefore it includes not only off diagonal terms describing the correlation between different triangles, but also the noise due to the fact that it is an estimated quantity.

Finally in the recent paper by \citet{Byun:2017fkz} similar improvement to ours have been obtained in parameter constraints (Table 3),  in particular for what concerns $b_1$ and $\sigma_8$ (proxy for $A_{\mathrm{s}}$).

The efficiency of the compression PCA + KL also drastically improves if ones considers the joint data-vector $\left[\mathrm{P}^{\mathrm{\mathrm{s}}}_{\mathrm{\mathrm{g}}}, \mathrm{B}^{\mathrm{\mathrm{s}}}_{\mathrm{\mathrm{g}}}\right]$ as can be seen in Figures \ref{fig:com_vs_or_pkbk_tr} and \ref{fig:pkbk_kl-pc-mc}. Indeed, the combination of power spectrum and bispectrum reduces the degeneracies between the considered parameters and the results obtained by running approximately 20 minutes long compression-pipeline almost perfectly match the ones given by a three days of MCMC sampling run in parallel on 64 processors (right panel Figure \ref{fig:err_comp}).

As is well known in the literature, the degeneracy between the bias parameters (in particular at the linear order $b_1$) with the amplitude of the dark matter perturbations $\sigma_8$ or the primordial perturbations scalar amplitude $A_{\mathrm{\mathrm{s}}}$ cannot be broken using only the power spectrum. These degeneracies are even larger when redshift-space distortions are considered. 
On the other hand, the bispectrum alone can (in theory) lift these degeneracies, even if it requires including the quadratic bias parameter $b_2$. In any case being a 3pt statistic,it is more difficult to measure and analyse from real surveys than 2pt statistics. Therefore combining power spectrum and bispectrum is of fundamental importance in order to obtain the best possible constraints, especially in light of the large data sets that are going to become available in the near future (DESI, Euclid, PFS, etc.).


\section{Conclusions}\label{conclusions}

In this paper we have shown that it is possible to compress the information about cosmological parameters contained in the galaxy power spectrum and bispectrum and to extract this information in an efficient way.
In particular we have shown:

\begin{itemize}
\item Compressing the data-vector using the Karhunen-Loève algorithm before running an MCMC sampling gives negligible loss of information in terms of parameters constraints (Figures \ref{fig:com_vs_or_bk_tr}, \ref{fig:com_vs_or_pkbk_tr}, \ref{fig:err_comp}).
In particular running an MCMC sampling on the compressed data-vector (MCMC + KL) returns $68\%$ confidence intervals less than $1\%$ larger in terms of relative difference than the MCMC's ones. This happens in both the cases of the bispectrum and power spectrum plus bispectrum ( $\mathrm{B}^{\mathrm{\mathrm{s}}}_{\mathrm{\mathrm{g}}}$ and $\left[\mathrm{P}^{\mathrm{\mathrm{s}}}_{\mathrm{\mathrm{g}}},\mathrm{B}^{\mathrm{\mathrm{s}}}_{\mathrm{\mathrm{g}}}\right]$).
For real surveys this would allow us to drastically reduce the number of simulations needed to numerically estimate the covariance matrix. This is because the dimension of the compressed covariace matrix corresponds to the number of model parameters, not the number of the original data-vector elements.

\item Orthogonalising the parameter space through the diagonalisation of the Fisher matrix before applying the compression (PCA + KL), proves to be competitive with the MCMC sampling with minimal loss of constraining power (Figures \ref{fig:bk_kl-pc-mc} and \ref{fig:pkbk_kl-pc-mc}).
PCA + KL returns for both $\mathrm{B}^{\mathrm{\mathrm{s}}}_{\mathrm{\mathrm{g}}}$ and  $\left[\mathrm{P}^{\mathrm{\mathrm{s}}}_{\mathrm{\mathrm{g}}},\mathrm{B}^{\mathrm{\mathrm{s}}}_{\mathrm{\mathrm{g}}}\right]$ $68\%$ confidence intervals less than $1\%$ different in terms of relative difference from the ones obtained running the MCMC sampling for the full data-vectors.
This method cannot be applied when the parameter space presents non-linear degeneracies, since a multidimensional Gaussian posterior distribution is no longer a valid approximation. Using the joint data-vector $\left[\mathrm{P}^{\mathrm{\mathrm{s}}}_{\mathrm{\mathrm{g}}},\mathrm{B}^{\mathrm{\mathrm{s}}}_{\mathrm{\mathrm{g}}}\right]$ lowers the minimum $\bm{k}_{\mathrm{max}}$ necessary for the only-compression method (PCA + KL) to work to $\bm{k}_{\mathrm{max}} \simeq 0.17\, \mathrm{Mpc}^{-1}h$ (while in case of just the bispectrum this threshold is $\bm{k}_{\mathrm{max}} \simeq 0.18\, \mathrm{Mpc}^{-1}h$). These values have been obtained considering the redshift bin of a DESI-like survey. Moreover these values can be further lowered if the parameter space we considered is reduced to a less degenerate one, for example ($b_1$,$b_2$,$f$, $\sigma_8$).

\item \citet{Byun:2017fkz} reduced the covariance  matrix dimension without significant loss of information by using a proxy that aggregates the matter bispectrum over a subset of Fourier configurations. In particular they obtained their best results in terms of constraints on $\Lambda$CDM parameters from combining the power spectrum with the modal decomposition of the bispectrum \citep{Fergusson:2010ia,PhysRevD.86.123524}. Their results demonstrate that the modal bispectrum performs as well as the Fourier bispectrum, even with considerably fewer modes ($\,10\,$) than Fourier configurations ($\,95\,$).

The main difference with the approach presented in this work is that we use the original full galaxy bispectrum data-vector in order to compress it. This does not have any limitations in terms of original size of the bispectrum data-vector. Therefore it allows us to access the full information content achievable through the bispectrum. Moreover the number of elements of the compressed data-vector in our case is independent of the number of elements of the original data-vector. In our case we considered $\sim1000$ triangle configurations and $7$ parameters which gives a compression of at least two orders of magnitude. In \citet{Byun:2017fkz} it is left for future work to check whether the achieved compression of the bispectrum data vector by an order of magnitude would improve if more triangle configurations are taken into consideration.

\item As already quantified using real data-sets like BOSS \citep{Gil-Marin:2016wya}, the use of the joint data-vector $\left[\mathrm{P}^{\mathrm{\mathrm{s}}}_{\mathrm{\mathrm{g}}},\mathrm{B}^{\mathrm{\mathrm{s}}}_{\mathrm{\mathrm{g}}}\right]$ significantly increases the constraining power on the cosmological parameters compared to using only the power spectrum or bispectrum individually. In the case of an ELGs sample for a DESI-like survey the improvement obtaining by combining power spectrum and bispectrum is quantified and can be visualised in the second panel of Figure \ref{fig:err_comp} and the second and third rows of Table \ref{tab:recap_table}. The constraints obtained considering power spectrum and bispectrum together are up to $52\%$ smaller than the constraints obtained using only the bispectrum. The difference is even greater with respect to the power spectrum alone. Together with lifting the degeneracies between amplitude-like parameters as it can be seen in Figure \ref{fig:3stats_mcmc}, the 68$\%$ confidence intervals of the marginalised 1D posteriors for the joint data-vector are up to $\sim5$ times smaller than the power spectrum ones.

\item  Using the compression PCA + KL as analysis method is much faster than MCMC and less computationally demanding (few minutes on a single processor compared to days using several processors working in parallel). Since it is relatively easy to implement, it can be used to "sample" different sets of parameters and obtain reliable constraints for a given model without having to wait days for each one of them as in the case of the MCMC.

\item The pre-compression PCA transformation allows us to better capture the nature of the degeneracies between the chosen parameters, returning realistic multidimensional posterior distributions that follow closely the MCMC ones. Also in the case of "strong" degeneracies, orthogonalising the parameter space before compressing returns contours qualitatively more realistic and closer to the MCMC than orthogonalising the weights after compressing.
\end{itemize}

Future work will include the study of how the compression method applied here is affected by the choice of the set of considered parameters. 
Finally we would like to test the methods presented here using BOSS data or simulations for the upcoming DESI, EUCLID and PFS surveys. Using MCMC + KL or PCA + KL compression methods has the potential of becoming a standard fast and reliable approach to adopt when dealing with large data-vector as in the case of higher order statistics.

\section*{Acknowledgement} 

The authors thank the anonymous referee for the useful comments. D.G. is grateful to William Jennings and Keir Rogers for carefully reading the manuscript.
D.G. is supported by the Perren and the IMPACT funds.
M.M. acknowledges funding from STFC Consolidated Grants RG84196 and
RG70655 LEAG/506 and has received funding from the European Union’s Horizon 2020 research and innovation programme under Marie Skłodowska-Curie grant agreement No 6655919.
B.J. acknowledges support by an STFC Ernest Rutherford Fellowship, grant reference ST/J004421/1. 
O.L. acknowledges support from a European Research Council Advanced Grant FP7/291329.
C \citep{Kernighan:1988:CPL:576122} and Python 2.7 \citep{Rossum:1995:PRM:869369} have been used together with many packages like IPython \citep{Perez:2007:ISI:1251563.1251831}, Numpy \citep{DBLP:journals/corr/abs-1102-1523}, Scipy \citep{jones} and Matplotlib \citep{Hunter:2007:MGE:1251563.1251845}.




\bibliographystyle{mnras}
\nocite{Zablocki:2015zcm}
\nocite{Tegmark:1996bz}
\nocite{Sefusatti:2006pa}
\nocite{Heavens:1998es}
\nocite{1984ApJ...279..499F}
\nocite{2009A&A...508.1193J}
\nocite{Gil-Marin:2016wya}
\nocite{Scoccimarro:1997st}
\nocite{Matarrese:1997sk}
\nocite{Verde:1998zr}
\nocite{Fry:1992vr}
\nocite{2013PASP..125..306F}


\bibliography{reference.bib}


\end{multicols*}

\appendix

\section{Redshift space kernels}
\label{sec:png}

\noindent The redshift space kernels 
derived from standard perturbation theory as reported in \citet{Matarrese:1997sk} and used in the expression for the power spectrum and bispectrum in this work are given by:

\begin{align}
&\mathrm{F}^{(1)}_{\mathrm{\mathrm{s}}}\left[\bm{k}\right]\,=\,b_1\,+\,f\mu^2;
\notag \\
\notag \\
&\mathrm{F}^{(2)}_{\mathrm{\mathrm{s}}}\left[\bm{k}_1,\bm{k}_2\right]\,=
\,b_1J^{(2)}_{\mathrm{\mathrm{s}}}\left[\bm{k}_1,\bm{k}_2\right]\,+\,f\mu^2\,K^{(2)}_{\mathrm{\mathrm{s}}}\left[\bm{k}_1,\bm{k}_2\right]\,+\,\dfrac{1}{2}b_2
+\dfrac{b_1f}{2}\left[\mu_1^2\,+\,\mu_2^2\,+\,\mu_1\mu_2\left(\dfrac{k_1}{k_2}\,+\,\dfrac{k_2}{k_1}\right)\right]
+ f^2\left[\mu_1^2\mu_2^2\,+\,\dfrac{\mu_1\mu_2}{2}\left(\mu_1^2\dfrac{k_1}{k_2}\,+\,\mu_2^2\dfrac{k_2}{k_1}\right)\right];
\notag\\
\notag \\
&\mathrm{F}^{(3)}_{\mathrm{\mathrm{s}}}\left[\bm{k}_1,\bm{k}_2,\bm{k}_3\right]\,
= b_1J_s^{(3)}\left[\bm{k}_1,\bm{k}_2,\bm{k}_3\right]+f\mu^2K^{(3)}\left[\bm{k}_1,\bm{k}_2,\bm{k}_3\right]
+\dfrac{b_2}{2}f\mu^2_3+\dfrac{b_3}{6}+\dfrac{b_2}{2}f\mu_1\mu_2\dfrac{k_2}{k_1}
+\dfrac{b_2}{2}f\mu_1\mu_3\dfrac{k_3}{k_1}
\notag \\ 
&+b_1f^2\mu_2^2\mu_3^2 
+2b_1f^2\mu_1\mu_2\mu_3^2\dfrac{k_1}{k_2}+b_1f^2\mu_2\mu_3^3\dfrac{k_3}{k_2}+\dfrac{b_1}{2}\mu_1^2\mu_2\mu_3\dfrac{k_1^2}{k_2k_3}
+f^3\mu^2_1\mu^2_2\mu^2_3+3f^3\mu_1\mu_2^2\mu_3^3\dfrac{k_3}{k_1}+\dfrac{1}{2}f^3\mu_1\mu_2\mu_3^4\dfrac{k_3^2}{k_1k_2} 
\notag\\ 
&+J^{(2)}_{\mathrm{\mathrm{s}}}\left[\bm{k}_2,\bm{k}_3\right]\left(b_2+b_1f\mu_1^2+b_1f\mu_1\mu_{2+3}\dfrac{k_{2+3}}{k_1}\right)
\notag\\ 
&+ K^{(2)}_{\mathrm{\mathrm{s}}}\left[\bm{k}_2,\bm{k}_3\right]\bigg(b_1f\mu_{2+3}^2+b_1f\mu_1\mu_{2+3}\dfrac{k_1}{k_{2+3}}+2f^2\mu_1^2\mu^2_{2+3}
+f^2\mu_1\mu^3_{2+3}\dfrac{k_{2+3}}{k_1}+f^2\mu_1^3\mu_{2+3}\dfrac{k_1}{k_{2+3}}\bigg),
\end{align}

\noindent where the $K$'s and $J$'s are given by:
 
\begin{align}
&J^{(2)}_{\mathrm{\mathrm{s}}}\left[\bm{k}_1,\bm{k}_2\right]\,=\,\dfrac{5}{7}\,+\,\dfrac{\bm{k}_1\cdot\bm{k}_2}{2k_1k_2}\left(\dfrac{k_1}{k_2}\,+\,\dfrac{k_2}{k_1}\right)\,+\,\dfrac{2}{7}\left(\dfrac{\bm{k}_1\cdot\bm{k}_2}{k_1k_2}\right)^2;
\notag \\
\notag\\
&K^{(2)}_{\mathrm{\mathrm{s}}}\left[\bm{k}_1,\bm{k}_2\right]\,=\,\dfrac{3}{7}\,+\,\dfrac{\bm{k}_1\cdot\bm{k}_2}{2k_1k_2}\left(\dfrac{k_1}{k_2}\,+\,\dfrac{k_2}{k_1}\right)\,+\,\dfrac{4}{7}\left(\dfrac{\bm{k}_1\cdot\bm{k}_2}{k_1k_2}\right)^2;
\notag\\
\end{align}

\begin{align}
&J^{(3)}_{\mathrm{\mathrm{s}}}\left[\bm{k}_1,\bm{k}_2,\bm{k}_3\right]\,=\,
J^{(2)}_{\mathrm{\mathrm{s}}}\left[\bm{k}_2,\bm{k}_3\right]\left[\dfrac{1}{3}+\dfrac{1}{3}\dfrac{\bm{k_1}\cdot\left(\bm{k}_2+\bm{k}_3\right)}{\left(\bm{k}_2+\bm{k}_3\right)^2}+\dfrac{4}{9}\dfrac{\bm{k}\cdot\bm{k}_1}{k_1^2}\dfrac{\bm{k}\cdot\left(\bm{k}_2+\bm{k}_3\right)}{\left(\bm{k}_2+\bm{k}_3\right)^2}\right]
\notag\\
&-\dfrac{2}{9}\dfrac{\bm{k}\cdot\bm{k}_1}{k_1^2}\dfrac{\bm{k}\cdot\left(\bm{k}_2+\bm{k}_3\right)}{\left(\bm{k}_2+\bm{k}_3\right)^2}\dfrac{\bm{k}_3\cdot\left(\bm{k}_2+\bm{k}_3\right)}{k_3^2}+\dfrac{1}{9}\dfrac{\bm{k}\cdot\bm{k}_2}{k_2^2}\dfrac{\bm{k}\cdot\bm{k}_3}{k_3^2};
\notag\\
\notag
\end{align}

\begin{align}
&K^{(3)}_{\mathrm{\mathrm{s}}}\left[\bm{k}_1,\bm{k}_2,\bm{k}_3\right]\,=\,3J^{(3)}_{\mathrm{\mathrm{s}}}\left[\bm{k}_1,\bm{k}_2,\bm{k}_3\right]-\dfrac{\bm{k}\cdot\bm{k}_1}{k_1^2}J^{(2)}_{\mathrm{\mathrm{s}}}\left[\bm{k}_2,\bm{k}_3\right]
-\dfrac{\bm{k}\cdot\left(\bm{k}_1+\bm{k}_2\right)}{\left(\bm{k}_1+\bm{k}_2\right)^2}K^{(2)}_{\mathrm{\mathrm{s}}}\left[\bm{k}_1,\bm{k}_2\right].
\end{align}

\noindent In the above expressions we defined $\mu = k_i^\parallel/k_i$, $\bm{k}=\bm{k}_1+\bm{k}_2+\bm{k}_3$ and $\mu_{2+3}\equiv(k_2^\parallel+k_3^\parallel)/|\bm{k}_2+\bm{k}_3|$.

\section{Estimators definition and unbiasedness check}
\label{sec:est_def}

In this appendix we present the definition of the power spectrum and the bispectrum estimators, as well as the definition of the trispectrum and the tetraspectrum, which are necessary for the computation of the full covariance  matrix for the joint data-vector $\left[\mathrm{P}^{\mathrm{\mathrm{s}}}_{\mathrm{\mathrm{g}}}, \mathrm{B}^{\mathrm{\mathrm{s}}}_{\mathrm{\mathrm{g}}}\right]$.

Including the effect of redshift-space distortions in our analysis requires enlarging the number of parameters needed to describe a particular configuration of the bispectrum. In this case, it is possible to see that, considering the distant observer approximation, which consists in assuming that all the line of sight vectors paired to each wave-vector are parallel between each others, the natural symmetry to exploit is the cylindrical one. In other words, what characterises the redshift effect on the bispectrum are the parallel to the line of sight components of the wave-vectors, from now on labeled as $k^\parallel$. Therefore the natural set of coordinates to describe each wave-vector  is the cylindrical one: $\left(k^\parallel, k^\perp, \phi\right)$ where $k^\perp$ and $\phi$ describe the component of the wave-vector laying on the perpendicular plane to the line of sight.
Hence, for what concerns the bispectrum in redshift-space, from the original 9 degrees of freedom, 3 are canceled by translational invariance given by the closed triangle condition $\delta_{\mathrm{D}}\left(\bm{k}_1+ \bm{k}_2+ \bm{k}_3\right)$.

 Moreover, from the chosen coordinates above it is possible to see that there is a further symmetry which is the rotation along the line of sight of a particular triangle of wave-vectors. Different configurations given simply by rotating the same triangle around the line of sight give the same value for the bispectrum. The coordinate $\phi$ describes these rotations and expresses this symmetry. 
As a consequence of the symmetries mentioned, we are left with five remaining degrees of freedom, describing all possible configurations, which are given by:
$\{k_1^\perp,k_2^\perp,k_3^\perp,k_1^\parallel,k_2^\parallel\}$, from which we can derive all the other quantities (e.g. $k_3^\parallel=-k_1^\parallel-k_2^\parallel$).
It is necessary then to define a new type of bin for our estimator: a cylindrical anulii defined by $\phi\in\left[0,\,2\pi\right]$, $\bar{k}^\perp\in\left[\bar{k}^\perp - \Delta k^\perp/2,\bar{k}^\perp + \Delta k^\perp/2\right]$ and $\bar{k}^\parallel \in\left[\bar{k}^\parallel - \Delta k^\parallel/2,\bar{k}^\parallel + \Delta k^\parallel/2\right]$. For thin anulii the surface area is given by $A_c = 2\pi\Delta k \,\bar{k}^\perp$ and hence the volume is given by $V^{\mathrm{c}}=2\pi\Delta k^2 \,\bar{k}^\perp $. The 3D Dirac's delta can be decomposed as
\begin{eqnarray}
\delta_{\mathrm{D}}\left(\bm{k}_1+\bm{k}_2+\bm{k}_3\right)\,=\,\delta_{\mathrm{D}}\left(k_1^\parallel+k_2^\parallel+k_3^\parallel\right)\,\delta_{\mathrm{D}}\left(\bm{k}_1^\perp+\bm{k}_2^\perp+\bm{k}_3^\perp\right).
\end{eqnarray}

\subsection{Power spectrum}

A standard way to define an estimator for the power spectrum is the following \citep{1980lssu.book.....P}:

\begin{align}
\label{pk_rsd_est}
 \hat{\mathrm{P}}^{\mathrm{\mathrm{s}}}_{\mathrm{\mathrm{g}}}\left(\bm{\bar{k}}_1\right) = 
 \dfrac{1}{V_{\mathrm{\mathrm{s}}}}\int_{V^{\mathrm{c}}_1}\dfrac{dV^{\mathrm{c}}_1}{V^{\mathrm{c}}_1}\int_{V^{\mathrm{c}}_1}dV^{\mathrm{c}}_2
 \delta_{\mathrm{D}}\left(\bm{k}_1+\bm{k}_2\right)\delta_{\mathrm{\mathrm{g}}}^{\mathrm{s}}\left(\bm{k}_1\right)\delta_{\mathrm{\mathrm{g}}}^{\mathrm{s}}\left(\bm{k}_2\right),
\end{align}

\noindent where $V_{\mathrm{\mathrm{s}}}$ is the survey volume. Notice that both cylindrical bins are centered on $\bm{\bar{k}}_1$ and therefore by definition $\bm{\bar{k}}_1=\bm{\bar{k}}_2$. We specify that the power spectrum for the redshift galaxy field depends on the wave-vector, precisely on its perpendicular and parallel components to the line of sight, and not on its module. Following the definition given in Eq. \eqref{pk_rsd_est}, it is straight forward to check whether the estimator is unbiased; one only needs to take the average

\begin{align}
\label{expectation_pk}
&\langle\hat{\mathrm{P}}^{\mathrm{\mathrm{s}}}_{\mathrm{\mathrm{g}}}\left(\bm{\bar{k}}_1\right)\rangle = \dfrac{1}{V_{\mathrm{\mathrm{s}}}}
\int_{V^{\mathrm{c}}(\bm{k}_1)}\dfrac{dV^{\mathrm{c}}_1}{V^{\mathrm{c}}(\bm{k}_1)}
\int_{V^{\mathrm{c}}(\bm{k}_1)}dV^{\mathrm{c}}_2 \,\delta_{\mathrm{D}}\left(\bm{k}_1+\bm{k}_2\right)
\langle\delta_{\mathrm{\mathrm{g}}}^{\mathrm{s}}\left(\bm{k}_1\right)\delta_{\mathrm{\mathrm{g}}}^{\mathrm{s}}\left(\bm{k}_2\right)\rangle
\notag \\
&=\dfrac{1}{V_{\mathrm{\mathrm{s}}}}
\int_{V^{\mathrm{c}}(\bm{k}_1)}\dfrac{dV^{\mathrm{c}}_1}{V^{\mathrm{c}}(\bm{k}_1)}
\int_{V^{\mathrm{c}}(\bm{k}_1)}dV^{\mathrm{c}}_2 \,\delta_{\mathrm{D}}\left(\bm{k}_1+\bm{k}_2\right)^2(2\pi)^3 \,\mathrm{P}^{\mathrm{\mathrm{s}}}_{\mathrm{\mathrm{g}}}(\bm{k}_1)
\notag \\
&= \int_{V^{\mathrm{c}}(\bm{k}_1)}\dfrac{dV^{\mathrm{c}}_1}{V^{\mathrm{c}}(\bm{k}_1)}
\int_{V^{\mathrm{c}}(\bm{k}_1)}dV^{\mathrm{c}}_2 \,\delta_{\mathrm{D}}\left(\bm{k}_1+\bm{k}_2\right) \,\mathrm{P}^{\mathrm{\mathrm{s}}}_{\mathrm{\mathrm{g}}}(\bm{k}_1)
\,=\, \int_{V^{\mathrm{c}}(\bm{k}_1)}\dfrac{dV^{\mathrm{c}}_1}{V^{\mathrm{c}}(\bm{k}_1)}
\,\mathrm{P}^{\mathrm{\mathrm{s}}}_{\mathrm{\mathrm{g}}}(\bm{k}_1)
\notag \\
&= \int_{\bar{k}^\perp_1-\frac{\Delta k }{2}}^{\bar{k}^\perp_1+\frac{\Delta k }{2}}\int_{\bar{k}^\parallel_1-\frac{\Delta k }{2}}^{\bar{k}^\parallel_i+\frac{\Delta k }{2}}\dfrac{dk^\perp_idk^\parallel_i\,k^\perp_i}{\bar{k}_i^\perp\Delta k^2} \,\mathrm{P}^{\mathrm{\mathrm{s}}}_{\mathrm{\mathrm{g}}}(\bm{k}_1) \,\approx\,\mathrm{P}^{\mathrm{\mathrm{s}}}_{\mathrm{\mathrm{g}}}(\bar{\bm{k}}_1).
\end{align}

\noindent 
where the standard expression for the galaxy power spectrum at leading order is given by

\begin{align}
\mathrm{P}_{\mathrm{\mathrm{g}}}^{\mathrm{\mathrm{s}}}(\bm{k}_1) = 
\mathrm{P}^{\mathrm{m}}(k_1)\,
\mathrm{F}_{\mathrm{s}}^{(1)}\left(\bm{k}_1\right)^2\,.
\end{align}

\subsection{Bispectrum estimator}

Proceeding now as shown in \citet{2009A&A...508.1193J}, we define the following estimator for the bispectrum

\begin{align}
\label{bk_rsd_est}
&\hat{\mathrm{B}}^{\mathrm{\mathrm{s}}}_{\mathrm{\mathrm{g}}}\left(\bar{k}_1^\perp,\bar{k}_2^\perp,\bar{k}_3^\perp,\bar{k}_1^\parallel,\bar{k}_2^\parallel\right)\,=\dfrac{(2\pi)^2\Delta k_3^\parallel}{V}\Lambda^{-1}\left(\bar{k}_1^\perp,\bar{k}_2^\perp,\bar{k}_3^\perp,\bar{k}_1^\parallel,\bar{k}_2^\parallel\right)
\prod_{i=1}^3\int_{V^{\mathrm{c}}_i}\dfrac{dV^{\mathrm{c}}_i}{V^{\mathrm{c}}_i}\,
\delta_{\mathrm{D}}\left(\bm{k}_1+\bm{k}_2+\bm{k}_3\right)\delta_{\mathrm{\mathrm{g}}}^{\mathrm{s}}\left(\bm{k}_1\right)\delta_{\mathrm{\mathrm{g}}}^{\mathrm{s}}\left(\bm{k}_2\right)\delta_{\mathrm{\mathrm{g}}}^{\mathrm{s}}\left(\bm{k}_3\right),
\end{align}

\noindent where $\Lambda$ is a function related to the fraction of wave-vectors triplets allowed by the triangle condition and defined such that the bispectrum estimator \eqref{bk_rsd_est} is unbiased, which can be checked by taking the average of the estimator, which also it is reported in the appendix. The expression for $\Lambda$ is

\begin{align}
&\Lambda\left(k^\perp_1,k^\perp_2,k^\perp_3\right)= 2\pi \int_0^\infty dr^\perp r^\perp \prod^{3}_{i=1}J_0\left(k_i^\perp r^\perp\right)
=\dfrac{4}{\sqrt[2]{2{k^{\perp}_1}^2{k^{\perp}_2}^2+2{k^{\perp}_1}^2{k^{\perp}_3}^2+2{k^{\perp}_2}^2{k^{\perp}_3}^2-{k^{\perp}_1}^4-{k^{\perp}_2}^4-{k^{\perp}_3}^4}} ,
\end{align}
\noindent if $|k^{\perp}_1 - k^{\perp}_2| < k^{\perp}_3 < k^{\perp}_1 + k^{\perp}_2$ or $0$ otherwise. $J_0$ is the zero-th order spherical Bessel function. It is also possible to check that for the bispectrum the estimator is unbiased

\begin{align}
\label{expectation_bk}
&\langle\hat{\mathrm{B}}^{\mathrm{\mathrm{s}}}_{\mathrm{\mathrm{g}}}\left(\bar{k}_1^\perp,\bar{k}_2^\perp,\bar{k}_3^\perp,\bar{k}_1^\parallel,\bar{k}_2^\parallel\right)\rangle 
\,=\, \dfrac{(2\pi)^2\Delta k_3^\parallel}{V_{\mathrm{\mathrm{s}}}}\Lambda^{-1}\left(\bar{k}_1^\perp,\bar{k}_2^\perp,\bar{k}_3^\perp,\bar{k}_1^\parallel,\bar{k}_2^\parallel\right)
\int_{V^{\mathrm{c}}_1}\int_{V^{\mathrm{c}}_2}\int_{V^{\mathrm{c}}_3}(2\pi)^3\left(\delta_{\mathrm{D}}\left(\bm{k}_1+\bm{k}_2+\bm{k}_3\right)\right)^2\mathrm{B}^{\mathrm{\mathrm{s}}}_{\mathrm{\mathrm{g}}}\left(k_1^\perp,k_2^\perp,k_3^\perp,k_1^\parallel,k_2^\parallel\right)
\notag\\
&=(2\pi)^2\Delta k_3^\parallel\Lambda^{-1}_{123}\int_{V^{\mathrm{c}}_1}\int_{V^{\mathrm{c}}_2}\int_{V^{\mathrm{c}}_3}\delta_{\mathrm{D}}\left(\bm{k}_1+\bm{k}_2+\bm{k}_3\right)\mathrm{B}_{g,123}^{\mathrm{s}},
\end{align}

\noindent where a shorthand notation for the bispectrum has been introduced and where we have used the approximation $\delta_{\mathrm{D}}^2 \simeq V_\mathrm{s}/(2\pi)^3 \delta_{\mathrm{D}}$ from \citet{2009A&A...508.1193J}. Since the bispectrum is invariant under rotation around the line of sight, we integrate now the angular part, namely over $\phi$

\begin{align}
\label{lambda}
&\int_0^{2\pi}d\phi_1\int_0^{2\pi}d\phi_2\int_0^{2\pi}d\phi_3\,\delta_{\mathrm{D}}\left(\bm{k}_1^\perp+\bm{k}_2^\perp+\bm{k}_3^\perp\right)
=\int d\phi_1d\phi_2d\phi_3\int\dfrac{d^2\bm{r}^\perp}{(2\pi)^2}\,e^{\left(\bm{k}_1^\perp+\bm{k}_2^\perp+\bm{k}_3^\perp\right)\bm{r}^\perp}
\notag\\
&=\int\dfrac{d^2\bm{r}^\perp}{(2\pi)^2}\int_0^{2\pi}d\phi_1\,e^{ik_1^\perp r^\perp \cos(\phi_1-\phi_r)}
\notag \\
&\times\int_0^{2\pi}d\phi_2e^{ik_2^\perp r^\perp \cos(\phi_2-\phi_r)}\int_0^{2\pi}d\phi_3e^{ik_3^\perp r^\perp \cos(\phi_3-\phi_r)}
\notag \\
&=\int\dfrac{d^2\bm{r}^\perp}{(2\pi)^2}(2\pi)^3\,J_0(k_1^\perp r^\perp)J_0(k_2^\perp r^\perp)J_0(k_3^\perp r^\perp)
=(2\pi)^2\int_0^\infty dr^\perp \,r^\perp\prod_{i=1}^{3}J_0(k_i^\perp r^\perp) \,=\, 2\pi\Lambda(k_1^\perp,k_2^\perp,k_3^\perp),
\end{align}

\noindent from which it is possible to see that in this case $\Lambda$ depends only on the perpendicular components of the wave-vectors. This agrees with the fact that on the orthogonal plane the wave-vectors components must form a closed triangle. In one of the steps above we used the argument that integrating a trigonometric function over an interval of $2\pi$ gives the same result no matter what are the extrema of integration as long as the interval is of $2\pi$. For completeness the Bessel functions of order 0 are defined as

\begin{eqnarray}
J_0(x)\,=\,\int_0^{2\pi}\dfrac{d\phi}{2\pi}\,e^{ix\cos\phi}.
\end{eqnarray}

\noindent Then, inserting the expression for $\Lambda$ \eqref{lambda} into \eqref{expectation_bk} it is possible to verify that our estimator is unbiased

\begin{align}
&\langle\hat{\mathrm{B}}^{\mathrm{\mathrm{s}}}_{\mathrm{g},123}\rangle=(2\pi)^2\Delta k_3^\parallel\Lambda^{-1}_{123}
\prod_{i=1}^3\int_{V^{\mathrm{c}}_i}\dfrac{d^2\bm{k}_i^\perp dk_i^\parallel}{2\pi\bar{k}_i^\perp\Delta k^2}
\delta_{\mathrm{D}}\left(\bm{k}_1+\bm{k}_2+\bm{k}_3\right)\mathrm{B}^{\mathrm{\mathrm{s}}}_{\mathrm{g}}\left(k^\perp_1,k^\perp_2,k^\perp_3,k^\parallel_1,k^\parallel_2\right) 
\notag\\
&=(2\pi)^2\Delta k_3^\parallel\Lambda^{-1}_{123}\prod_{i=1}^3\int_{\bar{k}^\perp_i-\frac{\Delta k }{2}}^{\bar{k}^\perp_i+\frac{\Delta k }{2}}\int_{\bar{k}^\parallel_i-\frac{\Delta k }{2}}^{\bar{k}^\parallel_i+\frac{\Delta k }{2}}\dfrac{dk^\perp_idk^\parallel_i\,k^\perp_i}{2\pi\bar{k}_i^\perp\Delta k^2}
,2\pi\Lambda_{123}\,\delta_{\mathrm{D}}\left(k^\parallel_1 + k^\parallel_2 + k^\parallel_3\right)\mathrm{B}^{\mathrm{\mathrm{s}}}_{\mathrm{g},123}
\notag\\
&\approx\,\delta^{\mathrm{K}}_{123}\mathrm{B}_{\mathrm{\mathrm{g}}}^{\mathrm{\mathrm{s}}}\left(\bar{k}_1^\perp,\bar{k}_2^\perp,\bar{k}_3^\perp,\bar{k}_1^\parallel,\bar{k}_2^\parallel\right),
\end{align}

\noindent where $\delta^{\mathrm{K}}$ is a Kronecker delta. In the last step the thin-shell approximation has been used in order to bring out the bispectrum from the integrals. 
The standard expression for the galaxy bispectrum is given by

\begin{align}
\mathrm{B}_{\mathrm{\mathrm{g}}}^{\mathrm{\mathrm{s}}}(\bm{k}_1,\bm{k}_2,\bm{k}_3) = 
2\,\mathrm{P}^{\mathrm{m}}(k_1)\,\mathrm{P}^{\mathrm{m}}(k_2)\,
\mathrm{F}_{\mathrm{s}}^{(1)}\left(\bm{k}_1\right)\,
\mathrm{F}_{\mathrm{s}}^{(1)}\left(\bm{k}_2\right)\,
\mathrm{F}_{\mathrm{s}}^{(2)}\left[\bm{k}_1,\bm{k}_2\right] \quad+\quad2\,\mathrm{p.} \,.
\end{align}

\subsection{Trispectrum definition}

The trispectrum is defined as

\begin{eqnarray}
&\langle \delta^{\mathrm{\mathrm{s}}}_{\mathrm{\mathrm{g}}}(\bm{k}_1)\delta^{\mathrm{\mathrm{s}}}_{\mathrm{\mathrm{g}}}(\bm{k}_2)\delta^{\mathrm{\mathrm{s}}}_{\mathrm{\mathrm{g}}}(\bm{k}_3)\delta^{\mathrm{\mathrm{s}}}_{\mathrm{\mathrm{g}}}(\bm{k}_4)\rangle_{\mathrm{c}}
\,=\,(2\pi)^3\,\delta_{\mathrm{D}}\left(\bm{k}_1+\bm{k}_2+\bm{k}_3+\bm{k}_4\right)\mathrm{T}^{\mathrm{\mathrm{s}}}_{\mathrm{\mathrm{g}}}\left(\bm{k}_1,\bm{k}_2,\bm{k}_3,\bm{k}_4\right),
\end{eqnarray}

\noindent where the subscript "c" indicates that trispectrum is the connected part of the four points correlation function. Therefore the leading order terms are of two types: $\mathrm{T}^{\mathrm{s}\,(2)}_{\mathrm{\mathrm{g}}}$ and $\mathrm{T}^{\mathrm{s}\,(3)}_{\mathrm{\mathrm{g}}}$ respectively characterised by the RSD perturbation kernels $\mathrm{F}_{\mathrm{s}}^{(2)}$ and $\mathrm{F}_{\mathrm{s}}^{(3)}$. Starting from the first one we have that:

\begin{align}
&(2\pi)^3\delta_{\mathrm{D}}\left(\bm{k}_1 +\bm{k}_2 + \bm{k}_3 +\bm{k}_4\right)\mathrm{T}^{\mathrm{s}\,(2)}_{\mathrm{\mathrm{g}}}
\left(\bm{k}_1,\bm{k}_2,\bm{k}_3,\bm{k}_4\right)= 
\notag\\
&=\langle \mathrm{F}^{(1)}_{\mathrm{\mathrm{s}}}\left[\bm{k}_1\right]\mathrm{F}^{(1)}_{\mathrm{\mathrm{s}}}\left[\bm{k}_2\right]\delta_{\bm{k}_1}\delta_{\bm{k}_2}
\dfrac{1}{(2\pi)^3}\int d^3k_a d^3k_b \,\delta_{\mathrm{D}}\left(\bm{k}_a+\bm{k}_b-\bm{k}_3 \right)\mathrm{F}_{\mathrm{s}}^{(2)}\left[\bm{k}_a,\bm{k}_b\right]\delta_{\bm{k}_a}\delta_{\bm{k}_b}
\notag \\
&\times\int d^3k_c d^3k_d \,\delta_{\mathrm{D}}\left(\bm{k}_c+\bm{k}_d-\bm{k}_4 \right)\mathrm{F}_{\mathrm{s}}^{(2)}\left[\bm{k}_c,\bm{k}_d\right]\delta_{\bm{k}_c}\delta_{\bm{k}_d}\rangle\,+\, 5\,\mathrm{p.}
\notag \\
&= \dfrac{1}{(2\pi)^6}\,\mathrm{F}^{(1)}_{\mathrm{\mathrm{s}}}\left[\bm{k}_1\right]\mathrm{F}^{(1)}_{\mathrm{\mathrm{s}}}\left[\bm{k}_2\right]
\int d^3k_a d^3k_b \,\delta_{\mathrm{D}}\left(\bm{k}_a+\bm{k}_b-\bm{k}_3 \right)\mathrm{F}_{\mathrm{s}}^{(2)}\left[\bm{k}_a,\bm{k}_b\right]
\int d^3k_c d^3k_d \,\delta_{\mathrm{D}}\left(\bm{k}_c+\bm{k}_d-\bm{k}_4 \right)\mathrm{F}_{\mathrm{s}}^{(2)}\left[\bm{k}_c,\bm{k}_d\right]
\notag \\
&\times\langle\delta_{\bm{k}_1}\delta_{\bm{k}_2}\delta_{\bm{k}_a}\delta_{\bm{k}_b}\delta_{\bm{k}_c}\delta_{\bm{k}_d} \rangle \,\,\, +\,\,\, 5\,\mathrm{p.}.
\end{align}

\noindent From the last line of the previous expression it follows that, when Wick's theorem is applied, there are three different ways to pair the wave-vectors.

\begin{align}
\mathrm{T}^{\mathrm{s}\,(2)}_{\mathrm{\mathrm{g}}} 
\begin{cases}
\mathrm{T}^{\mathrm{s}\,(2a)}_{\mathrm{\mathrm{g}}} &\Leftrightarrow \langle\delta_{\bm{k}_1}\delta_{\bm{k}_2}\rangle\langle\delta_{\bm{k}_a}\delta_{\bm{k}_c}\rangle\langle\delta_{\bm{k}_b}\delta_{\bm{k}_d}\rangle \times 2 \quad (\mathrm{sym.}) \notag \\
\mathrm{T}^{\mathrm{s}\,(2b)}_{\mathrm{\mathrm{g}}} &\Leftrightarrow \langle\delta_{\bm{k}_1}\delta_{\bm{k}_a}\rangle\langle\delta_{\bm{k}_2}\delta_{\bm{k}_c}\rangle\langle\delta_{\bm{k}_b}\delta_{\bm{k}_d}\rangle \times 4 \quad (\mathrm{sym.}) \notag \\
\mathrm{T}^{\mathrm{s}\,(2c)}_{\mathrm{\mathrm{g}}} &\Leftrightarrow \langle\delta_{\bm{k}_1}\delta_{\bm{k}_c}\rangle\langle\delta_{\bm{k}_2}\delta_{\bm{k}_a}\rangle\langle\delta_{\bm{k}_b}\delta_{\bm{k}_d}\rangle \times 4 \quad (\mathrm{sym.}) 
\end{cases}
\end{align}

\noindent The first term represents a non connected 1-loop correction to the power spectrum covariance matrix \citep{1984ApJ...279..499F, 2017MNRAS.466..780M}. For completeness, we  just show below that the simplified  expression makes explicit the fact that it is an unconnected term of the full 4-points correlator

\begin{align}
\label{t2a_term}
&(2\pi)^3\delta_{\mathrm{D}}\left(\bm{k}_1 +\bm{k}_2 + \bm{k}_3 +\bm{k}_4\right) \mathrm{T}^{\mathrm{s}\,(2a)}_{\mathrm{\mathrm{g}}}
\left(\bm{k}_1,\bm{k}_2,\bm{k}_3,\bm{k}_4\right)=
\notag \\
&=\dfrac{2}{(2\pi)^6}\mathrm{F}^{(1)}_{\mathrm{\mathrm{s}}}\left[\bm{k}_1\right]\mathrm{F}^{(1)}_{\mathrm{\mathrm{s}}}\left[\bm{k}_2\right]\int  d^3k_a d^3k_b d^3k_c d^3k_d \delta_{\mathrm{D}}\left(\bm{k}_a+\bm{k}_b-\bm{k}_3 \right)
\delta_{\mathrm{D}}\left(\bm{k}_c+\bm{k}_d-\bm{k}_4 \right)
\mathrm{F}_{\mathrm{s}}^{(2)}\left[\bm{k}_a,\bm{k}_b\right]\mathrm{F}_{\mathrm{s}}^{(2)}\left[\bm{k}_c,\bm{k}_d\right]
\notag \\
&\times (2\pi)^9 \,\delta_{\mathrm{D}}\left(\bm{k}_1+\bm{k}_2\right)
\delta_{\mathrm{D}}\left(\bm{k}_a+\bm{k}_c\right)\delta_{\mathrm{D}}\left(\bm{k}_b+\bm{k}_d\right)\,\mathrm{P}^{\mathrm{m}}_{k_1}\mathrm{P}^{\mathrm{m}}_{k_a}\mathrm{P}^{\mathrm{m}}_{k_b} 
\,\,\, +\,\,\, 5\,\mathrm{p.}
\notag \\
&=2(2\pi)^3\delta_{\mathrm{D}}\left(\bm{k}_1+\bm{k}_2\right)\mathrm{P}^{\mathrm{s}}_{\mathrm{g}}\left(\bm{k}_1\right)
\int d^3k_a d^3k_b\, \mathrm{F}^{(2)}_s\left[\bm{k}_a,\bm{k}_b\right]\mathrm{F}^{(2)}_s\left[-\bm{k}_a,-\bm{k}_b\right]\delta_{\mathrm{D}}\left(\bm{k}_a+\bm{k}_b-\bm{k}_3 \right)\delta_{\mathrm{D}}\left(\bm{k}_3+\bm{k}_4 \right)\mathrm{P}^{\mathrm{m}}_{k_a}\mathrm{P}^{\mathrm{m}}_{k_b} 
\,\,\, +\,\,\, 5\,\mathrm{p.}
\notag \\
& = 2(2\pi)^3\delta_{\mathrm{D}}\left(\bm{k}_1+\bm{k}_2\right)\delta_{\mathrm{D}}\left(\bm{k}_3+\bm{k}_4\right)
\mathrm{P}^{\mathrm{s}}_{\mathrm{g}}\left(\bm{k}_1\right)
\int d^3k_a \,\mathrm{F}^{(2)}_s\left[\bm{k}_a,\bm{k}_3-\bm{k}_a\right]^2
\mathrm{P}^{\mathrm{m}}_{k_a}\mathrm{P}^{\mathrm{m}}_{|\bm{k}_3-\bm{k}_a|} 
\,\,\, +\,\,\, 5\,\mathrm{p.}
\notag \\
& = 2(2\pi)^3\delta_{\mathrm{D}}\left(\bm{k}_1+\bm{k}_2 + \bm{k}_3+\bm{k}_4\right)\delta_{\mathrm{D}}\left(\bm{k}_3+\bm{k}_4\right)
\mathrm{P}^{\mathrm{s}}_{\mathrm{g}}\left(\bm{k}_1\right)
\int d^3k_a \,\mathrm{F}^{(2)}_s\left[\bm{k}_a,\bm{k}_3-\bm{k}_a\right]^2
\mathrm{P}^{\mathrm{m}}_{k_a}\mathrm{P}^{\mathrm{m}}_{|\bm{k}_3-\bm{k}_a|} 
\,\,\, +\,\,\, 5\,\mathrm{p.}
.
\end{align}

\noindent We then look at the first connected tree level term $\mathrm{T}^{\mathrm{g}\,(2b)}$

\begin{align}
&(2\pi)^3\delta_{\mathrm{D}}\left(\bm{k}_1 +\bm{k}_2 + \bm{k}_3 +\bm{k}_4\right) \mathrm{T}^{\mathrm{s}\,(2b)}_{\mathrm{\mathrm{g}}}
\left(\bm{k}_1,\bm{k}_2,\bm{k}_3,\bm{k}_4\right)=
\notag \\
&=\dfrac{4}{(2\pi)^6}\mathrm{F}^{(1)}_{\mathrm{\mathrm{s}}}\left[\bm{k}_1\right]\mathrm{F}^{(1)}_{\mathrm{\mathrm{s}}}\left[\bm{k}_2\right]\int  d^3k_a d^3k_b d^3k_c d^3k_d \delta_{\mathrm{D}}\left(\bm{k}_a+\bm{k}_b-\bm{k}_3 \right)
\delta_{\mathrm{D}}\left(\bm{k}_c+\bm{k}_d-\bm{k}_4 \right)
\mathrm{F}_{\mathrm{s}}^{(2)}\left[\bm{k}_a,\bm{k}_b\right]\mathrm{F}_{\mathrm{s}}^{(2)}\left[\bm{k}_c,\bm{k}_d\right]
\notag \\
&\times (2\pi)^9 \,\delta_{\mathrm{D}}\left(\bm{k}_1+\bm{k}_a\right)
\delta_{\mathrm{D}}\left(\bm{k}_2+\bm{k}_c\right)\delta_{\mathrm{D}}\left(\bm{k}_b+\bm{k}_d\right)\,\mathrm{P}^{\mathrm{m}}_{k_1}\mathrm{P}^{\mathrm{m}}_{k_a}\mathrm{P}^{\mathrm{m}}_{k_b} 
\,\,\, +\,\,\, 5\,\mathrm{p.}
\notag \\
&= 4\times(2\pi)^3\,{\mathrm{F}^{(1)}_{\mathrm{\mathrm{s}}}\left[\bm{k}_1\right]}{\mathrm{F}^{(1)}_{\mathrm{\mathrm{s}}}\left[\bm{k}_2\right]} \mathrm{P}^{\mathrm{m}}_{k_1}\mathrm{P}^{\mathrm{m}}_{k_2}
\int d^3k_b d^3k_d\,\delta_{\mathrm{D}}\left(\bm{k}_b-\bm{k}_1-\bm{k}_3\right)\delta_{\mathrm{D}}\left(\bm{k}_d-\bm{k}_2-\bm{k}_4\right) 
\notag \\
&\times\,\mathrm{F}_{\mathrm{s}}^{(2)}\left[-\bm{k}_1,\bm{k}_1 + \bm{k}_3\right] \mathrm{F}_{\mathrm{s}}^{(2)}\left[-\bm{k}_2,\bm{k}_2 + \bm{k}_4\right] \mathrm{P}^{\mathrm{m}}_{k_b} \delta_{\mathrm{D}}\left(\bm{k}_b+\bm{k}_d\right)
\,\,\, +\,\,\, 5\,\mathrm{p.}
\notag \\
&= 4\times(2\pi)^3\,{\mathrm{F}^{(1)}_{\mathrm{\mathrm{s}}}\left[\bm{k}_1\right]}{\mathrm{F}^{(1)}_{\mathrm{\mathrm{s}}}\left[\bm{k}_2\right]}\mathrm{F}_{\mathrm{s}}^{(2)}\left[-\bm{k}_1,\bm{k}_1 + \bm{k}_3\right]
 \mathrm{F}_{\mathrm{s}}^{(2)}\left[-\bm{k}_2,-\bm{k}_1 - \bm{k}_3\right]
\delta_{\mathrm{D}}\left(\bm{k}_1 +\bm{k}_2 + \bm{k}_3 +\bm{k}_4\right) \mathrm{P}^{\mathrm{m}}_{k_1}\mathrm{P}^{\mathrm{m}}_{k_2}\mathrm{P}^{\mathrm{m}}_{|\bm{k}_1+\bm{k}_3|},
\,\,\, +\,\,\, 5\,\mathrm{p.}.
\end{align}

\noindent from which it is possible to directly write the third term which result to be exactly the same

\begin{align}
&(2\pi)^3\delta_{\mathrm{D}}\left(\bm{k}_1 +\bm{k}_2 + \bm{k}_3 +\bm{k}_4\right)\mathrm{T}^{\mathrm{s}\,(2c)}_{\mathrm{\mathrm{g}}}
\left(\bm{k}_1,\bm{k}_2,\bm{k}_3,\bm{k}_4\right)=
\notag \\
&\,=\,4\times(2\pi)^3\,{\mathrm{F}^{(1)}_{\mathrm{\mathrm{s}}}\left[\bm{k}_1\right]}{\mathrm{F}^{(1)}_{\mathrm{\mathrm{s}}}\left[\bm{k}_2\right]}\mathrm{F}_{\mathrm{s}}^{(2)}\left[-\bm{k}_2,\bm{k}_2 + \bm{k}_3\right]  \mathrm{F}_{\mathrm{s}}^{(2)}\left[-\bm{k}_1,-\bm{k}_2 - \bm{k}_3\right]
\delta_{\mathrm{D}}\left(\bm{k}_1 +\bm{k}_2 + \bm{k}_3 +\bm{k}_4\right) \mathrm{P}^{\mathrm{m}}_{k_1}\mathrm{P}^{\mathrm{m}}_{k_2}\mathrm{P}^{\mathrm{m}}_{|\bm{k}_2+\bm{k}_3|}
\,\,\, +\,\,\, 5\,\mathrm{p.}.
\end{align}

\noindent Finally it is possible to consider the last term $T_s^{g\,(3)}$

\begin{align}
&(2\pi)^3\delta_{\mathrm{D}}\left(\bm{k}_1 +\bm{k}_2 + \bm{k}_3 +\bm{k}_4\right) \mathrm{T}^{\mathrm{s}\,(3)}_{\mathrm{\mathrm{g}}}
\left(\bm{k}_1,\bm{k}_2,\bm{k}_3,\bm{k}_4\right)=
\notag \\
&\,=\,\langle \mathrm{F}^{(1)}_{\mathrm{\mathrm{s}}}\left[\bm{k}_1\right]\mathrm{F}^{(1)}_{\mathrm{\mathrm{s}}}\left[\bm{k}_2\right]\mathrm{F}^{(1)}_{\mathrm{\mathrm{s}}}\left[\bm{k}_3\right]\delta_{\bm{k}_1}\delta_{\bm{k}_2}\delta_{\bm{k}_3}\dfrac{1}{(2\pi)^6}\,
\int d^3k_a d^3k_b d^3k_c \,\delta_{\mathrm{D}}\left(\bm{k}_a+\bm{k}_b+\bm{k}_c-\bm{k}_4 \right)
\mathrm{F}_{\mathrm{s}}^{(3)}\left[\bm{k}_a,\bm{k}_b,\bm{k}_c\right]\delta_{\bm{k}_a}\delta_{\bm{k}_b}\delta_{\bm{k}_c}\rangle
\quad +\quad 3\,\mathrm{p.}\notag \\
&=\dfrac{1}{(2\pi)^6}\mathrm{F}^{(1)}_{\mathrm{\mathrm{s}}}\left[\bm{k}_1\right]\mathrm{F}^{(1)}_{\mathrm{\mathrm{s}}}\left[\bm{k}_2\right]\mathrm{F}^{(1)}_{\mathrm{\mathrm{s}}}\left[\bm{k}_3\right]
\int d^3k_a d^3k_b d^3k_c \,\delta_{\mathrm{D}}\left(\bm{k}_a+\bm{k}_b+\bm{k}_c-\bm{k}_4 \right) 
\mathrm{F}_{\mathrm{s}}^{(3)}\left[\bm{k}_a,\bm{k}_b,\bm{k}_c\right] \langle \delta_{\bm{k}_1}\delta_{\bm{k}_2}\delta_{\bm{k}_3}\delta_{\bm{k}_a}\delta_{\bm{k}_b}\delta_{\bm{k}_c}\rangle\,
\quad +\quad 3\,\mathrm{p.}.
\end{align}

\noindent Like we did before, we apply now the Wick's theorem. In this case there are two different ways to pair the wave-vectors.

\begin{align}
\mathrm{T}^{\mathrm{s}\,(3)}_{\mathrm{\mathrm{g}}} 
\begin{cases}
\mathrm{T}^{\mathrm{s}\,(3a)}_{\mathrm{\mathrm{g}}} &\Leftrightarrow \left(\langle\delta_{\bm{k}_1}\delta_{\bm{k}_2}\rangle\langle\delta_{\bm{k}_3}\delta_{\bm{k}_a}\rangle\langle\delta_{\bm{k}_b}\delta_{\bm{k}_c}\rangle + 2\,\mathrm{p.}\right) \times 3 \, (\mathrm{sym.}) \notag \\
\mathrm{T}^{\mathrm{s}\,(3b)}_{\mathrm{\mathrm{g}}} &\Leftrightarrow \langle\delta_{\bm{k}_1}\delta_{\bm{k}_a}\rangle\langle\delta_{\bm{k}_2}\delta_{\bm{k}_b}\rangle\langle\delta_{\bm{k}_3}\delta_{\bm{k}_c}\rangle \times 6 \, (\mathrm{sym.}) 
\end{cases}
\end{align}

\noindent The first option represents again a non connected 1-loop correction to the power spectrum covariance matrix. 

\begin{align}
&(2\pi)^3\delta_{\mathrm{D}}\left(\bm{k}_1 +\bm{k}_2 + \bm{k}_3 +\bm{k}_4\right)  \mathrm{T}^{\mathrm{s}\,(3a)}_{\mathrm{\mathrm{g}}}
\left(\bm{k}_1,\bm{k}_2,\bm{k}_3,\bm{k}_4\right)=
\notag \\
&\,=\,\dfrac{3}{(2\pi)^6}\mathrm{F}^{(1)}_{\mathrm{\mathrm{s}}}\left[\bm{k}_1\right]\mathrm{F}^{(1)}_{\mathrm{\mathrm{s}}}\left[\bm{k}_2\right]\mathrm{F}^{(1)}_{\mathrm{\mathrm{s}}}\left[\bm{k}_2\right]
 \int d^3k_a d^3k_b d^3k_c  \delta_{\mathrm{D}}\left(\bm{k}_a+\bm{k}_b+\bm{k}_c-\bm{k}_4 \right) \mathrm{F}_{\mathrm{s}}^{(3)}\left[\bm{k}_a,\bm{k}_b,\bm{k}_c\right]
 \notag\\
 &\,\times(2\pi)^9 \,\delta_{\mathrm{D}}\left(\bm{k}_1+\bm{k}_2\right)\delta_{\mathrm{D}}\left(\bm{k}_3+\bm{k}_a\right)\delta_{\mathrm{D}}\left(\bm{k}_b+\bm{k}_c\right)\,\mathrm{P}^{\mathrm{m}}_{k_1}\mathrm{P}^{\mathrm{m}}_{k_3}\mathrm{P}^{\mathrm{m}}_{k_b} 
 \,\,\, +\,\,\, 2\,\mathrm{p.}
\notag\\
&\,=\, 3\times(2\pi)^3\,\mathrm{P}^{\mathrm{s}}_{\mathrm{g}}\left(\bm{k}_1\right)\mathrm{F}^{(1)}_{\mathrm{\mathrm{s}}}\left[\bm{k}_3\right]\mathrm{P}^{\mathrm{m}}_{k_3}
\delta_{\mathrm{D}}\left(\bm{k}_1+\bm{k}_2\right)\,
\int d^3k_a d^3k_b d^3k_c 
\delta_{\mathrm{D}}\left(\bm{k}_a+\bm{k}_b+\bm{k}_c-\bm{k}_4 \right)
\mathrm{F}_{\mathrm{s}}^{(3)}\left[\bm{k}_a,\bm{k}_b,\bm{k}_c\right]
\notag \\
&\,\times\delta_{\mathrm{D}}\left(\bm{k}_3+\bm{k}_a\right)\delta_{\mathrm{D}}\left(\bm{k}_b+\bm{k}_c\right)\,
\mathrm{P}^{\mathrm{m}}_{k_b} 
\,\,\, +\,\,\, 2\,\mathrm{p.}
\notag \\
&\,=\, 3\times(2\pi)^3\,\mathrm{P}^{\mathrm{s}}_{\mathrm{g}}\left(\bm{k}_1\right)\mathrm{F}^{(1)}_{\mathrm{\mathrm{s}}}\left[\bm{k}_3\right]\mathrm{P}^{\mathrm{m}}_{k_3}
\delta_{\mathrm{D}}\left(\bm{k}_1+\bm{k}_2\right)\,
\int d^3k_b d^3k_c 
\delta_{\mathrm{D}}\left(\bm{k}_b+\bm{k}_c-\bm{k}_3-\bm{k}_4 \right)
\mathrm{F}_{\mathrm{s}}^{(3)}\left[-\bm{k}_3,\bm{k}_b,\bm{k}_c\right]
\delta_{\mathrm{D}}\left(\bm{k}_b+\bm{k}_c\right)\,
\mathrm{P}^{\mathrm{m}}_{k_b} 
\,\,\, +\,\,\, 2\,\mathrm{p.}
\notag \\
&\,=\, 3\times(2\pi)^3\,\mathrm{P}^{\mathrm{s}}_{\mathrm{g}}\left(\bm{k}_1\right)\mathrm{F}^{(1)}_{\mathrm{\mathrm{s}}}\left[\bm{k}_3\right]\mathrm{P}^{\mathrm{m}}_{k_3}
\delta_{\mathrm{D}}\left(\bm{k}_1+\bm{k}_2\right)
\delta_{\mathrm{D}}\left(\bm{k}_3+\bm{k}_4\right)\,
\int d^3k_b\, \mathrm{F}_{\mathrm{s}}^{(3)}\left[-\bm{k}_3,\bm{k}_b,-\bm{k}_b\right]\mathrm{P}^{\mathrm{m}}_{k_b}
\,\,\, +\,\,\, 2\,\mathrm{p.}
\notag \\
&\,=\, 3\times(2\pi)^3\,\mathrm{P}^{\mathrm{s}}_{\mathrm{g}}\left(\bm{k}_1\right)\mathrm{F}^{(1)}_{\mathrm{\mathrm{s}}}\left[\bm{k}_3\right]\mathrm{P}^{\mathrm{m}}_{k_3}
\delta_{\mathrm{D}}\left(\bm{k}_1+\bm{k}_2+\bm{k}_3+\bm{k}_4\right)\,\delta_{\mathrm{D}}\left(\bm{k}_3+\bm{k}_4\right)
\int d^3k_b\, \mathrm{F}_{\mathrm{s}}^{(3)}\left[-\bm{k}_3,\bm{k}_b,-\bm{k}_b\right]\mathrm{P}^{\mathrm{m}}_{k_b}
\,\,\, +\,\,\, 2\,\mathrm{p.}.
\end{align}

\noindent From the last line it is clear that this is also an un-connencted term. The second term gives the tree level contribution

\begin{align}
&(2\pi)^3\delta_{\mathrm{D}}\left(\bm{k}_1 +\bm{k}_2 + \bm{k}_3 +\bm{k}_4\right)  \mathrm{T}^{\mathrm{s}\,(3b)}_{\mathrm{\mathrm{g}}}
\left(\bm{k}_1,\bm{k}_2,\bm{k}_3,\bm{k}_4\right)=
\notag \\
&\,=\,\dfrac{6}{(2\pi)^6}\mathrm{F}^{(1)}_{\mathrm{\mathrm{s}}}\left[\bm{k}_1\right]\mathrm{F}^{(1)}_{\mathrm{\mathrm{s}}}\left[\bm{k}_2\right]\mathrm{F}^{(1)}_{\mathrm{\mathrm{s}}}\left[\bm{k}_2\right]
 \int d^3k_a d^3k_b d^3k_c  \delta_{\mathrm{D}}\left(\bm{k}_a+\bm{k}_b+\bm{k}_c-\bm{k}_4 \right) \mathrm{F}_{\mathrm{s}}^{(3)}\left[\bm{k}_a,\bm{k}_b,\bm{k}_c\right]
 \notag\\
 &\,\times(2\pi)^9 \,\delta_{\mathrm{D}}\left(\bm{k}_1+\bm{k}_a\right)\delta_{\mathrm{D}}\left(\bm{k}_2+\bm{k}_b\right)\delta_{\mathrm{D}}\left(\bm{k}_3+\bm{k}_c\right)\,\mathrm{P}^{\mathrm{m}}_{k_1}\mathrm{P}^{\mathrm{m}}_{k_2}\mathrm{P}^{\mathrm{m}}_{k_3} 
 \,\,\, +\,\,\, 3\,\mathrm{p.}
\notag\\
&\,=\, 6\times(2\pi)^3\,\mathrm{F}^{(1)}_{\mathrm{\mathrm{s}}}\left[\bm{k}_1\right]\mathrm{F}^{(1)}_{\mathrm{\mathrm{s}}}\left[\bm{k}_2\right]\mathrm{F}^{(1)}_{\mathrm{\mathrm{s}}}\left[\bm{k}_3\right]
\delta_{\mathrm{D}}\left(\bm{k}_1+\bm{k}_2+\bm{k}_3+\bm{k}_4\right)\,\mathrm{F}_{\mathrm{s}}^{(3)}\left[-\bm{k}_1,-\bm{k}_2,-\bm{k}_3\right]\mathrm{P}^{\mathrm{m}}_{k_1}\mathrm{P}^{\mathrm{m}}_{k_2} \mathrm{P}^{\mathrm{m}}_{k_3}
\,\,\, +\,\,\, 3\,\mathrm{p.}.
\end{align}

\noindent Therefore the full trispectrum at tree level is given by

\begin{align}
\mathrm{T}^{\mathrm{s}}_{\mathrm{\mathrm{g}}}&=  \,\mathrm{T}^{\mathrm{s}\,(2b)}_{\mathrm{\mathrm{g}}}+\mathrm{T}^{\mathrm{s}\,(2c)}_{\mathrm{\mathrm{g}}} + \mathrm{T}^{\mathrm{s}\,(3b)}_{\mathrm{\mathrm{g}}}.
\end{align}

\subsection{Tetraspectrum definition}

The tetraspectrum is defined as

\begin{eqnarray}
&\langle \delta^{\mathrm{\mathrm{s}}}_{\mathrm{\mathrm{g}}}(\bm{k}_1)\delta^{\mathrm{\mathrm{s}}}_{\mathrm{\mathrm{g}}}(\bm{k}_2)\delta^{\mathrm{\mathrm{s}}}_{\mathrm{\mathrm{g}}}(\bm{k}_3)\delta^{\mathrm{\mathrm{s}}}_{\mathrm{\mathrm{g}}}(\bm{k}_4)\delta^{\mathrm{\mathrm{s}}}_{\mathrm{\mathrm{g}}}(\bm{k}_5)\rangle_{\mathrm{c}}\,= (2\pi)^3\,\delta_{\mathrm{D}}\left(\bm{k}_1+\bm{k}_2+\bm{k}_3+\bm{k}_4+\bm{k}_5\right)
\mathrm{T}^{\mathrm{s}}_{\mathrm{e,g}}\left(\bm{k}_1,\bm{k}_2,\bm{k}_3,\bm{k}_4,\bm{k}_5\right),
\end{eqnarray}

\noindent where the subscript "c" stands for the connected part of the five points correlation function in Fourier space.

Drawing the diagrams as shown in \citet{1984ApJ...279..499F}, it is possible to see that the leading term of the connected part has order  $\mathrm{O}\left(\delta_m^8\right)$. There are two possible ways through which $\mathrm{O}\left(\delta_m^8\right)$ can be reached, either by having ${\mathrm{F}^{(2)}_{\mathrm{s}}}^3$ or $\mathrm{F}^{(2)}_{\mathrm{s}}\mathrm{F}^{(3)}_{\mathrm{s}}$ terms:

\begin{align}
& (2\pi)^3\,\delta_{\mathrm{D}}\left(\bm{k}_1+\bm{k}_2+\bm{k}_3+\bm{k}_4+\bm{k}_5\right)\mathrm{T}^{\mathrm{s}}_{\mathrm{e,g}}\left(\bm{k}_1,\bm{k}_2,\bm{k}_3,\bm{k}_4,\bm{k}_5\right) =
 \dfrac{1}{(2\pi)^9} \mathrm{F}^{(1)}_{\mathrm{s}}\left[\bm{k}_1\right]\,\mathrm{F}^{(1)}_{\mathrm{s}}\left[\bm{k}_2\right]\,
 \int d\bm{k}_a^3d\bm{k}_b^3 \,\delta_{\mathrm{D}}\left(\bm{k}_a+\bm{k}_b-\bm{k}_3\right) \mathrm{F}^{(2)}_{\mathrm{\mathrm{s}}}\left[\bm{k}_a,\bm{k}_b\right] 
 \notag \\ 
 &\times\int d\bm{k}_c^3d\bm{k}_d^3 \,\delta_{\mathrm{D}}\left(\bm{k}_c+\bm{k}_d-\bm{k}_4\right) \mathrm{F}^{(2)}_{\mathrm{\mathrm{s}}}\left[\bm{k}_c,\bm{k}_d\right] 
 \int d\bm{k}_e^3d\bm{k}_f^3 \,\delta_{\mathrm{D}}\left(\bm{k}_e+\bm{k}_f-\bm{k}_5\right) \mathrm{F}^{(2)}_{\mathrm{\mathrm{s}}}\left[\bm{k}_e,\bm{k}_f\right] 
\langle\delta_1\delta_2\delta_a\delta_b\delta_c\delta_d\delta_e\delta_f\rangle\quad+\quad59\,\mathrm{p.}
\notag\\
&+\dfrac{1}{(2\pi)^9} \mathrm{F}^{(1)}_{\mathrm{s}}\left[\bm{k}_1\right]\,\mathrm{F}^{(1)}_{\mathrm{s}}\left[\bm{k}_2\right]\,
\mathrm{F}^{(1)}_{\mathrm{s}}\left[\bm{k}_3\right]\,
 \int d\bm{k}_a^3d\bm{k}_b^3 \,\delta_{\mathrm{D}}\left(\bm{k}_a+\bm{k}_b-\bm{k}_4\right) \mathrm{F}^{(2)}_{\mathrm{\mathrm{s}}}\left[\bm{k}_a,\bm{k}_b\right] 
 \notag \\ 
 &\times  \int d\bm{k}_c^3d\bm{k}_d^3d\bm{k}_e^3 \,\delta_{\mathrm{D}}\left(\bm{k}_c+\bm{k}_d+\bm{k}_e-\bm{k}_5\right) \mathrm{F}^{(3)}_{\mathrm{\mathrm{s}}}\left[\bm{k}_c,\bm{k}_d,\bm{k}_e\right]
 \langle\delta_1\delta_2\delta_3\delta_a\delta_b\delta_c\delta_d\delta_e\rangle\quad+\quad19\,\mathrm{p.}.
\end{align}

\noindent Starting from the first term that appears in the above expansion:

\begin{align}
&(2\pi)^3\,\delta_{\mathrm{D}}\left(\bm{k}_1+\bm{k}_2+\bm{k}_3+\bm{k}_4+\bm{k}_5\right)\mathrm{T}^{\mathrm{s}(a)}_{\mathrm{e,g}}\left(\bm{k}_1,\bm{k}_2,\bm{k}_3,\bm{k}_4,\bm{k}_5\right) =    
\notag \\
& = \dfrac{1}{(2\pi)^9} \mathrm{F}^{(1)}_{\mathrm{s}}\left[\bm{k}_1\right]\,\mathrm{F}^{(1)}_{\mathrm{s}}\left[\bm{k}_2\right]\,
 \int d\bm{k}_a^3d\bm{k}_b^3d\bm{k}_c^3d\bm{k}_d^3  d\bm{k}_e^3d\bm{k}_f^3 \,
\delta_{\mathrm{D}}\left(\bm{k}_a+\bm{k}_b-\bm{k}_3\right)\,
\delta_{\mathrm{D}}\left(\bm{k}_c+\bm{k}_d-\bm{k}_4\right) \,
\delta_{\mathrm{D}}\left(\bm{k}_e+\bm{k}_f-\bm{k}_5\right) 
 \notag \\ 
 &\times \,
\mathrm{F}^{(2)}_{\mathrm{\mathrm{s}}}\left[\bm{k}_a,\bm{k}_b\right]  
\mathrm{F}^{(2)}_{\mathrm{\mathrm{s}}}\left[\bm{k}_c,\bm{k}_d\right]  
\mathrm{F}^{(2)}_{\mathrm{\mathrm{s}}}\left[\bm{k}_e,\bm{k}_f\right] 
\langle\delta_1\delta_2\delta_a\delta_b\delta_c\delta_d\delta_e\delta_f\rangle\quad+\quad59\,\mathrm{p.}
\notag\\
&= 8\,(2\pi)^3 \mathrm{F}^{(1)}_{\mathrm{s}}\left[\bm{k}_1\right]\,\mathrm{F}^{(1)}_{\mathrm{s}}\left[\bm{k}_2\right]\,
 \int d\bm{k}_a^3d\bm{k}_b^3d\bm{k}_c^3d\bm{k}_d^3  d\bm{k}_e^3d\bm{k}_f^3 \,
\delta_{\mathrm{D}}\left(\bm{k}_a+\bm{k}_b-\bm{k}_3\right)\,
\delta_{\mathrm{D}}\left(\bm{k}_c+\bm{k}_d-\bm{k}_4\right) \,
\delta_{\mathrm{D}}\left(\bm{k}_e+\bm{k}_f-\bm{k}_5\right) \,
 \notag \\ 
 &\times \,
\delta_{\mathrm{D}}\left(\bm{k}_1+\bm{k}_f\right)\,
\delta_{\mathrm{D}}\left(\bm{k}_2+\bm{k}_a\right)\,
\delta_{\mathrm{D}}\left(\bm{k}_b+\bm{k}_c\right)\,
\delta_{\mathrm{D}}\left(\bm{k}_d+\bm{k}_e\right)\,
\mathrm{F}^{(2)}_{\mathrm{\mathrm{s}}}\left[\bm{k}_a,\bm{k}_b\right]  
\mathrm{F}^{(2)}_{\mathrm{\mathrm{s}}}\left[\bm{k}_c,\bm{k}_d\right]  
\mathrm{F}^{(2)}_{\mathrm{\mathrm{s}}}\left[\bm{k}_e,\bm{k}_f\right] 
\mathrm{P}^{\mathrm{m}}\left(k_1\right)
\mathrm{P}^{\mathrm{m}}\left(k_2\right)
\mathrm{P}^{\mathrm{m}}\left(k_b\right)
\mathrm{P}^{\mathrm{m}}\left(k_c\right)
\quad+\quad359\,\mathrm{p.}
\notag\\
&= 8\,(2\pi)^3 \mathrm{F}^{(1)}_{\mathrm{s}}\left[\bm{k}_1\right]\,\mathrm{F}^{(1)}_{\mathrm{s}}\left[\bm{k}_2\right]\,
 \int d\bm{k}_b^3d\bm{k}_d^3\,
\delta_{\mathrm{D}}\left(-\bm{k}_2+\bm{k}_b-\bm{k}_3\right)\,
\delta_{\mathrm{D}}\left(-\bm{k}_b+\bm{k}_d-\bm{k}_4\right) \,
\delta_{\mathrm{D}}\left(-\bm{k}_d-\bm{k}_1-\bm{k}_5\right) \,
 \notag \\ 
 &\times \,
\mathrm{F}^{(2)}_{\mathrm{\mathrm{s}}}\left[-\bm{k}_2,\bm{k}_b\right]  
\mathrm{F}^{(2)}_{\mathrm{\mathrm{s}}}\left[-\bm{k}_b,\bm{k}_d\right]  
\mathrm{F}^{(2)}_{\mathrm{\mathrm{s}}}\left[-\bm{k}_d,-\bm{k}_1\right] 
\mathrm{P}^{\mathrm{m}}\left(k_1\right)
\mathrm{P}^{\mathrm{m}}\left(k_2\right)
\mathrm{P}^{\mathrm{m}}\left(k_b\right)
\mathrm{P}^{\mathrm{m}}\left(k_b\right) \quad+\quad359\,\mathrm{p.}
\notag\\
&= (2\pi)^3 \delta_{\mathrm{D}}\left(\bm{k}_1+\bm{k}_2+\bm{k}_3+\bm{k}_4+\bm{k}_5\right) \,
8\,\mathrm{F}^{(1)}_{\mathrm{s}}\left[\bm{k}_1\right]\,\mathrm{F}^{(1)}_{\mathrm{s}}\left[\bm{k}_2\right]
 \notag \\ 
 &\times \,
\mathrm{F}^{(2)}_{\mathrm{\mathrm{s}}}\left[-\bm{k}_2,\bm{k}_2+\bm{k}_3\right]  
\mathrm{F}^{(2)}_{\mathrm{\mathrm{s}}}\left[-\bm{k}_2-\bm{k}_3,-\bm{k}_1-\bm{k}_5\right]  
\mathrm{F}^{(2)}_{\mathrm{\mathrm{s}}}\left[\bm{k}_1+\bm{k}_5,-\bm{k}_1\right] 
\mathrm{P}^{\mathrm{m}}\left(k_1\right)
\mathrm{P}^{\mathrm{m}}\left(k_2\right)
\mathrm{P}^{\mathrm{m}}\left(|\bm{k}_2+\bm{k}_3|\right)
\mathrm{P}^{\mathrm{m}}\left(|\bm{k}_2+\bm{k}_3|\right)\quad+\quad359\,\mathrm{p.}.
\end{align}

\noindent where the factor of $8$ comes from permutations which do not influence the relations between the starting five wave vectors. On the contrary for each the original $60$ permutations, there are $6$ for which the relation between the initial wave vectors varies.
Considering now the second term:

\begin{align}
&(2\pi)^3\,\delta_{\mathrm{D}}\left(\bm{k}_1+\bm{k}_2+\bm{k}_3+\bm{k}_4+\bm{k}_5\right)\mathrm{T}^{\mathrm{s}(b)}_{\mathrm{e,g}}\left(\bm{k}_1,\bm{k}_2,\bm{k}_3,\bm{k}_4,\bm{k}_5\right) =    
\notag \\   
&=\dfrac{1}{(2\pi)^9} \mathrm{F}^{(1)}_{\mathrm{s}}\left[\bm{k}_1\right]\,\mathrm{F}^{(1)}_{\mathrm{s}}\left[\bm{k}_2\right]\,
\mathrm{F}^{(1)}_{\mathrm{s}}\left[\bm{k}_3\right]\,
\int d\bm{k}_a^3d\bm{k}_b^3 d\bm{k}_c^3 d\bm{k}_d^3d\bm{k}_e^3 \,
\delta_{\mathrm{D}}\left(\bm{k}_a+\bm{k}_b-\bm{k}_4\right) \,
\delta_{\mathrm{D}}\left(\bm{k}_c+\bm{k}_d+\bm{k}_e-\bm{k}_5\right) 
\notag \\ 
&\times
\mathrm{F}^{(2)}_{\mathrm{\mathrm{s}}}\left[\bm{k}_a,\bm{k}_b\right] \,
\mathrm{F}^{(3)}_{\mathrm{\mathrm{s}}}\left[\bm{k}_c,\bm{k}_d,\bm{k}_e\right]
\langle\delta_1\delta_2\delta_3\delta_a\delta_b\delta_c\delta_d\delta_e\rangle\quad+\quad19\,\mathrm{p.}
\notag\\
&=
6\,(2\pi)^3 \mathrm{F}^{(1)}_{\mathrm{s}}\left[\bm{k}_1\right]\,\mathrm{F}^{(1)}_{\mathrm{s}}\left[\bm{k}_2\right]\,
\mathrm{F}^{(1)}_{\mathrm{s}}\left[\bm{k}_3\right]\,
\int d\bm{k}_a^3d\bm{k}_b^3 d\bm{k}_c^3 d\bm{k}_d^3d\bm{k}_e^3 \,
\delta_{\mathrm{D}}\left(\bm{k}_a+\bm{k}_b-\bm{k}_4\right) \,
\delta_{\mathrm{D}}\left(\bm{k}_c+\bm{k}_d+\bm{k}_e-\bm{k}_5\right)\,
\notag \\ 
&\times
\delta_{\mathrm{D}}\left(\bm{k}_1 + \bm{k}_e \right)\,
\delta_{\mathrm{D}}\left(\bm{k}_2 + \bm{k}_d \right)\,
\delta_{\mathrm{D}}\left(\bm{k}_3 + \bm{k}_b \right)\,
\delta_{\mathrm{D}}\left(\bm{k}_a + \bm{k}_c \right)\,
\mathrm{F}^{(2)}_{\mathrm{\mathrm{s}}}\left[\bm{k}_a,\bm{k}_b\right] \,
\mathrm{F}^{(3)}_{\mathrm{\mathrm{s}}}\left[\bm{k}_c,\bm{k}_d,\bm{k}_e\right]\,
\mathrm{P}^{\mathrm{m}}\left(k_1\right)
\mathrm{P}^{\mathrm{m}}\left(k_2\right)
\mathrm{P}^{\mathrm{m}}\left(k_3\right)
\mathrm{P}^{\mathrm{m}}\left(k_a\right) \quad+\quad119\,\mathrm{p.}
\notag\\
&=
6\,(2\pi)^3 \mathrm{F}^{(1)}_{\mathrm{s}}\left[\bm{k}_1\right]\,\mathrm{F}^{(1)}_{\mathrm{s}}\left[\bm{k}_2\right]\,
\mathrm{F}^{(1)}_{\mathrm{s}}\left[\bm{k}_3\right]\,
\int d\bm{k}_a^3\,
\delta_{\mathrm{D}}\left(\bm{k}_a-\bm{k}_3-\bm{k}_4\right) \,
\delta_{\mathrm{D}}\left(-\bm{k}_a-\bm{k}_2-\bm{k}_1-\bm{k}_5\right)\,
\notag \\ 
&\times
\mathrm{F}^{(2)}_{\mathrm{\mathrm{s}}}\left[\bm{k}_a,-\bm{k}_3\right] \,
\mathrm{F}^{(3)}_{\mathrm{\mathrm{s}}}\left[-\bm{k}_a,-\bm{k}_2,-\bm{k}_1\right]\,
\mathrm{P}^{\mathrm{m}}\left(k_1\right)
\mathrm{P}^{\mathrm{m}}\left(k_2\right)
\mathrm{P}^{\mathrm{m}}\left(k_3\right)
\mathrm{P}^{\mathrm{m}}\left(k_a\right) \quad+\quad119\,\mathrm{p.}
\notag\\
&=
(2\pi)^3 \delta_{\mathrm{D}}\left(\bm{k}_1+\bm{k}_2+\bm{k}_3+\bm{k}_4+\bm{k}_5\right) \,
6\,\mathrm{F}^{(1)}_{\mathrm{s}}\left[\bm{k}_1\right]\,\mathrm{F}^{(1)}_{\mathrm{s}}\left[\bm{k}_2\right]\,
\mathrm{F}^{(1)}_{\mathrm{s}}\left[\bm{k}_3\right]\,
\notag \\ 
&\times
\mathrm{F}^{(2)}_{\mathrm{\mathrm{s}}}\left[\bm{k}_3 + \bm{k}_4,-\bm{k}_3\right] \,
\mathrm{F}^{(3)}_{\mathrm{\mathrm{s}}}\left[-\bm{k}_3 - \bm{k}_4,-\bm{k}_2,-\bm{k}_1\right]\,
\mathrm{P}^{\mathrm{m}}\left(k_1\right)
\mathrm{P}^{\mathrm{m}}\left(k_2\right)
\mathrm{P}^{\mathrm{m}}\left(k_3\right)
\mathrm{P}^{\mathrm{m}}\left(|\bm{k}_3 + \bm{k}_4|\right) \quad+\quad119\,\mathrm{p.}.
\end{align}

\noindent  where the factor of $6$ comes from permutations which do not influence the relations between the starting five wave vectors. On the contrary for each the original $19$ permutations, there are $6$ for which the relation between the initial wave vectors varies.
From the above expressions of both $\mathrm{T}^{\mathrm{s}(a)}_{\mathrm{e,g}}$ and $\mathrm{T}^{\mathrm{s}(b)}_{\mathrm{e,g}}$ it is possible to see that at leading order the tetraspectrum has order $\mathrm{O}\left(\delta_m^8\right)$. In other words it is proportional to the fourth power of the linear matter power spectrum.

\subsection{Unconnected part of the five points correlation function}

If one looks at the unconnected part of the five point correlation function, the leading term has order $\mathrm{O}\left(\delta_m^6\right)$.
Below are derived the analytical expression for this unconnected part.


\begin{align}
&\mathrm{U}^{\rm{s}}_{\rm{5p,g}}\left(\bm{k}_1,\bm{k}_2,\bm{k}_3,\bm{k}_4,\bm{k}_5\right) =
&\dfrac{1}{(2\pi)^3}\prod_{i=1}^{4}\mathrm{F}^{(1)}_{\mathrm{\mathrm{s}}}\left[\bm{k}_i\right]\,\int d\bm{k}_a^3d\bm{k}_b^3 \,\delta_{\mathrm{D}}\left(\bm{k}_a+\bm{k}_b-\bm{k}_5\right) \mathrm{F}^{(2)}_{\mathrm{\mathrm{s}}}\left[\bm{k}_a,\bm{k}_b\right] 
\langle\delta_1\delta_2\delta_3\delta_4\delta_a\delta_b\rangle\quad+\quad5\,\mathrm{p.}.
\end{align}

\noindent Using Wick's theorem the six-points correlator can be approximated using in to products of two points correlators. For example, for the first permutation one obtains

\begin{align}
\langle\delta_1\delta_2\delta_3\delta_4\delta_a\delta_b\rangle 
&= 2 \times \langle\delta_1\delta_a\rangle\langle\delta_2\delta_b\rangle\langle\delta_3\delta_4\rangle \leftrightarrow \mathrm{U}_{\rm{5p}}^1
\notag \\
&+ 2 \times \langle\delta_1\delta_a\rangle\langle\delta_3\delta_b\rangle\langle\delta_2\delta_4\rangle \leftrightarrow \mathrm{U}_{\rm{5p}}^2
\notag \\
&+ 2 \times \langle\delta_1\delta_a\rangle\langle\delta_4\delta_b\rangle\langle\delta_2\delta_3\rangle \leftrightarrow \mathrm{U}_{\rm{5p}}^3
\notag \\
&+ 2 \times \langle\delta_2\delta_a\rangle\langle\delta_3\delta_b\rangle\langle\delta_1\delta_4\rangle \leftrightarrow \mathrm{U}_{\rm{5p}}^4
\notag \\
&+ 2 \times \langle\delta_2\delta_a\rangle\langle\delta_4\delta_b\rangle\langle\delta_1\delta_3\rangle \leftrightarrow \mathrm{U}_{\rm{5p}}^5
\notag \\
&+ 2 \times \langle\delta_3\delta_a\rangle\langle\delta_4\delta_b\rangle\langle\delta_1\delta_2\rangle \leftrightarrow \mathrm{U}_{\rm{5p}}^6.
\end{align}

\noindent Since all these terms are similar to each other, we compute below only the first one of them as an example on how to derive the others.

\begin{align}
&(2\pi)^3\,\delta_{\mathrm{D}}\left(\bm{k}_1+\bm{k}_2+\bm{k}_3+\bm{k}_4+\bm{k}_5\right) \,\mathrm{U}_{\rm{5p}}^1 =
\notag \\
&=\dfrac{2}{2(\pi)^3}\prod_{i=1}^{4}\mathrm{F}^{(1)}_{\mathrm{\mathrm{s}}}\left(\bm{k}_i\right) \,\int d\bm{k}_a^3d\bm{k}_b^3 \,\delta_{\mathrm{D}}\left(\bm{k}_a+\bm{k}_b-\bm{k}_5\right) \mathrm{F}^{(2)}_{\mathrm{\mathrm{s}}}\left[\bm{k}_a,\bm{k}_b\right] 
 (2\pi)^9\delta_{\mathrm{D}}\left(\bm{k}_1+\bm{k}_a\right)\delta_{\mathrm{D}}\left(\bm{k}_2+\bm{k}_b\right)\delta_{\mathrm{D}}\left(\bm{k}_3+\bm{k}_4\right)
 \mathrm{P}^{\mathrm{m}}\left(k_1\right)\mathrm{P}^{\mathrm{m}}\left(k_2\right)\mathrm{P}^{\mathrm{m}}\left(k_3\right)
\notag \\
& = 2 \times (2\pi)^6 \mathrm{F}^{(1)}_{\mathrm{\mathrm{s}}}\left[\bm{k}_1\right]\mathrm{F}^{(1)}_{\mathrm{\mathrm{s}}}\left[\bm{k}_2\right]\mathrm{F}^{(1)}_{\mathrm{\mathrm{s}}}\left[\bm{k}_3\right]\mathrm{F}^{(1)}_{\mathrm{\mathrm{s}}}\left[\bm{k}_4\right]\mathrm{F}^{(2)}_{\mathrm{\mathrm{s}}}\left[\bm{k}_1,\bm{k}_2\right]
 \delta_{\mathrm{D}}\left(\bm{k}_3+\bm{k}_4\right)\delta_{\mathrm{D}}\left(\bm{k}_1+\bm{k}_2+\bm{k}_5\right) 
\mathrm{P}^{\mathrm{m}}\left(k_1\right)\mathrm{P}^{\mathrm{m}}\left(k_2\right)\mathrm{P}^{\mathrm{m}}\left(k_3\right)
\notag\\
& = 2 \times (2\pi)^6 \mathrm{F}^{(1)}_{\mathrm{\mathrm{s}}}\left[\bm{k}_1\right]\mathrm{F}^{(1)}_{\mathrm{\mathrm{s}}}\left[\bm{k}_2\right]\mathrm{F}^{(1)}_{\mathrm{\mathrm{s}}}\left[\bm{k}_3\right]\mathrm{F}^{(1)}_{\mathrm{\mathrm{s}}}\left[\bm{k}_4\right]\mathrm{F}^{(2)}_{\mathrm{\mathrm{s}}}\left[\bm{k}_1,\bm{k}_2\right]
\delta^{\mathrm{K}}_{125}\, \delta_{\mathrm{D}}\left(\bm{k}_3+\bm{k}_4\right)\delta_{\mathrm{D}}\left(\bm{k}_1+\bm{k}_2+\bm{k}_3+\bm{k}_4+\bm{k}_5\right)
\,\mathrm{P}^{\mathrm{m}}\left(k_1\right)\mathrm{P}^{\mathrm{m}}\left(k_2\right)\mathrm{P}^{\mathrm{m}}\left(k_3\right),
\end{align}

\noindent where it is important to note that, in the last line, when substituting the Dirac's delta for the sum of three wave-vectors with the sum of all five of them,  a Kronecker's delta must be added to keep track of the original relation between the three $\bm{k}$'s. From the last line one can immediately write another one of the six terms

\begin{align}
&(2\pi)^3\,\delta_{\mathrm{D}}\left(\bm{k}_1+\bm{k}_2+\bm{k}_3+\bm{k}_4+\bm{k}_5\right)  \mathrm{U}_{\rm{5p}}^5 
\left(\bm{k}_1,\bm{k}_2,\bm{k}_3,\bm{k}_4,\bm{k}_5\right)= 
\notag \\
&= 2 (2\pi)^6 \mathrm{F}^{(1)}_{\mathrm{\mathrm{s}}}\left[\bm{k}_1\right]\mathrm{F}^{(1)}_{\mathrm{\mathrm{s}}}\left[\bm{k}_2\right]\mathrm{F}^{(1)}_{\mathrm{\mathrm{s}}}\left[\bm{k}_3\right]\mathrm{F}^{(1)}_{\mathrm{\mathrm{s}}}\left[\bm{k}_4\right]\mathrm{F}^{(2)}_{\mathrm{\mathrm{s}}}\left[\bm{k}_2,\bm{k}_4\right]
\delta^{\mathrm{K}}_{245}\,\delta_{\mathrm{D}}\left(\bm{k}_1+\bm{k}_3\right)\delta_{\mathrm{D}}\left(\bm{k}_1+\bm{k}_2+\bm{k}_3+\bm{k}_4+\bm{k}_5\right)
\mathrm{P}^{\mathrm{m}}\left(k_2\right)\mathrm{P}^{\mathrm{m}}\left(\bm{k}_4\right)\mathrm{P}^{\mathrm{m}}\left(k_3\right).
\end{align}

\noindent Therefore, it is possible to write down the complete expression that takes into account all the permutations and terms as follows

\begin{eqnarray}
\label{tetraspectrum}
\mathrm{U}^{\rm{s}}_{\rm{5p,g}}\left(\bm{k}_1,\bm{k}_2,\bm{k}_3,\bm{k}_4,\bm{k}_5\right) =
\sum_{i=1}^5 \mathrm{U}^{\rm{s}}_{\rm{5p,g}}\left(\delta^{(2)}_i\right)
\mathrm{where}\quad \mathrm{U}^{\rm{s}}_{\rm{5p,g}}\left(\delta^{(2)}_i\right) = \mathrm{U}_{\rm{5p}}^1 + \mathrm{U}_{\rm{5p}}^2 +\mathrm{U}_{\rm{5p}}^3+\mathrm{U}_{\rm{5p}}^4+\mathrm{U}_{\rm{5p}}^5+\mathrm{U}_{\rm{5p}}^6 .
\end{eqnarray}

\noindent The argument $\delta^{(2)}_i$ identifies which wave-vectors corresponds to the second order perturbation expansion for each term.

\section{Covariance terms derivation}

\label{sec:cov_terms}

In this Appendix we present the derivation of the covariance terms, starting from the power spectrum one.

\subsection{Covariance term: $\mathrm{\textbf{C}^{\mathrm{PP}}}$}
\noindent As stated in the main text, the covariance  matrix element can be computed as

\begin{align}
\label{cov_exp_pk}
&\mathrm{C}^{\mathrm{P}^{\mathrm{\mathrm{s}}}_{\mathrm{\mathrm{g}}}}\equiv\mathrm{\textbf{Cov}}\left[\mathrm{P}^{\mathrm{\mathrm{s}}}_{\mathrm{\mathrm{g}}}\left(\bar{\bm{k}}_1\right),\mathrm{P}^{\mathrm{\mathrm{s}}}_{\mathrm{\mathrm{g}}}\left(\bar{\bm{k}}_3\right)\right]=
\bigg\langle\left(\hat{\mathrm{P}}^{\mathrm{\mathrm{s}}}_{\mathrm{g},1}-\langle\hat{\mathrm{P}}^{\mathrm{\mathrm{s}}}_{\mathrm{g},1}\rangle\right)\bigg\rangle\bigg\langle\left(\hat{\mathrm{P}}^{\mathrm{\mathrm{s}}}_{\mathrm{g},3}-\langle\hat{\mathrm{P}}^{\mathrm{\mathrm{s}}}_{\mathrm{g},3}\rangle\right)\bigg\rangle
=\langle\hat{\mathrm{P}}^{\mathrm{\mathrm{s}}}_{\mathrm{g},1}\hat{\mathrm{P}}^{\mathrm{\mathrm{s}}}_{\mathrm{g},3}\rangle-\hat{\mathrm{P}}^{\mathrm{\mathrm{s}}}_{\mathrm{g},1}\hat{\mathrm{P}}^{\mathrm{\mathrm{s}}}_{\mathrm{g},3}.
\end{align}

\noindent The computation of the covariance  matrix involves a four points correlator of $\delta^{\mathrm{\mathrm{s}}}_{\mathrm{\mathrm{g}}}$ that can be expanded into its connected parts

\begin{align}
\langle\delta^{\mathrm{\mathrm{s}}}_{\mathrm{\mathrm{g}}}(\bm{k}_1)\delta^{\mathrm{\mathrm{s}}}_{\mathrm{\mathrm{g}}}(\bm{k}_2)\delta^{\mathrm{\mathrm{s}}}_{\mathrm{\mathrm{g}}}(\bm{k}_3)\delta^{\mathrm{\mathrm{s}}}_{\mathrm{\mathrm{g}}}(\bm{k}_4) \rangle &=
\langle\delta^{\mathrm{\mathrm{s}}}_{\mathrm{\mathrm{g}}}(\bm{k}_1)\delta^{\mathrm{\mathrm{s}}}_{\mathrm{\mathrm{g}}}(\bm{k}_2)\rangle_{\mathrm{c}} \langle\delta^{\mathrm{\mathrm{s}}}_{\mathrm{\mathrm{g}}}(\bm{k}_3)\delta^{\mathrm{\mathrm{s}}}_{\mathrm{\mathrm{g}}}(\bm{k}_4)\rangle_{\mathrm{c}}
 +\,2\,\mathrm{perms.}\notag\\
&+\langle\delta^{\mathrm{\mathrm{s}}}_{\mathrm{\mathrm{g}}}(\bm{k}_1)\delta^{\mathrm{\mathrm{s}}}_{\mathrm{\mathrm{g}}}(\bm{k}_2)\delta^{\mathrm{\mathrm{s}}}_{\mathrm{\mathrm{g}}}(\bm{k}_3)\delta^{\mathrm{\mathrm{s}}}_{\mathrm{\mathrm{g}}}(\bm{k}_4)\rangle_{\mathrm{c}}.
\end{align}

\noindent In this work, the following short hand for the integrals will be also adopted

\begin{eqnarray}
\int_{V^{\mathrm{c}}_{\bar{k}_i}}\dfrac{d\phi_idk^\perp_idk^\parallel_i\,k^\perp_i}{2\pi\bar{k}_i^\perp\Delta k^2}\equiv \int_i.
\end{eqnarray}

\subsubsection{$\mathrm{\textbf{C}^{\mathrm{PP}}_{P}}$ term:}
Starting from the first term  $\mathrm{C^{\mathrm{PP}}_{P}}$ in the particular case in which the pairs are made of identical wave vectors (in this case, from the power spectrum estimator definition: $\bm{k}_1=\bm{k}_2$ and $\bm{k}_3 = \bm{k}_4$ ) we have that

\begin{align}
&\mathrm{C^{\mathrm{PP}}_{P}}\left(\bar{\bm{k}}_1; \bar{\bm{k}}_3\right) =  \dfrac{1}{V^2_{\mathrm{s}}\,V^{\mathrm{c}}_1\,V^{\mathrm{c}}_3}\,\prod_{i=0}^4\int\,dV_i^{\mathrm{c}}\, \delta_{\mathrm{D}}\left(\bm{k}_1+\bm{k}_2\right)\delta_{\mathrm{D}}\left(\bm{k}_3+\bm{k}_4\right)
\langle\delta^{\mathrm{\mathrm{s}}}_{\mathrm{\mathrm{g}}}\left(\bm{k}_1\right)\delta^{\mathrm{\mathrm{s}}}_{\mathrm{\mathrm{g}}}\left(\bm{k}_2\right)\rangle\langle\delta^{\mathrm{\mathrm{s}}}_{\mathrm{\mathrm{g}}}\left(\bm{k}_3\right)\delta^{\mathrm{\mathrm{s}}}_{\mathrm{\mathrm{g}}}\left(\bm{k}_4\right)\rangle,
\end{align}

\noindent it is straightforward to see that  this terms cancels the second term of  Eq. \eqref{cov_exp_pk}. Considering now the other  two possibilities for  $\mathrm{C^{\mathrm{PP}}_{P}}$

\begin{align}
\label{eq:pp_gauss}
&\mathrm{C^{\mathrm{PP}}_{P}}\left(\bar{\bm{k}}_1; \bar{\bm{k}}_3\right) =  \dfrac{1}{V^2_{\mathrm{s}}\,V^{\mathrm{c}}_1\,V^{\mathrm{c}}_3}\,\prod_{i=0}^4\int dV^{\mathrm{c}}_i\, \delta_{\mathrm{D}}\left(\bm{k}_1+\bm{k}_2\right)\delta_{\mathrm{D}}\left(\bm{k}_3+\bm{k}_4\right)
\langle\delta^{\mathrm{\mathrm{s}}}_{\mathrm{\mathrm{g}}}\left(\bm{k}_1\right)\delta^{\mathrm{\mathrm{s}}}_{\mathrm{\mathrm{g}}}\left(\bm{k}_3\right)\rangle\langle\delta^{\mathrm{\mathrm{s}}}_{\mathrm{\mathrm{g}}}\left(\bm{k}_2\right)\delta^{\mathrm{\mathrm{s}}}_{\mathrm{\mathrm{g}}}\left(\bm{k}_4\right)\rangle\,\,+\,1\mathrm{p.}
\notag \\
& =   \dfrac{(2\pi)^6}{V^2_{\mathrm{s}}\,V^{\mathrm{c}}_1\,V^{\mathrm{c}}_3}\,\prod_{i=0}^4\int dV^{\mathrm{c}}_i\, \delta_{\mathrm{D}}\left(\bm{k}_1+\bm{k}_2\right)
 \delta_{\mathrm{D}}\left(\bm{k}_3+\bm{k}_4\right)\delta_{\mathrm{D}}\left(\bm{k}_1+\bm{k}_3\right)\delta_{\mathrm{D}}\left(\bm{k}_2+\bm{k}_4\right)
\mathrm{P}^{\mathrm{\mathrm{s}}}_{\mathrm{\mathrm{g}}}\left(\bm{k}_1\right)\mathrm{P}^{\mathrm{\mathrm{s}}}_{\mathrm{\mathrm{g}}}\left(\bm{k}_2\right)\,+\,1\mathrm{p.}
\notag \\
& =    \dfrac{(2\pi)^6}{V^2_{\mathrm{s}}\,V^{\mathrm{c}}_1\,V^c_2} \left(\delta^{\mathrm{K}}_{13}\delta^{\mathrm{K}}_{24} + \delta^{\mathrm{K}}_{14}\delta^{\mathrm{K}}_{23} \right) \int dV^{\mathrm{c}}_1 \int dV^{\mathrm{c}}_2 \,\delta_{\mathrm{D}}\left(\bm{k}_1+\bm{k}_2\right)^2 
 \mathrm{P}^{\mathrm{\mathrm{s}}}_{\mathrm{\mathrm{g}}}\left(\bm{k}_1\right)\mathrm{P}^{\mathrm{\mathrm{s}}}_{\mathrm{\mathrm{g}}}\left(\bm{k}_2\right)
\notag \\
& =    \dfrac{(2\pi)^3}{V_{\mathrm{\mathrm{s}}}\,V^{\mathrm{c}}_1\,V^c_2} \mathrm{D}^\mathrm{P}_{1234}\int dV^{\mathrm{c}}_1 \int dV^{\mathrm{c}}_2 \,\delta_{\mathrm{D}}\left(\bm{k}_1+\bm{k}_2\right) \mathrm{P}^{\mathrm{\mathrm{s}}}_{\mathrm{\mathrm{g}}}\left(\bm{k}_1\right)\mathrm{P}^{\mathrm{\mathrm{s}}}_{\mathrm{\mathrm{g}}}\left(\bm{k}_2\right)
\notag \\
& =    \dfrac{(2\pi)^3}{V_{\mathrm{\mathrm{s}}}\,V^{\mathrm{c}}_1} \mathrm{D}^\mathrm{P}_{1234} \int_{\bar{k}^\perp_1-\frac{\Delta k }{2}}^{\bar{k}^\perp_1+\frac{\Delta k }{2}}\int_{\bar{k}^\parallel_1-\frac{\Delta k }{2}}^{\bar{k}^\parallel_1+\frac{\Delta k }{2}}\dfrac{dk^\perp_1 dk^\parallel_1\,k^\perp_1}{\bar{k}_1^\perp\Delta k^2}  \, \mathrm{P}^{\mathrm{\mathrm{s}}}_{\mathrm{\mathrm{g}}}\left(\bm{k}_1\right)^2
\notag \\
& \approx \dfrac{(2\pi)^2}{V_{\mathrm{\mathrm{s}}} \bar{k}_1^\perp \Delta k^2}\mathrm{D}^\mathrm{P}_{1234} \, \mathrm{P}^{\mathrm{\mathrm{s}}}_{\mathrm{\mathrm{g}}}\left(\bar{\bm{k}}_1\right)^2
\,=\,\dfrac{2\times (2\pi)^2}{V_{\mathrm{\mathrm{s}}} \bar{k}_1^\perp \Delta k^2}\delta^{\mathrm{K}}_{13} \, \mathrm{P}^{\mathrm{\mathrm{s}}}_{\mathrm{\mathrm{g}}}\left(\bar{\bm{k}}_1\right)^2,
\end{align}

\noindent where $\mathrm{D}^\mathrm{P}_{1234} = \delta^{\mathrm{K}}_{13}\delta^{\mathrm{K}}_{24} + \delta^{\mathrm{K}}_{14}\delta^{\mathrm{K}}_{23}$ has been defined  in order to take in account the additional permutation  from the initial  Dirac's deltas conditions, which now reduces to  $\mathrm{D}^\mathrm{P}_{1234} = 2 \times \delta^{\mathrm{K}}_{13} $.
If, when obtaining an expression for the trispectrum, one considers also 1loop corrections, in addition to this Gaussian term on the diagonal of the power spectrum covariance matrix, one should add the unconnected terms encountered in Appendix \ref{sec:est_def}, $ \mathrm{T}^{\mathrm{s}\,(2a)}_{\mathrm{\mathrm{g}}}$ and $ \mathrm{T}^{\mathrm{s}\,(3a)}_{\mathrm{\mathrm{g}}}$. This has been recently well  described in \citet{2017MNRAS.466..780M}. As stated in Sec. \ref{sub_sec:rsd_formalism} the error made with this tree level approximation for the diagonal term \ref{eq:pp_gauss} is around $\sim1\%$ for $k_{\mathrm{max}} = 0.2 \, \mathrm{Mpc}^{-1}h$ and up to $\sim9\%$  for $k_{\mathrm{max}} = 0.3\,  \mathrm{Mpc}^{-1}h$ \citep{2008PhRvD..78l3534T}.

\subsubsection{$\mathrm{\textbf{C}^{\mathrm{PP}}_{T}}$ term:}
Proceeding in the same way it is possible to compute as well the other term of the cross-correlation  matrix, the one containing the trispectrum contribution.

\begin{align}
&\mathrm{C^{\mathrm{PP}}_{T}}\left(\bar{\bm{k}}_1; \bar{\bm{k}}_3\right) =  \dfrac{1}{V^2_{\mathrm{s}}\,V^{\mathrm{c}}_1\,V^{\mathrm{c}}_3}\,\prod_{i=0}^4\int dV^{\mathrm{c}}_i\, \delta_{\mathrm{D}}\left(\bm{k}_1+\bm{k}_2\right)\delta_{\mathrm{D}}\left(\bm{k}_3+\bm{k}_4\right)
\langle\delta^{\mathrm{\mathrm{s}}}_{\mathrm{\mathrm{g}}}\left(\bm{k}_1\right)\delta^{\mathrm{\mathrm{s}}}_{\mathrm{\mathrm{g}}}\left(\bm{k}_3\right)\delta^{\mathrm{\mathrm{s}}}_{\mathrm{\mathrm{g}}}\left(\bm{k}_2\right)\delta^{\mathrm{\mathrm{s}}}_{\mathrm{\mathrm{g}}}\left(\bm{k}_4\right)\rangle
\notag \\
& =   \dfrac{(2\pi)^3}{V^2_{\mathrm{s}}\,V^{\mathrm{c}}_1\,V^{\mathrm{c}}_3}\,\prod_{i=0}^4\int dV^{\mathrm{c}}_i\, \delta_{\mathrm{D}}\left(\bm{k}_1+\bm{k}_2\right)\delta_{\mathrm{D}}\left(\bm{k}_3+\bm{k}_4\right)
\delta_{\mathrm{D}}\left(\bm{k}_1+\bm{k}_2 + \bm{k}_3 + \bm{k}_4\right) \mathrm{T}^{\mathrm{\mathrm{s}}}_{\mathrm{\mathrm{g}}}\left(\bm{k}_1,\bm{k}_2,\bm{k}_3,\bm{k}_4\right)
\notag \\
& =  \dfrac{(2\pi)^3}{V^2_{\mathrm{s}}\,V^{\mathrm{c}}_1\,V^{\mathrm{c}}_3} \delta^{\mathrm{K}}_{34}\,\int dV^{\mathrm{c}}_1 dV^{\mathrm{c}}_2 dV^{\mathrm{c}}_3 \delta_{\mathrm{D}}\left(\bm{k}_1+\bm{k}_2\right)^2\mathrm{T}^{\mathrm{\mathrm{s}}}_{\mathrm{\mathrm{g}}}\left(\bm{k}_1,\bm{k}_2,\bm{k}_3,-\bm{k}_3\right)
\notag \\
& =  \dfrac{1}{V_{\mathrm{\mathrm{s}}}\,V^{\mathrm{c}}_1\,V^{\mathrm{c}}_3} \delta^{\mathrm{K}}_{34}\,\int dV^{\mathrm{c}}_1 dV^{\mathrm{c}}_2 dV^{\mathrm{c}}_3 \delta_{\mathrm{D}}\left(\bm{k}_1+\bm{k}_2\right)\mathrm{T}^{\mathrm{\mathrm{s}}}_{\mathrm{\mathrm{g}}}\left(\bm{k}_1,\bm{k}_2,\bm{k}_3,-\bm{k}_3\right)
\notag \\
& =  \dfrac{1}{V_{\mathrm{\mathrm{s}}}} \delta^{\mathrm{K}}_{12}\delta^{\mathrm{K}}_{34}\,\int \dfrac{dV^{\mathrm{c}}_1}{V^{\mathrm{c}}_1} \int \dfrac{dV^{\mathrm{c}}_3}{V^{\mathrm{c}}_3} \,\mathrm{T}^{\mathrm{\mathrm{s}}}_{\mathrm{\mathrm{g}}}\left(\bm{k}_1,-\bm{k}_1,\bm{k}_3,-\bm{k}_3\right)
\notag \\
&\approx \dfrac{1}{2\pi \, V_{\mathrm{\mathrm{s}}}}\,\int d\phi_{13}\,
\mathrm{T}^{\mathrm{\mathrm{s}}}_{\mathrm{\mathrm{g}}}\left(\bar{k}^\perp_1, \bar{k}^\parallel_1,\bar{k}^\perp_3,\bar{k}^\parallel_3, \phi^-_{13}\right).
\end{align}

\noindent Following what was done in \citet{2010A&A...514A..79P}, i.e, substituting in here the previously derived expression for the trispectrum, this expression can be "simplified" at tree level to

\begin{align}
&\mathrm{C^{\mathrm{PP}}_{T}}\left(\bar{\bm{k}}_1; \bar{\bm{k}}_3\right) \approx \dfrac{1}{2\pi \, V_{\mathrm{\mathrm{s}}}}\,\int d\phi_{13}\,
\mathrm{T}^{\mathrm{\mathrm{s}}}_{\mathrm{\mathrm{g}}}\left(\bar{k}^\perp_1, \bar{k}^\parallel_1,\bar{k}^\perp_3,\bar{k}^\parallel_3, \phi_{13}\right)
\notag \\
&=  \dfrac{1}{2\pi \, V_{\mathrm{\mathrm{s}}}}\int d\phi_{13} 
\bigg\{ 4\,\mathrm{P}^{\mathrm{\mathrm{s}}}_{\mathrm{\mathrm{g}}}(\bm{k}_1)^2\bigg[ \mathrm{F}_{\mathrm{s}}^{(2)}\left[\bm{k}_1, -\bm{k}_+\right]^2 \mathrm{P}^{\mathrm{m}}(\bm{k}_+) + 
\mathrm{F}_{\mathrm{s}}^{(2)}\left[\bm{k}_1, \bm{k}_-\right]^2 \mathrm{P}^{\mathrm{m}}(\bm{k}_-) \bigg]
+ 4\,\mathrm{P}^{\mathrm{\mathrm{s}}}_{\mathrm{\mathrm{g}}}(\bm{k}_3)^2 \bigg[ \mathrm{F}_{\mathrm{s}}^{(2)}\left[\bm{k}_3, -\bm{k}_+\right]^2 \mathrm{P}^{\mathrm{m}}(\bm{k}_+) + 
\mathrm{F}_{\mathrm{s}}^{(2)}\left[\bm{k}_3, -\bm{k}_-\right]^2 \mathrm{P}^{\mathrm{m}}(\bm{k}_-) \bigg]
\notag \\
& + 8\,\mathrm{P}^{\mathrm{\mathrm{s}}}_{\mathrm{\mathrm{g}}}(\bm{k}_1)\mathrm{P}^{\mathrm{\mathrm{s}}}_{\mathrm{\mathrm{g}}}(\bm{k}_3) \bigg[ \mathrm{F}_{\mathrm{s}}^{(2)}\left[\bm{k}_1, -\bm{k}_+\right]\mathrm{F}_{\mathrm{s}}^{(2)}\left[\bm{k}_3, -\bm{k}_+\right]\mathrm{P}^{\mathrm{m}}(\bm{k}_+)
+ \mathrm{F}_{\mathrm{s}}^{(2)}\left[\bm{k}_1, \bm{k}_-\right]\mathrm{F}_{\mathrm{s}}^{(2)}\left[\bm{k}_3, - \bm{k}_-\right]\mathrm{P}^{\mathrm{m}}(\bm{k}_-) \bigg]
\notag\\
&+ 12 \mathrm{P}^{\mathrm{\mathrm{s}}}_{\mathrm{\mathrm{g}}}(\bm{k}_1)\mathrm{P}^{\mathrm{\mathrm{s}}}_{\mathrm{\mathrm{g}}}(\bm{k}_3)
\bigg[\mathrm{P}^{\mathrm{\mathrm{s}}}_{\mathrm{\mathrm{g}}}(\bm{k}_1)\mathrm{F}_{\mathrm{s}}^{(3)}\left[\bm{k}_1, -\bm{k}_1, \bm{k}_3\right]
+\mathrm{P}^{\mathrm{\mathrm{s}}}_{\mathrm{\mathrm{g}}}(\bm{k}_3)\mathrm{F}_{\mathrm{s}}^{(3)}\left[\bm{k}_1, \bm{k}_3,-\bm{k}_3\right]\bigg]\bigg\},
\end{align}

\noindent where $\bm{k}_+ = \bm{k}_1 + \bm{k}_3$ and $\bm{k}_- = \bm{k}_3 - \bm{k}_1$ and $\mathrm{P}^{\mathrm{m}}$ is the linear matter power spectrum and $\phi_{13}$ is the difference between the two azimuthal angles of the two wave-vectors. Therefore this is the only term that requires an integration since this angular dependence defines the relation between the two wave-vectors involved.
\subsection{Covariance term: $\mathrm{\textbf{C}^{\mathrm{BB}}}$}

By definition, the covariance  matrix element can be computed as

\begin{align}
\label{cov_exp}
\mathrm{C}^{\mathrm{B}^{\mathrm{\mathrm{s}}}_{\mathrm{\mathrm{g}}}}&\equiv\mathrm{\textbf{Cov}}\left[\mathrm{B}^{\mathrm{\mathrm{s}}}_{\mathrm{\mathrm{g}}}\left(\bar{k}_1^\perp,\bar{k}_2^\perp,\bar{k}_3^\perp,\bar{k}_1^\parallel,\bar{k}_2^\parallel\right),\mathrm{B}^{\mathrm{\mathrm{s}}}_{\mathrm{\mathrm{g}}}\left(\bar{k}_4^\perp,\bar{k}_5^\perp,\bar{k}_6^\perp,\bar{k}_4^\parallel,\bar{k}_5^\parallel\right)\right]
=\bigg\langle\left(\hat{\mathrm{B}}^{\mathrm{\mathrm{s}}}_{\mathrm{g},123}-\langle\hat{\mathrm{B}}^{\mathrm{\mathrm{s}}}_{\mathrm{g},123}\rangle\right)\bigg\rangle\bigg\langle\left(\hat{\mathrm{B}}^{\mathrm{\mathrm{s}}}_{\mathrm{g},456}-\langle\hat{\mathrm{B}}^{\mathrm{\mathrm{s}}}_{\mathrm{g},456}\rangle\right)\bigg\rangle
=\langle\hat{\mathrm{B}}^{\mathrm{\mathrm{s}}}_{\mathrm{g},123}\hat{\mathrm{B}}^{\mathrm{\mathrm{s}}}_{\mathrm{g},456}\rangle-\hat{\mathrm{B}}^{\mathrm{\mathrm{s}}}_{\mathrm{g},123}\hat{\mathrm{B}}^{\mathrm{\mathrm{s}}}_{\mathrm{g},456}.
\end{align}

\noindent The six points correlation function can be decomposed into all the possible combinations of connected parts

\begin{align}
\langle\delta^{\mathrm{\mathrm{s}}}_{\mathrm{\mathrm{g}}}(\bm{k}_1)\delta^{\mathrm{\mathrm{s}}}_{\mathrm{\mathrm{g}}}(\bm{k}_2)\delta^{\mathrm{\mathrm{s}}}_{\mathrm{\mathrm{g}}}(\bm{k}_3)\delta^{\mathrm{\mathrm{s}}}_{\mathrm{\mathrm{g}}}(\bm{k}_4)\delta^{\mathrm{\mathrm{s}}}_{\mathrm{\mathrm{g}}}(\bm{k}_5)\delta^{\mathrm{\mathrm{s}}}_{\mathrm{\mathrm{g}}}(\bm{k}_6) \rangle
&= \langle\delta^{\mathrm{\mathrm{s}}}_{\mathrm{\mathrm{g}}}(\bm{k}_1)\delta^{\mathrm{\mathrm{s}}}_{\mathrm{\mathrm{g}}}(\bm{k}_2)\rangle_{\mathrm{c}} \langle\delta^{\mathrm{\mathrm{s}}}_{\mathrm{\mathrm{g}}}(\bm{k}_3)\delta^{\mathrm{\mathrm{s}}}_{\mathrm{\mathrm{g}}}(\bm{k}_4)\rangle_{\mathrm{c}}
\langle\delta^{\mathrm{\mathrm{s}}}_{\mathrm{\mathrm{g}}}(\bm{k}_5)\delta^{\mathrm{\mathrm{s}}}_{\mathrm{\mathrm{g}}}(\bm{k}_6)\rangle_{\mathrm{c}} +\,14\,\mathrm{perms.}\notag\\
&+\langle\delta^{\mathrm{\mathrm{s}}}_{\mathrm{\mathrm{g}}}(\bm{k}_1)\delta^{\mathrm{\mathrm{s}}}_{\mathrm{\mathrm{g}}}(\bm{k}_2)\delta^{\mathrm{\mathrm{s}}}_{\mathrm{\mathrm{g}}}(\bm{k}_3)\rangle_{\mathrm{c}}
\langle\delta^{\mathrm{\mathrm{s}}}_{\mathrm{\mathrm{g}}}(\bm{k}_4)\delta^{\mathrm{\mathrm{s}}}_{\mathrm{\mathrm{g}}}(\bm{k}_5)\delta^{\mathrm{\mathrm{s}}}_{\mathrm{\mathrm{g}}}(\bm{k}_6)\rangle_{\mathrm{c}} +\,9\,\mathrm{perms.}\notag\\
&+\langle\delta^{\mathrm{\mathrm{s}}}_{\mathrm{\mathrm{g}}}(\bm{k}_1)\delta^{\mathrm{\mathrm{s}}}_{\mathrm{\mathrm{g}}}(\bm{k}_2)\delta^{\mathrm{\mathrm{s}}}_{\mathrm{\mathrm{g}}}(\bm{k}_3)\delta^{\mathrm{\mathrm{s}}}_{\mathrm{\mathrm{g}}}(\bm{k}_4)\rangle_{\mathrm{c}}\langle\delta^{\mathrm{\mathrm{s}}}_{\mathrm{\mathrm{g}}}(\bm{k}_5)\delta^{\mathrm{\mathrm{s}}}_{\mathrm{\mathrm{g}}}(\bm{k}_6)\rangle_{\mathrm{c}}+\,14\,\mathrm{perms.}\notag\\
&+\langle\delta^{\mathrm{\mathrm{s}}}_{\mathrm{\mathrm{g}}}(\bm{k}_1)\delta^{\mathrm{\mathrm{s}}}_{\mathrm{\mathrm{g}}}(\bm{k}_2)\delta^{\mathrm{\mathrm{s}}}_{\mathrm{\mathrm{g}}}(\bm{k}_3)\delta^{\mathrm{\mathrm{s}}}_{\mathrm{\mathrm{g}}}(\bm{k}_4)\delta^{\mathrm{\mathrm{s}}}_{\mathrm{\mathrm{g}}}(\bm{k}_5)\delta^{\mathrm{\mathrm{s}}}_{\mathrm{\mathrm{g}}}(\bm{k}_6)\rangle_{\mathrm{c}}.
\end{align}

\noindent Below we present the full derivation of the of the bispectrum Covariance  matrix to leading order in Gaussian terms

\begin{align}
&\mathrm{C}^{\mathrm{BB}}\left(\bar{\bm{k}}_1, \bar{\bm{k}}_2,\bar{\bm{k}}_3;\bar{\bm{k}}_4, \bar{\bm{k}}_5,\bar{\bm{k}}_6\right)= 
\notag \\
&=\dfrac{(2\pi)^4\Delta k^\parallel_3\Delta k^\parallel_6}{V^2_{\mathrm{s}}}\Lambda^{-1}_{123}\,\Lambda^{-1}_{456}\prod_{i=1}^6\int_i\dfrac{dV^{\mathrm{c}}_i}{V^{\mathrm{c}}_i}\mathrm{F}_{\mathrm{s},i}^{(1)}\delta_{\mathrm{D}}\left(\bm{k}_1+\bm{k}_2+\bm{k}_3\right)
\delta_{\mathrm{D}}\left(\bm{k}_4+\bm{k}_5+\bm{k}_6\right)
\langle\delta_{\mathrm{m}}(\bm{k}_1)\delta_{\mathrm{m}}(\bm{k}_4)\rangle\langle\delta_{\mathrm{m}}(\bm{k}_2)\delta_{\mathrm{m}}(\bm{k}_5)\rangle\langle\delta_{\mathrm{m}}(\bm{k}_3)\delta_{\mathrm{m}}(\bm{k}_6)\rangle \,+\,5\,\mathrm{p.}
\notag \\
&=\dfrac{(2\pi)^{13}{\Delta k^\parallel_3}^2}{V^2_{\mathrm{s}}V^{\mathrm{c}}_1V^{\mathrm{c}}_2V^{\mathrm{c}}_3}\Lambda^{-2}_{123}\,\delta^{\mathrm{K}}_{14}\,\delta^{\mathrm{K}}_{25}\,\delta^{\mathrm{K}}_{36}
\prod_{i=1}^3\int_i\mathrm{F}_{\mathrm{s},i}^{(1)2}\mathrm{P}^{\mathrm{m}}(k_i)\,\delta_{\mathrm{D}}\left(\bm{k}_1+\bm{k}_2+\bm{k}_3\right)^2 \,+\,5\,\mathrm{p.}
\notag\\
&=\dfrac{(2\pi)^{10}{\Delta k^\parallel_3}^2}{V_{\mathrm{\mathrm{s}}}\,V^{\mathrm{c}}_1V^{\mathrm{c}}_2V^{\mathrm{c}}_3}\Lambda^{-2}_{123}\,\mathrm{D}_{123456}\prod_{i=1}^3\int_i\mathrm{P}^{\mathrm{\mathrm{s}}}_{\mathrm{\mathrm{g}}}(\bm{k}_i)\,\delta_{\mathrm{D}}\left(\bm{k}_1+\bm{k}_2+\bm{k}_3\right)
\notag \\ 
&=\dfrac{(2\pi)^{11}{\Delta k^\parallel_3}^2}{V_{\mathrm{\mathrm{s}}}\,V^{\mathrm{c}}_1V^{\mathrm{c}}_2V^{\mathrm{c}}_3}\Lambda^{-1}_{123}\,\mathrm{D}_{123456}\prod_{i=1}^3\int_{\bar{k}^\perp - \frac{\Delta k}{2}}^{\bar{k}^\perp + \frac{\Delta k}{2}}\int_{\bar{k}^\parallel - \frac{\Delta k}{2}}^{\bar{k}^\parallel + \frac{\Delta k}{2}}\dfrac{dk^\perp_i dk^\parallel_i\,k_i^\perp}{V^{\mathrm{c}}_i}
\mathrm{P}^{\mathrm{\mathrm{s}}}_{\mathrm{\mathrm{g}}}(\bm{k}_i)\delta_{\mathrm{D}}\left(k^\parallel_1+k^\parallel_2+k^\parallel_3\right)
\notag\\
&=\dfrac{(2\pi)^{5}\Delta k^{\parallel}_3}{V_{\mathrm{s}}\,\bar{k}^\perp_1\bar{k}^\perp_2\bar{k}^\perp_3\Delta k^6}\Lambda^{-1}_{123}\,\mathrm{D}_{123456}\prod_{i=1}^3\int_{\bar{k}^\perp - \frac{\Delta k}{2}}^{\bar{k}^\perp + \frac{\Delta k}{2}}\dfrac{dk^\perp_i \,k_i^\perp}{\bar{k}^\perp_i\Delta k^\perp_i}
\prod_{j=1}^2\int_{\bar{k}^\parallel - \frac{\Delta k}{2}}^{\bar{k}^\parallel + \frac{\Delta k}{2}}\dfrac{ dk^\parallel_j}{\Delta k^\parallel_j}\mathrm{P}^{\mathrm{\mathrm{s}}}_{\mathrm{\mathrm{g}}}(\bm{k}_i),
\notag\\
\end{align}

\noindent where $\mathrm{D}_{123456}$ has been used as a shorthand notation for all the possible permutations allowed by the fact that only pairs formed by wave-vectors from different triplets survive, which are in total 6. For all the other steps we have used relations described previously. Finally, defining a bin average by using the thin shell approximation, it is possible to write down the final result

\begin{eqnarray}
&\mathrm{C}^{\mathrm{BB}}=\dfrac{(2\pi)^{5}\Delta k^\parallel_3}{V\,\bar{k}^\perp_1\bar{k}^\perp_2\bar{k}^\perp_3\Delta k^6}\Lambda^{-1}_{123}\,\mathrm{D}_{123456}\prod_{i=1}\mathrm{F}_{\mathrm{s}}^{(1)2}\left[\bar{k}_i^\perp, \bar{k}_i^\parallel\right]\mathrm{P}^{\mathrm{m}}(\bar{k}^\perp_i,\bar{k}^\parallel_i).
\end{eqnarray}

\subsection{Cross - variance term: $\mathrm{\textbf{C}^{\mathrm{BP}}}$}

\noindent The cross-correlation part of the covariance matrix of the joint data-vector $\left[\mathrm{P}^{\mathrm{\mathrm{s}}}_{\mathrm{\mathrm{g}}}, \mathrm{B}^{\mathrm{\mathrm{s}}}_{\mathrm{\mathrm{g}}}\right]$ can be computed as

\begin{eqnarray}
\label{cov_exp_pk_bk}
\mathrm{C^{\mathrm{BP}}}&\equiv\mathrm{\textbf{Cov}}\left[\mathrm{P}^{\mathrm{\mathrm{s}}}_{\mathrm{\mathrm{g}}}\left(\bar{\bm{k}}_1\right),\mathrm{B}^{\mathrm{\mathrm{s}}}_{\mathrm{\mathrm{g}}}\left(\bar{\bm{k}}_3,\bar{\bm{k}}_4,\bar{\bm{k}}_5\right)\right] 
=\langle\hat{\mathrm{P}}^{\mathrm{\mathrm{s}}}_{\mathrm{g},1}\hat{\mathrm{B}}^{\mathrm{\mathrm{s}}}_{\mathrm{g},345}\rangle-\hat{\mathrm{P}}^{\mathrm{\mathrm{s}}}_{\mathrm{g},1}\hat{\mathrm{B}}^{\mathrm{\mathrm{s}}}_{\mathrm{g},345}.
\end{eqnarray}

\noindent The computation of the covariance  matrix involves a five points correlator of $\delta^{\mathrm{\mathrm{s}}}_{\mathrm{\mathrm{g}}}$ that can be expanded in its connected parts

\begin{align}
\langle\delta^{\mathrm{\mathrm{s}}}_{\mathrm{\mathrm{g}}}(\bm{k}_1)\delta^{\mathrm{\mathrm{s}}}_{\mathrm{\mathrm{g}}}(\bm{k}_2)\delta^{\mathrm{\mathrm{s}}}_{\mathrm{\mathrm{g}}}(\bm{k}_3)\delta^{\mathrm{\mathrm{s}}}_{\mathrm{\mathrm{g}}}(\bm{k}_4)\delta^{\mathrm{\mathrm{s}}}_{\mathrm{\mathrm{g}}}(\bm{k}_5) \rangle 
&=\langle\delta^{\mathrm{\mathrm{s}}}_{\mathrm{\mathrm{g}}}(\bm{k}_1)\delta^{\mathrm{\mathrm{s}}}_{\mathrm{\mathrm{g}}}(\bm{k}_2)\rangle_{\mathrm{c}} \langle\delta^{\mathrm{\mathrm{s}}}_{\mathrm{\mathrm{g}}}(\bm{k}_3)\delta^{\mathrm{\mathrm{s}}}_{\mathrm{\mathrm{g}}}(\bm{k}_4)\delta^{\mathrm{\mathrm{s}}}_{\mathrm{\mathrm{g}}}(\bm{k}_5)\rangle_{\mathrm{c}}
\notag\\
&+\langle\delta^{\mathrm{\mathrm{s}}}_{\mathrm{\mathrm{g}}}(\bm{k}_1)\delta^{\mathrm{\mathrm{s}}}_{\mathrm{\mathrm{g}}}(\bm{k}_3)\rangle_{\mathrm{c}}\langle\delta^{\mathrm{\mathrm{s}}}_{\mathrm{\mathrm{g}}}(\bm{k}_2)\delta^{\mathrm{\mathrm{s}}}_{\mathrm{\mathrm{g}}}(\bm{k}_4)\delta^{\mathrm{\mathrm{s}}}_{\mathrm{\mathrm{g}}}(\bm{k}_5)\rangle_{\mathrm{c}} \quad + \quad 5\,\mathrm{p.}
\notag \\
&+\langle\delta^{\mathrm{\mathrm{s}}}_{\mathrm{\mathrm{g}}}(\bm{k}_3)\delta^{\mathrm{\mathrm{s}}}_{\mathrm{\mathrm{g}}}(\bm{k}_4)\rangle_{\mathrm{c}}\langle\delta^{\mathrm{\mathrm{s}}}_{\mathrm{\mathrm{g}}}(\bm{k}_1)\delta^{\mathrm{\mathrm{s}}}_{\mathrm{\mathrm{g}}}(\bm{k}_2)\delta^{\mathrm{\mathrm{s}}}_{\mathrm{\mathrm{g}}}(\bm{k}_5)\rangle_{\mathrm{c}} \quad + \quad 2\,\mathrm{p.}
\notag \\
&+\langle\delta^{\mathrm{\mathrm{s}}}_{\mathrm{\mathrm{g}}}(\bm{k}_3)\delta^{\mathrm{\mathrm{s}}}_{\mathrm{\mathrm{g}}}(\bm{k}_4)\delta^{\mathrm{\mathrm{s}}}_{\mathrm{\mathrm{g}}}(\bm{k}_1)\delta^{\mathrm{\mathrm{s}}}_{\mathrm{\mathrm{g}}}(\bm{k}_2)\delta^{\mathrm{\mathrm{s}}}_{\mathrm{\mathrm{g}}}(\bm{k}_5)\rangle_{\mathrm{c}}
\notag \\
&+\langle\delta^{\mathrm{\mathrm{s}}}_{\mathrm{\mathrm{g}}}(\bm{k}_1)\delta^{\mathrm{\mathrm{s}}}_{\mathrm{\mathrm{g}}}(\bm{k}_2)\delta^{\mathrm{\mathrm{s}}}_{\mathrm{\mathrm{g}}}(\bm{k}_3)\delta^{\mathrm{\mathrm{s}}}_{\mathrm{\mathrm{g}}}(\bm{k}_4)\delta^{\mathrm{\mathrm{s}}}_{\mathrm{\mathrm{g}}}(\bm{k}_5) \rangle_{\rm{unc.}}
\notag\\
&= \mathrm{C^{\mathrm{BP}}_{m0}} + \mathrm{C^{\mathrm{BP}}_{m1}} + \mathrm{C^{\mathrm{BP}}_{m2}}+ \mathrm{C^{\mathrm{BP}}_{Te}} + \mathrm{C^{\mathrm{BP}}_{U_{\rm{5p}}}}.
\end{align}

\noindent In the case of the five points correlation function it has to be included also the unconnected part in the expansion. This is because while for an even number $2n$ the leading term of unconnected part of the $2n$-points correlation function is proportional to the product of $n$ power spectra, for an odd number $m$ this is no longer the case. As shown in last subsection of Appendix \ref{sec:est_def}, the unconnected part of the 5-points correlation function has a leading term proportional to the product of three power spectra. Therefore in this case the unconnected part needs to be considered since it has the same order of the other leading terms of the above expansion.
From the last equation, it is possible to immediately that $\mathrm{C^{\mathrm{BP}}}_{\mathrm{m0}}$ cancels out with the last term o Eq. \eqref{cov_exp_pk_bk} and that  $\mathrm{C^{\mathrm{BP}}}_\mathrm{{m2}} = 0$ because the Dirac's deltas combination leads to terms having $\delta_{\mathrm{D}}\left(\bm{k}_5\right) = 0$.

\subsubsection{$\mathrm{\textbf{C}^{\mathrm{BP}}_{m1}}$ term:}
Starting from the first term written previously, below we report the full derivation

\begin{align}
&C^{\mathrm{BP}}_{\mathrm{m1}}\left(\bar{\bm{k}}_1; \bar{\bm{k}}_3,\bar{\bm{k}}_4,\bar{\bm{k}}_5\right) = 
\notag\\
& = \dfrac{(2\pi)^2\Delta k_5^\parallel}{V^2_{\mathrm{s}}}\Lambda^{-1}_{345}\,\int dV_1^{\rm{c}}\prod_{i=2}^5\int \dfrac{dV^{\mathrm{c}}_i}{V^{\mathrm{c}}_i}\,\delta_{\mathrm{D}}\left(\bm{k}_1+\bm{k}_2\right)
\delta_{\mathrm{D}}\left(\bm{k}_3 + \bm{k}_4 + \bm{k}_5\right)
 \langle\delta^{\mathrm{\mathrm{s}}}_{\mathrm{\mathrm{g}}}(\bm{k}_1)\delta^{\mathrm{\mathrm{s}}}_{\mathrm{\mathrm{g}}}(\bm{k}_3)\rangle_{\mathrm{c}}\langle\delta^{\mathrm{\mathrm{s}}}_{\mathrm{\mathrm{g}}}(\bm{k}_2)
\delta^{\mathrm{\mathrm{s}}}_{\mathrm{\mathrm{g}}}(\bm{k}_4) \delta^{\mathrm{\mathrm{s}}}_{\mathrm{\mathrm{g}}}(\bm{k}_5)\rangle_{\mathrm{c}}\, +\, 5\,\mathrm{p.} 
\notag \\
& = \dfrac{(2\pi)^8\Delta k_5^\parallel}{V^2_{\mathrm{s}}}\Lambda^{-1}_{345}\,\int dV_1^{\rm{c}}\prod_{i=2}^5\int \dfrac{dV^{\mathrm{c}}_i}{V^{\mathrm{c}}_i}\,\delta_{\mathrm{D}}\left(\bm{k}_1+\bm{k}_2\right)
\delta_{\mathrm{D}}\left(\bm{k}_3 + \bm{k}_4 + \bm{k}_5\right)
\delta_{\mathrm{D}}\left(\bm{k}_1+\bm{k}_3\right)\delta_{\mathrm{D}}\left(\bm{k}_2 + \bm{k}_4 + \bm{k}_5\right)
\mathrm{P}^{\mathrm{\mathrm{s}}}_{\mathrm{\mathrm{g}}}\left(\bm{k}_3\right)\mathrm{B}^{\mathrm{\mathrm{s}}}_{\mathrm{\mathrm{g}}}\left(\bm{k}_2,\bm{k}_4,\bm{k}_5\right)\, +\, 5\,\mathrm{p.} 
\notag \\
& = \dfrac{(2\pi)^5\Delta k_5^\parallel}{V_{\mathrm{s}}}\Lambda^{-1}_{345}\,\int dV_1^{\rm{c}}\prod_{i=2}^5\int \dfrac{dV^{\mathrm{c}}_i}{V^{\mathrm{c}}_i}\,\delta_{\mathrm{D}}\left(\bm{k}_1+\bm{k}_2\right)\delta_{\mathrm{D}}\left(\bm{k}_1+\bm{k}_3\right) 
 \delta_{\mathrm{D}}\left(\bm{k}_3 + \bm{k}_4 + \bm{k}_5\right)
\mathrm{P}^{\mathrm{\mathrm{s}}}_{\mathrm{\mathrm{g}}}\left(\bm{k}_3\right)\mathrm{B}^{\mathrm{\mathrm{s}}}_{\mathrm{\mathrm{g}}}\left(\bm{k}_2,\bm{k}_4,\bm{k}_5\right)\, +\, 5\,\mathrm{p.}
\notag \\
& = \dfrac{(2\pi)^5\Delta k_5^\parallel}{V_{\mathrm{s}}}\Lambda^{-1}_{345}\,\delta^{\mathrm{K}}_{12}\,\prod_{i=2}^5\int \dfrac{dV^{\mathrm{c}}_i}{V^{\mathrm{c}}_i}\,\delta_{\mathrm{D}}\left(-\bm{k}_2+\bm{k}_3\right) 
\delta_{\mathrm{D}}\left(\bm{k}_3 + \bm{k}_4 + \bm{k}_5\right)
\mathrm{P}^{\mathrm{\mathrm{s}}}_{\mathrm{\mathrm{g}}}\left(\bm{k}_3\right)\mathrm{B}^{\mathrm{\mathrm{s}}}_{\mathrm{\mathrm{g}}}\left(\bm{k}_2,\bm{k}_4,\bm{k}_5\right)\, +\, 5\,\mathrm{p.}
\notag \\
& = \dfrac{(2\pi)^5\Delta k_5^\parallel}{V_{\mathrm{\mathrm{s}}}\,V_2^{\rm{c}}}\Lambda^{-1}_{345}\,\delta^{\mathrm{K}}_{12}\,\delta^{\mathrm{K}}_{-23}\,\prod_{i=3}^5\int \dfrac{dV^{\mathrm{c}}_i}{V^{\mathrm{c}}_i}\,
\delta_{\mathrm{D}}\left(\bm{k}_3 + \bm{k}_4 + \bm{k}_5\right)
\mathrm{P}^{\mathrm{\mathrm{s}}}_{\mathrm{\mathrm{g}}}\left(\bm{k}_3\right)\mathrm{B}^{\mathrm{\mathrm{s}}}_{\mathrm{\mathrm{g}}}\left(\bm{k}_3,\bm{k}_4,\bm{k}_5\right)\, +\, 5\,\mathrm{p.}
\notag \\
& =  \dfrac{(2\pi)^5\Delta k_5^\parallel}{V_{\mathrm{\mathrm{s}}}\,V_2^{\rm{c}}}\Lambda^{-1}_{345}\,\delta^{\mathrm{K}}_{12}\,\delta^{\mathrm{K}}_{-23}\,\prod_{i=3}^5\int \dfrac{dk_i^\parallel dk_i^\perp k_i^\perp}{2\pi\bar{k}_i^\perp\Delta k^2}
\,2\pi\, \Lambda_{345}
\delta_{\mathrm{D}}\left(k_3^\parallel+k_4^\parallel+k_5^\parallel\right) \mathrm{P}^{\mathrm{\mathrm{s}}}_{\mathrm{\mathrm{g}}}\left(\bm{k}_3\right)\mathrm{B}^{\mathrm{\mathrm{s}}}_{\mathrm{\mathrm{g}}}\left(\bm{k}_3,\bm{k}_4,\bm{k}_5\right)\,+\, 5\,\mathrm{p.}
\notag\\
& \approx \dfrac{(2\pi)^3}{V_{\mathrm{\mathrm{s}}}\,V_2^{\rm{c}}}\,\delta^{\mathrm{K}}_{-23}\,
\mathrm{P}^{\mathrm{\mathrm{s}}}_{\mathrm{\mathrm{g}}}\left(\bar{\bm{k}}_3\right)\mathrm{B}^{\mathrm{\mathrm{s}}}_{\mathrm{\mathrm{g}}}\left(\bar{\bm{k}}_3,\bar{\bm{k}}_4,\bar{\bm{k}}_5\right)\, +\,5\,\mathrm{p.}.
\end{align}

\noindent Note that only three of the six permutations are different terms, since the other three are just obtained by switching $\bm{k}_1$ with $\bm{k}_2$ which does not change the final result. Therefore the final expression for this term can be written as

\begin{align}
C^{\mathrm{BP}}_{\mathrm{m1}}& =2\times  \dfrac{(2\pi)^3}{V_{\mathrm{\mathrm{s}}}\,V_2^{\rm{c}}}\,\left(\delta^{\mathrm{K}}_{-13} + \delta^{\mathrm{K}}_{-14} + \delta^{\mathrm{K}}_{-15}\right)\,
\mathrm{P}^{\mathrm{\mathrm{s}}}_{\mathrm{\mathrm{g}}}\left(\bar{\bm{k}}_1\right)\mathrm{B}^{\mathrm{\mathrm{s}}}_{\mathrm{\mathrm{g}}}\left(\bar{\bm{k}}_3,\bar{\bm{k}}_4,\bar{\bm{k}}_5\right).
\end{align}

\noindent Note that the argument of the power spectrum depend on the wave-vector selected by the Kronecker Delta.

\subsubsection{$\mathrm{\textbf{C}^{\mathrm{BP}}_{Te}}$ term:}

The other contribution at tree level to the connected five points correlator resulting in the cross covariance term between power spectrum and bispectrum is the one proportional to the tetraspectrum defined before, we get

\begin{align}
&C^{\mathrm{BP}}_{\mathrm{\mathrm{U}_{\rm{5p}}}} \left(\bar{\bm{k}}_2; \bar{\bm{k}}_3,\bar{\bm{k}}_4,\bar{\bm{k}}_5\right) = \dfrac{(2\pi)^2\Delta k_5^\parallel}{V^2_{\mathrm{s}}}\Lambda^{-1}_{345}\,\int dV_1^{\rm{c}}\prod_{i=2}^5\int \dfrac{dV^{\mathrm{c}}_i}{V^{\mathrm{c}}_i}\,\delta_{\mathrm{D}}\left(\bm{k}_1+\bm{k}_3\right)\delta_{\mathrm{D}}\left(\bm{k}_3 + \bm{k}_4 + \bm{k}_5\right)
\langle\delta^{\mathrm{\mathrm{s}}}_{\mathrm{\mathrm{g}}}(\bm{k}_1)\delta^{\mathrm{\mathrm{s}}}_{\mathrm{\mathrm{g}}}(\bm{k}_3)\delta^{\mathrm{\mathrm{s}}}_{\mathrm{\mathrm{g}}}(\bm{k}_2)
\delta^{\mathrm{\mathrm{s}}}_{\mathrm{\mathrm{g}}}(\bm{k}_4) \delta^{\mathrm{\mathrm{s}}}_{\mathrm{\mathrm{g}}}(\bm{k}_5)\rangle_{\mathrm{c}} 
\notag \\
& = \dfrac{(2\pi)^5\Delta k_5^\parallel}{V^2_{\mathrm{s}}}\Lambda^{-1}_{345}\,\int dV_1^{\rm{c}}\prod_{i=2}^5\int \dfrac{dV^{\mathrm{c}}_i}{V^{\mathrm{c}}_i}\,\delta_{\mathrm{D}}\left(\bm{k}_1+\bm{k}_2\right)\delta_{\mathrm{D}}\left(\bm{k}_3 + \bm{k}_4 + \bm{k}_5\right)
\delta_{\mathrm{D}}\left(\bm{k}_1+\bm{k}_2 +\bm{k}_3 + \bm{k}_4 + \bm{k}_5\right)
\mathrm{T}^{\mathrm{\mathrm{s}}}_{e,g}\left(\bm{k}_1,\bm{k}_2,\bm{k}_3,\bm{k}_4,\bm{k}_5\right)
\notag \\
& = \dfrac{(2\pi)^5\Delta k_5^\parallel}{V^2_{\mathrm{s}}}\Lambda^{-1}_{345}\,\delta^{\mathrm{K}}_{12}\prod_{i=2}^5\int \dfrac{dV^{\mathrm{c}}_i}{V^{\mathrm{c}}_i}\,\delta_{\mathrm{D}}\left(\bm{k}_3 + \bm{k}_4 + \bm{k}_5\right)^2\mathrm{T}^{\mathrm{\mathrm{s}}}_{e,g}\left(\bm{k}_2,\bm{k}_3,\bm{k}_4,\bm{k}_5\right)
\notag \\
& = \dfrac{(2\pi)^2\Delta k_5^\parallel}{V_{\mathrm{\mathrm{s}}}}\Lambda^{-1}_{345}\,\prod_{i=2}^5\int \dfrac{dV^{\mathrm{c}}_i}{V^{\mathrm{c}}_i}\,\delta_{\mathrm{D}}\left(\bm{k}_3 + \bm{k}_4 + \bm{k}_5\right)\mathrm{T}^{\mathrm{\mathrm{s}}}_{e,g}\left(\bm{k}_2,\bm{k}_3,\bm{k}_4,\bm{k}_5\right).
\end{align}

\noindent Since the leading order of the tetraspectrum is higher than the one considered in this work, the above covariance term has not been included in the numerical computation. For what concerns the unconnected part of the five point correlation function described in the end of Appendix \ref{sec:est_def}, it represents loop correction to the standard power spectrum times bispectrum cross covariance $C^{\mathrm{BP}}_{\mathrm{m1}}$. Coherently with what done in the same case for the trispectrum it has also not been included in the numerical computations. 

\section{Compressed Covariance  matrix}

\label{sec:comp_cov}
We run the MCMC not on the original data-vector, e.g. $\mathrm{B}^{\mathrm{\mathrm{s}}}_{\mathrm{\mathrm{g}}}$ or $\left[\mathrm{P}^{\mathrm{\mathrm{s}}}_{\mathrm{\mathrm{g}}} + \mathrm{B}^{\mathrm{\mathrm{s}}}_{\mathrm{\mathrm{g}}}\right]$ but on their compressed version obtained through Karhunen-Loève weighting.
Given a set of parameters $\left\{\theta_1, \dots  \theta_m\right\}$, a data-vector $\bm{x} = \left\{x_1,\dots x_n\right\}$ with mean $\langle\bm{x}\rangle$ the set of weights with respect to each parameter $\theta_i$ is given by

\begin{eqnarray}
\bm{b}_{i}\,=\,\mathrm{\textbf{Cov}}^{-1}\langle\bm{x}\rangle_{,i},
\end{eqnarray}

\noindent where $\mathrm{\textbf{Cov}}(\bm{x})$ is the covariance  matrix relative to the data-vector $\bm{x}$. Assuming the covariance  matrix to be approximately independent of the cosmology, then for each considered parameter the original data-vector is compressed to a single scalar

\begin{eqnarray}
y_{\theta_i} \, = \, \bm{b}_{i} \cdot \bm{x}.
\end{eqnarray}

\noindent Therefore stacking all the weighting vectors as rows of an $m \times n$  matrix $\bm{B}$ the new $m$-dimensional data-vector will be

\begin{eqnarray}
\bm{y} \, =\, \bm{B} \bm{x}.
\end{eqnarray}

\noindent For what concerns the covariance  matrix for the new data-vector, it will be related to the original one by

\begin{eqnarray}
\label{comp_cov}
\mathrm{\textbf{Cov}}_{\bm{y},ij}\,=\,\mathrm{\textbf{Cov}}\left[y_i,y_j\right]
&\,=\,\mathrm{\textbf{Cov}}\left[\sum_k^n b^k_ix_k\,,\,\sum_j^n b^l_j x_l\right]
\notag \\
&\,=\,\sum_k^n\sum_j^n b^k_ib^l_j \mathrm{\textbf{Cov}}\left[x_k, x_l\right]
\notag \\
&\,=\, \bm{b}_i^\intercal \cdot \mathrm{\textbf{Cov}}_{\bm{x}} \cdot \bm{b}_j,
\end{eqnarray}

\noindent Therefore when running the MCMC using $\bm{y}$ as data-vector the natural logarithm of likelihood will be proportional to 

\begin{eqnarray}
\log \mathcal{L} \,\propto \, - \dfrac{1}{2} \left(\bm{y} - \bm{y}_{\mathrm{fid.}}\right)^\intercal
\mathrm{\textbf{Cov}}_{\bm{y}}^{-1} \left(\bm{y} - \bm{y}_{\mathrm{fid.}}\right).
\end{eqnarray}

\noindent Where $\bm{y}_{\mathrm{fid.}}$ is the compressed data-vector obtained using the fiducial values of the cosmological parameters. In our case of interest this will be applied to $\mathrm{B}^{\mathrm{\mathrm{s}}}_{\mathrm{\mathrm{g}}}$ and $\left[\mathrm{P}^{\mathrm{\mathrm{s}}}_{\mathrm{\mathrm{g}}},\mathrm{B}^{\mathrm{\mathrm{s}}}_{\mathrm{\mathrm{g}}}\right]$. An additional option consists in leaving $\mathrm{P}^{\mathrm{\mathrm{s}}}_{\mathrm{\mathrm{g}}}$ uncompressed, compressing only the bispectrum. 
In that case the cross-variance term would be given by

\begin{eqnarray}
\mathrm{\textbf{Cov}}^{Py_B}_{ij}\,=\,\mathrm{\textbf{Cov}}\left[\mathrm{P}^{\mathrm{\mathrm{s}}}_{\mathrm{\mathrm{g}}}(\bm{k}_i), y_j \right]
\,=\,\mathrm{\textbf{Cov}}\left[\mathrm{P}^{\mathrm{\mathrm{s}}}_{\mathrm{\mathrm{g}}}(\bm{k}_i), \bm{\mathrm{B}}^{\mathrm{\mathrm{s}}}_{\mathrm{\mathrm{g}}} \right] \cdot \bm{b}_j.
\end{eqnarray}

\noindent Putting everything together the full covariance  matrix is

\begin{eqnarray}
\mathrm{\textbf{Cov}} \,&= \,
\begin{bmatrix}
\mathrm{\textbf{Cov}}^{\bm{P}\bm{P}}  & \mathrm{\textbf{Cov}}^{\bm{P}\bm{y}_B}\\
\mathrm{\textbf{Cov}}^{\bm{P}\bm{y}_B}  & \mathrm{\textbf{Cov}}^{\bm{y}_B\bm{y}_B} \\
\end{bmatrix},
\end{eqnarray}

\noindent where $\mathrm{\textbf{Cov}}^{\bm{y}_B\bm{y}_B}$ is the covariance  matrix for the compressed data-vector derived in eq. \eqref{comp_cov} in the case of $\bm{x}$ being the Bispectrum.

\section{Weights orthogonalisation}
\label{sec:weights_ort}

As stated in \citet{Zablocki:2015zcm} each compressed data set $y_i$ contains all the information regarding the parameter $\theta_i$ but at the same time it will have some sensitivity to the other parameters. It is possible to remove this sensitivity by marginalising at linear level over these other parameters. This is done by orthogonalising the weighting vectors for all the individual parameters. If the model is parametrised by $m$ parameters $\theta_i$, compressing with respect to all these will return $m$ different linear combinations $y_i$ (or in other words $m$ weighting vectors). Therefore it is possible to define another set of new scalars $y_i'$ given by a linear combination of the original ones, for example:

\begin{eqnarray}
\label{new_y}
y_1' \,=\,c_1y_1\,+\,c_2y_2\,+\,...\,+\,c_my_m,
\end{eqnarray}

\noindent with the constants $c_i$ such that $y_1'$ contains all the information regarding $\theta_1$ while at the same time removing all the sensitivity with respect to the other parameters at linear level. This means that when taking the derivative with respect to the second parameter $\theta_2$ we are looking for $c$ coefficients such that:

\begin{align}
\dfrac{dy_1'}{d\theta_2} & = c_1 \dfrac{dy_1}{d\theta_2} +  c_2 \dfrac{dy_2}{d\theta_2} + ... + c_m \dfrac{dy_m}{d\theta_2}
\notag \\
& = c_1 \sum_j b^j_1 \dfrac{dx_j}{d\theta_2} + c_2 \sum_j b^j_2 \dfrac{dx_j}{d\theta_2} + ... + c_m \sum_j b^j_m \dfrac{dx_j}{d\theta_2}
\notag \\
& = c_1 \mathrm{F}_{12} + c_2 \mathrm{F}_{22} + ... + c_m \mathrm{F}_{m2} = 0,
\end{align}

\noindent where for the Fisher information matrix elements it has been used Eq. \ref{fmatrix_comp} together with the expression for the weights  $\bm{b}_i = \mathrm{\textbf{Cov}}^{-1} \langle\bm{x}\rangle_{,i}$ . Imposing this for the derivative with respect to all the $m-1$ parameters beside $\theta_1$ it gives the following matrix problem to be solved in order to get the $m-1$ coefficients needed to compute $y_1'$:

\begin{eqnarray}
\begin{bmatrix}
\mathrm{F}_{22} & \mathrm{F}_{23} & . & . & . &\mathrm{F}_{2m} \\
\mathrm{F}_{32} & \mathrm{F}_{33} & . & . & . &\mathrm{F}_{3m} \\
.               &              .  & . &   &   &  . \\
.               &              .  &   & . &   &  . \\
.               &              .  &   &   & . &  . \\
\mathrm{F}_{m2} & \mathrm{F}_{m3} & . & . & . &\mathrm{F}_{mm} \\
\end{bmatrix}
\,\,
\begin{bmatrix}
c_2 \\
c_3 \\
. \\
. \\
. \\ 
c_m \\
\end{bmatrix}
\,=\,
\begin{bmatrix}
-\mathrm{F}_{12} \\
-\mathrm{F}_{13} \\
. \\
. \\
. \\ 
-\mathrm{F}_{1m} \\
\end{bmatrix}
.
\end{eqnarray}

\noindent Solving this equation returns $m-1$ unique constants for the $c_i$ with $i>1$. $c_1$ can be set equal to one.
Once the $c$'s are known, $y_1'$ can be obtained using Eq. \ref{new_y}, which corresponds to defining the new weighting vector for the parameter $\theta_1$:

\begin{eqnarray}
\bm{b}_1' = \bm{b}_1 + c_2 \bm{b}_2 + ... + c_m\bm{b}_m,
\end{eqnarray}

\noindent where $\bm{b}_1'$ is the orthogonalised weighting vector for $\theta_1$.  The same procedure can be repeated for all the other parameters and relative weighting vectors. 
In general, for the mode with $i=\alpha$, the coefficients are determined by the general equation:

\begin{eqnarray}
\sum_j\mathrm{F}_{\alpha,ij}'c_j\,=\,-\mathrm{F}_{\alpha i},
\end{eqnarray}

\noindent where $\mathrm{F}_\alpha'$ is the Fisher  matrix with row and column $\alpha$ removed.

\label{lastpage}

\end{document}